%% file: main_elsarticle-template-harv_.tex
\documentclass[final,3p,times,twocolumn,authoryear]{elsarticle}

\usepackage{makecell}
\usepackage{float}
\usepackage{xurl}
\usepackage{booktabs}
\usepackage{amssymb}
\usepackage{amsmath}
\usepackage{eurosym}
\usepackage{listings}
\usepackage{xcolor}

\newif\ifrevising
\revisingfalse

\newcommand{\rv}[1]{%
  \ifrevising
    \textcolor{blue}{#1}%
  \else
    #1%
  \fi
}

\newcommand{\rvdel}[1]{}

\newcommand{\rvreplace}[2]{%
  \ifrevising
    \textcolor{blue}{#2}%
  \else
    #2%
  \fi
}

\journal{International Journal of Human - Computer Studies}

\begin{document}

\begin{frontmatter}



\title{Grounding GUI Design in Computational Psychology} 

\author[aff1]{Xianni Wang\corref{cor1}}
\cortext[cor1]{Corresponding author.}
\ead{xianni.x.wang@jyu.fi}

\author[aff2]{Javier Romero Davila}
\author[aff1]{Saku Sourulahti}
\author[aff2]{Torsten Schaub}
\author[aff1]{Jussi P. P. Jokinen}

\affiliation[aff1]{organization={University of Jyväskylä, Faculty of Information Technology},
            addressline={Mattilanniemi 2}, 
            city={Jyväskylä},
            country={Finland}}

\affiliation[aff2]{organization={University of Potsdam, Institute of Computer Science},
            addressline={An der Bahn 2, D-14476}, 
            city={Potsdam},
            country={Germany}}
            
\begin{abstract}
  Creating visually appealing user interfaces often requires extensive manual iteration.
  We propose an approach that applies answer set programming (ASP) to automatically generate and optimize UI layouts while satisfying \rvreplace{aesthetic rules}{design objectives} such as grid alignment, grouping, color harmony, and whitespace, along with designer-specified preferences.
  Our method encodes constraints on element properties and relative positioning, producing layouts that balance functional and aesthetic goals.
  We evaluate this approach in \rvreplace{two}{three} studies.
  \rvreplace{In a user study,}{Across two user studies,} participants rated \rvreplace{fully ASP-generated layouts}{layouts generated with the full ASP model} higher than both random designs and those based on simple heuristics.  
  \rvreplace{Professional designers}{Designers} reported that ASP-generated layouts supported early-stage sketching and exploration.
\end{abstract}





\end{frontmatter}


\section{Introduction}
\input{01-introduction}

\section{Related Work}
\input{02-relatedwork}

\section{Generating and Optimizing UIs with ASP}
\input{03-definition}

\section{End-User Evaluation}
\input{04-experiment1}

\section{Designer Evaluation}
\input{05-experiment2}

\section{Applications}
\input{06-applications}

\section{Discussion}
\input{07-discussion}

\section*{Acknowledgments}
This research was supported by the Research Council of Finland 
(grant number: 362080).

\bibliographystyle{elsarticle-harv}
\bibliography{references}

\clearpage
\onecolumn
\appendix

\clearpage
\section{End-User Evaluation}
\label{appendix:exp1}
\setcounter{figure}{0}

\subsection{30 Stimuli}
\label{appendix:exp1_stimuli}

\newcommand{\expstimulus}[3]{%
    \begin{figure}[H]
        \centering
        \includegraphics[
            width=0.72\linewidth,
            height=0.245\textheight,
            keepaspectratio
        ]{figures/#1}
        \caption{#2}
        \label{#3}
    \end{figure}
}

\expstimulus
    {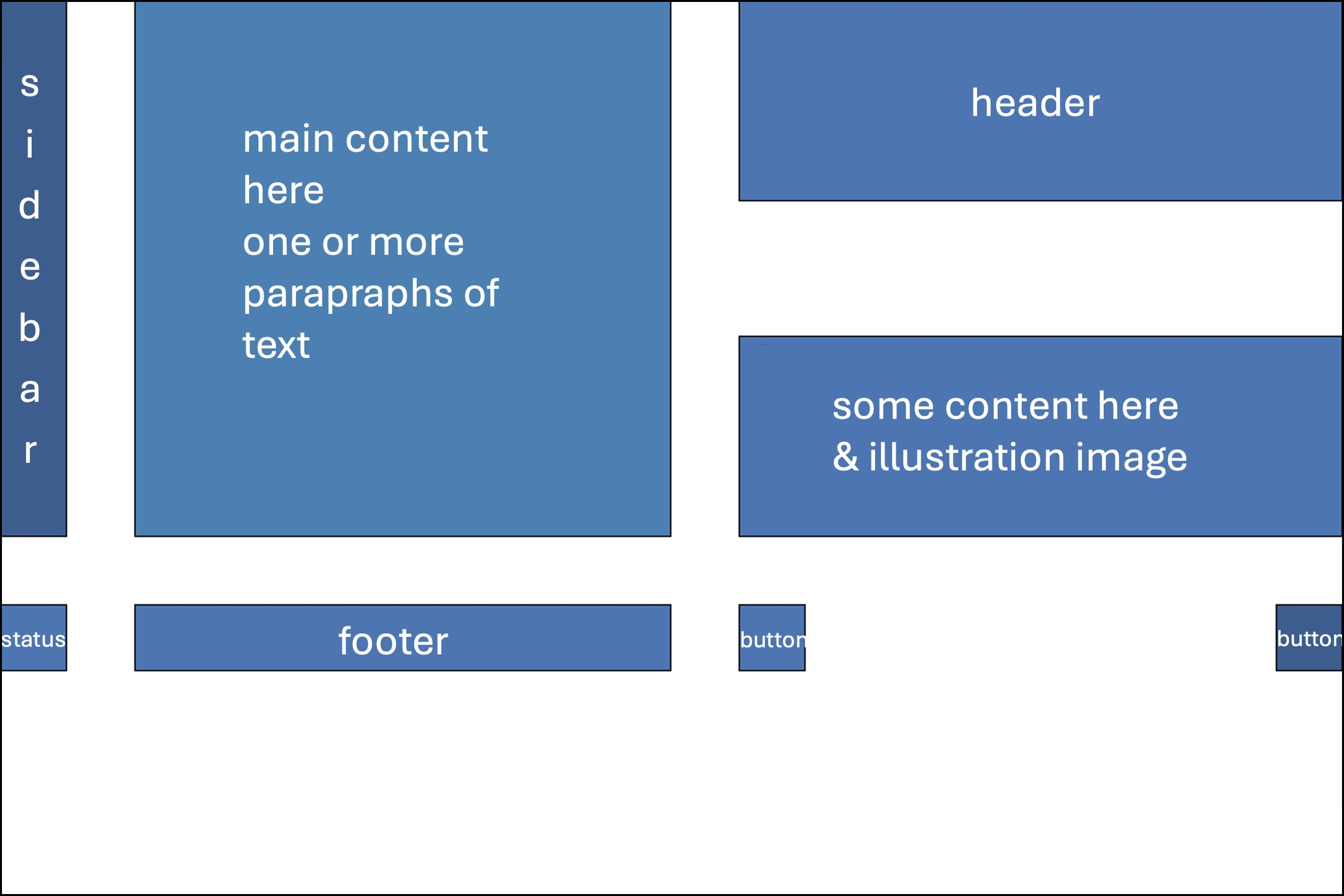}
    {Full--Complex stimulus 1 used in the end-user evaluation. Used also in the validation experiment.}
    {fig:exp1_full_complex_1}
\expstimulus
    {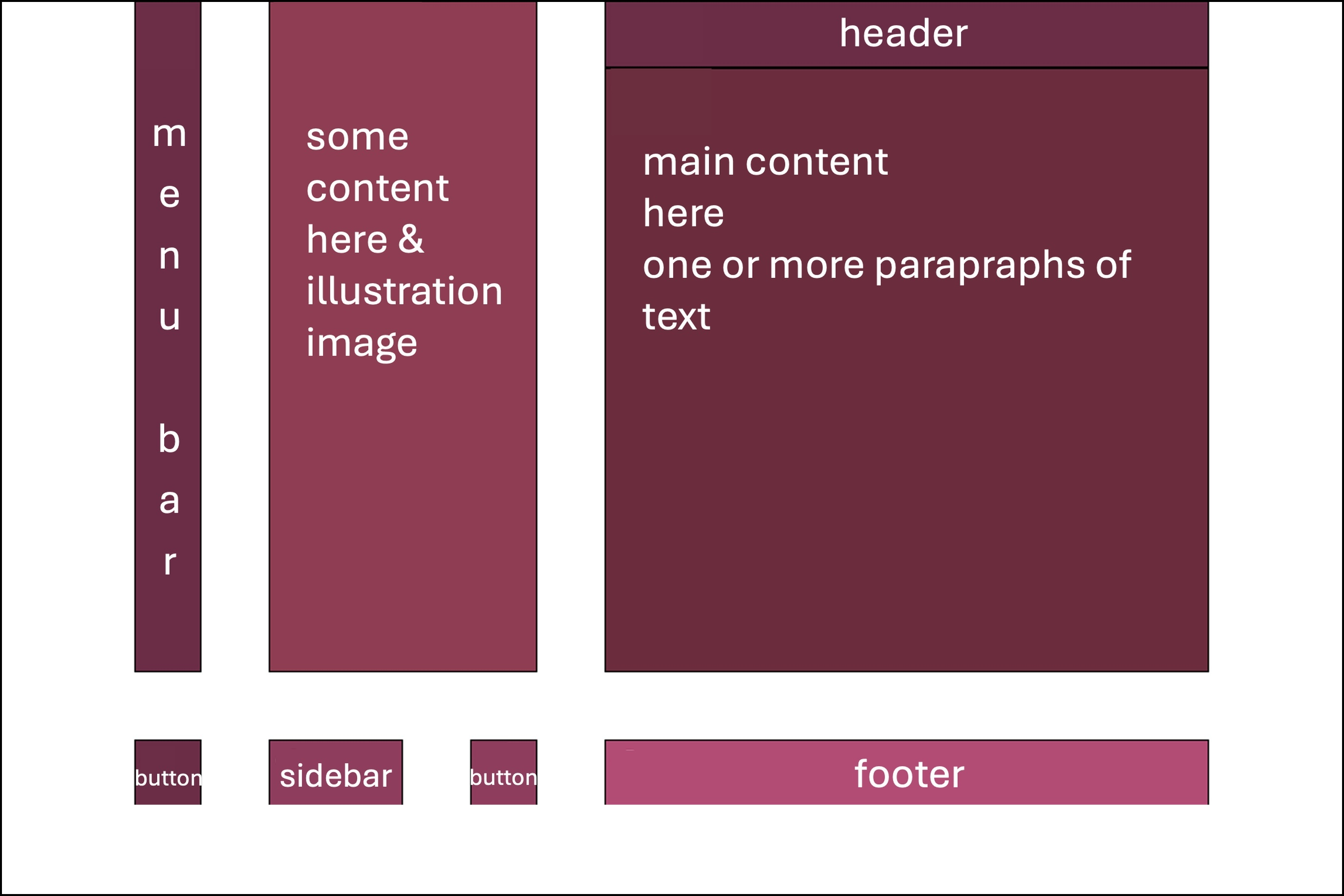}
    {Full--Complex stimulus 2 used in the end-user evaluation.}
    {fig:exp1_full_complex_2}
\expstimulus
    {fig_exp1_Full_Complex_3.png}
    {Full--Complex stimulus 3 used in the end-user evaluation. Used also in the validation experiment.}
    {fig:exp1_full_complex_3}

\expstimulus
    {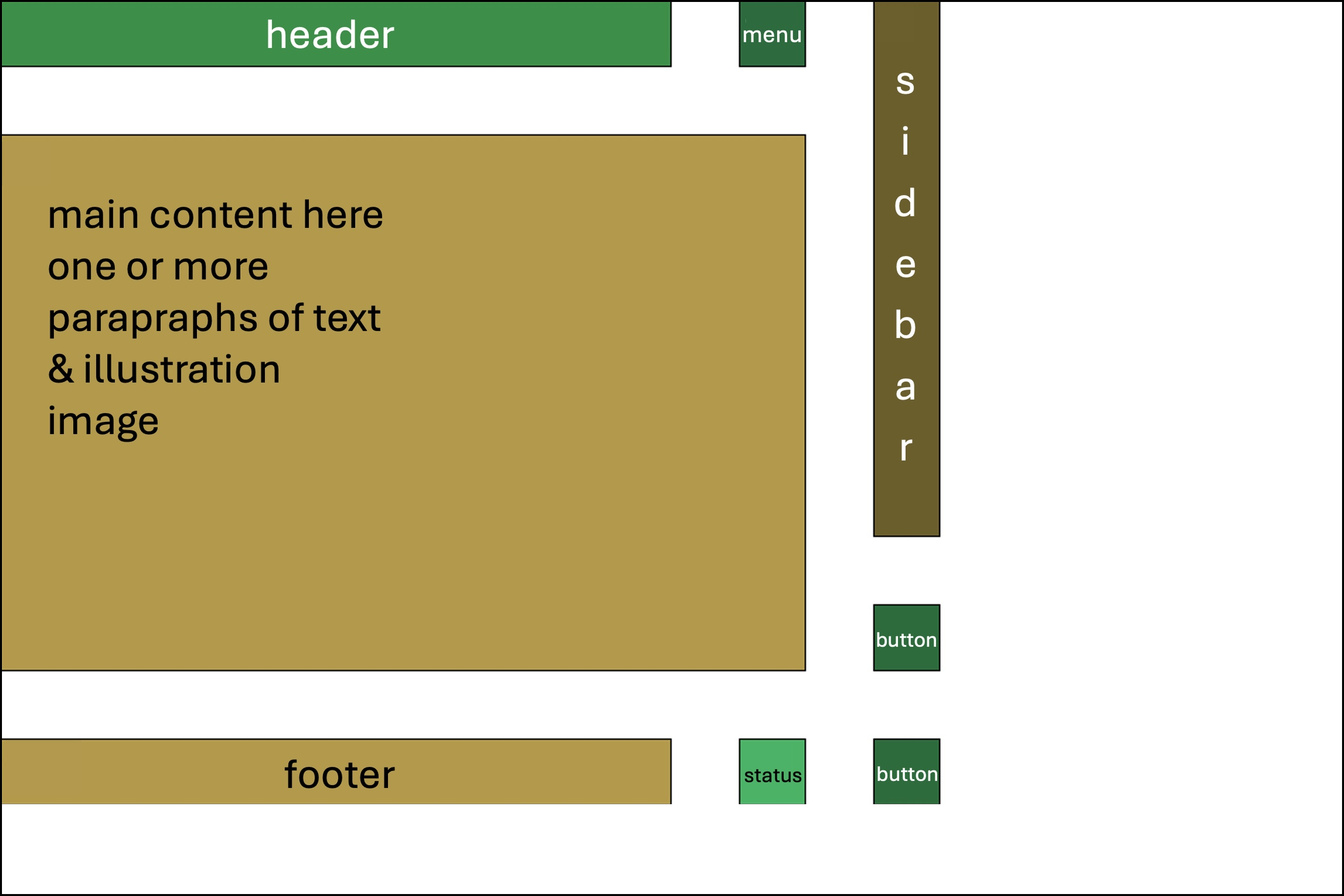}
    {Full--Complex stimulus 4 used in the end-user evaluation.}
    {fig:exp1_full_complex_4}
\expstimulus
    {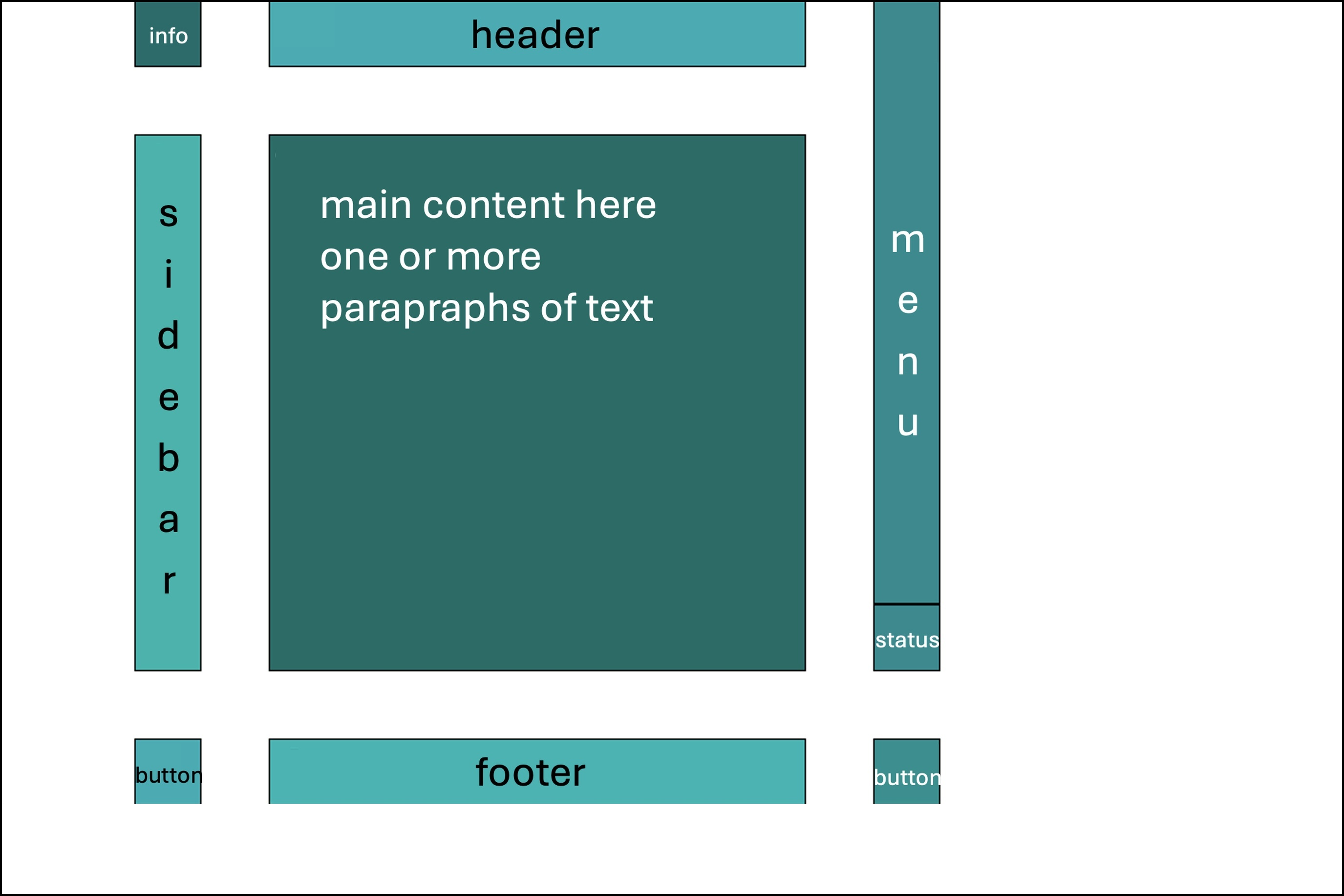}
    {Full--Complex stimulus 5 used in the end-user evaluation.}
    {fig:exp1_full_complex_5}
\expstimulus
    {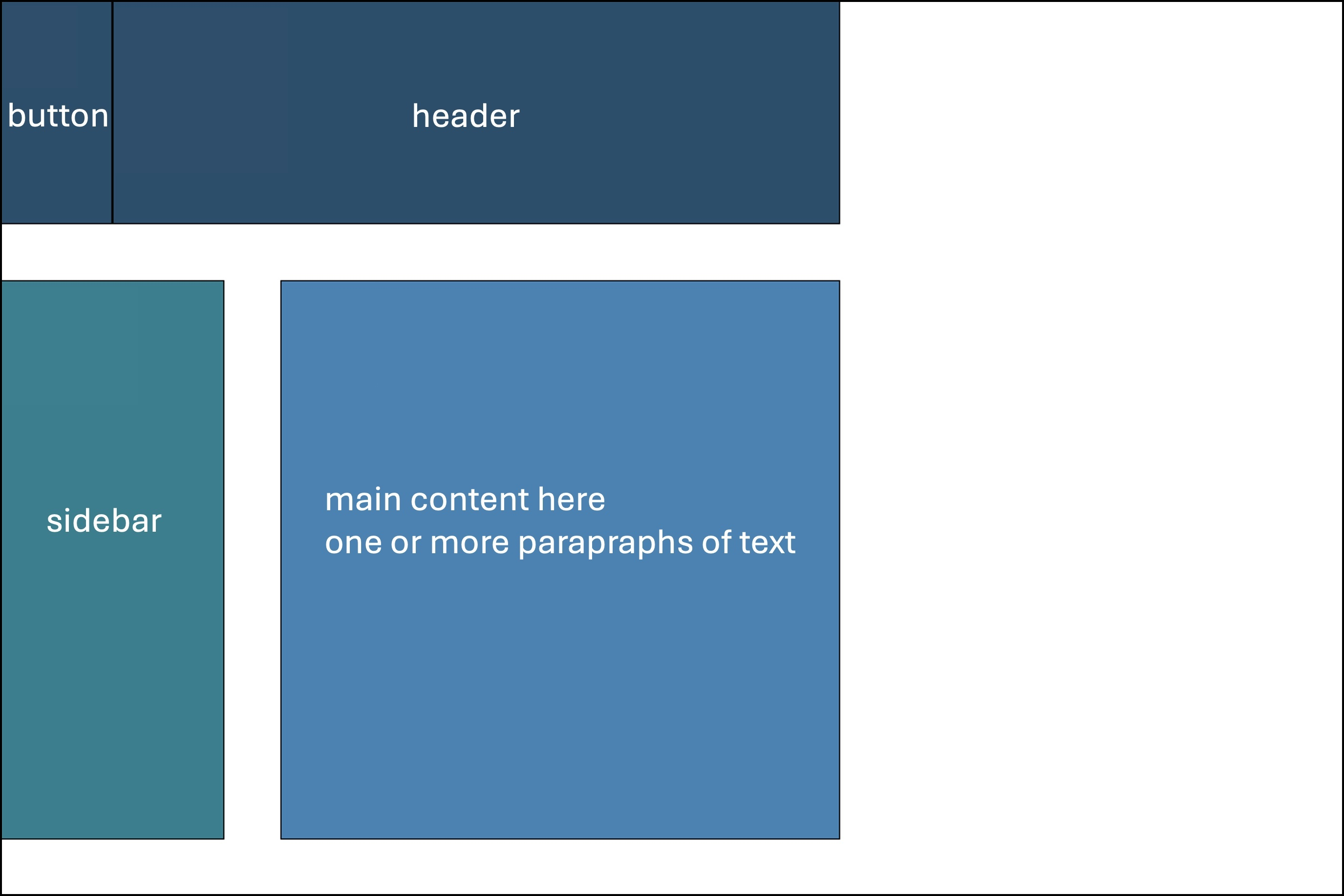}
    {Full--Simple stimulus 1 used in the end-user evaluation. Used also in the validation experiment.}
    {fig:exp1_full_simple_1}

\expstimulus
    {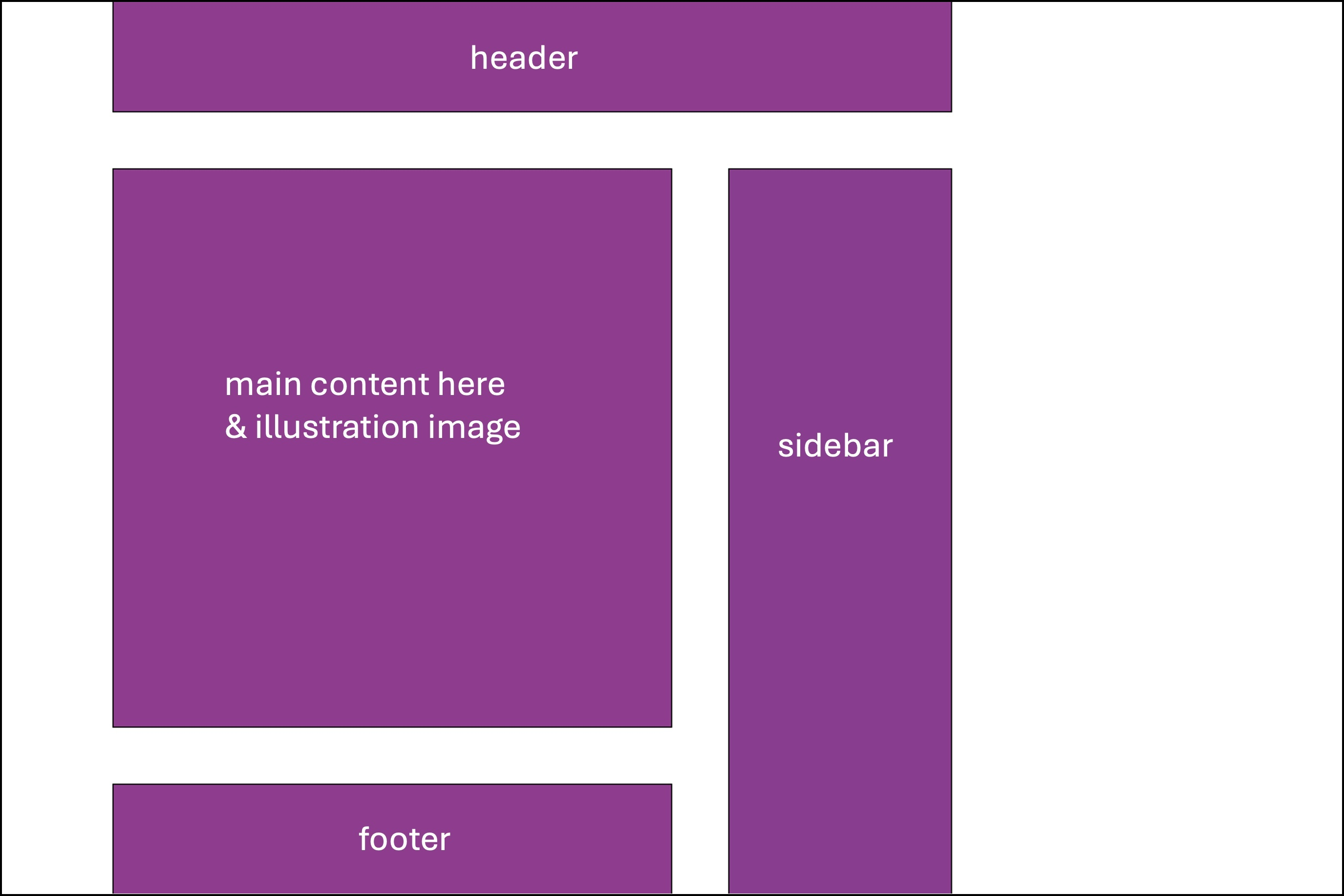}
    {Full--Simple stimulus 2 used in the end-user evaluation. Used also in the validation experiment.}
    {fig:exp1_full_simple_2}
\expstimulus
    {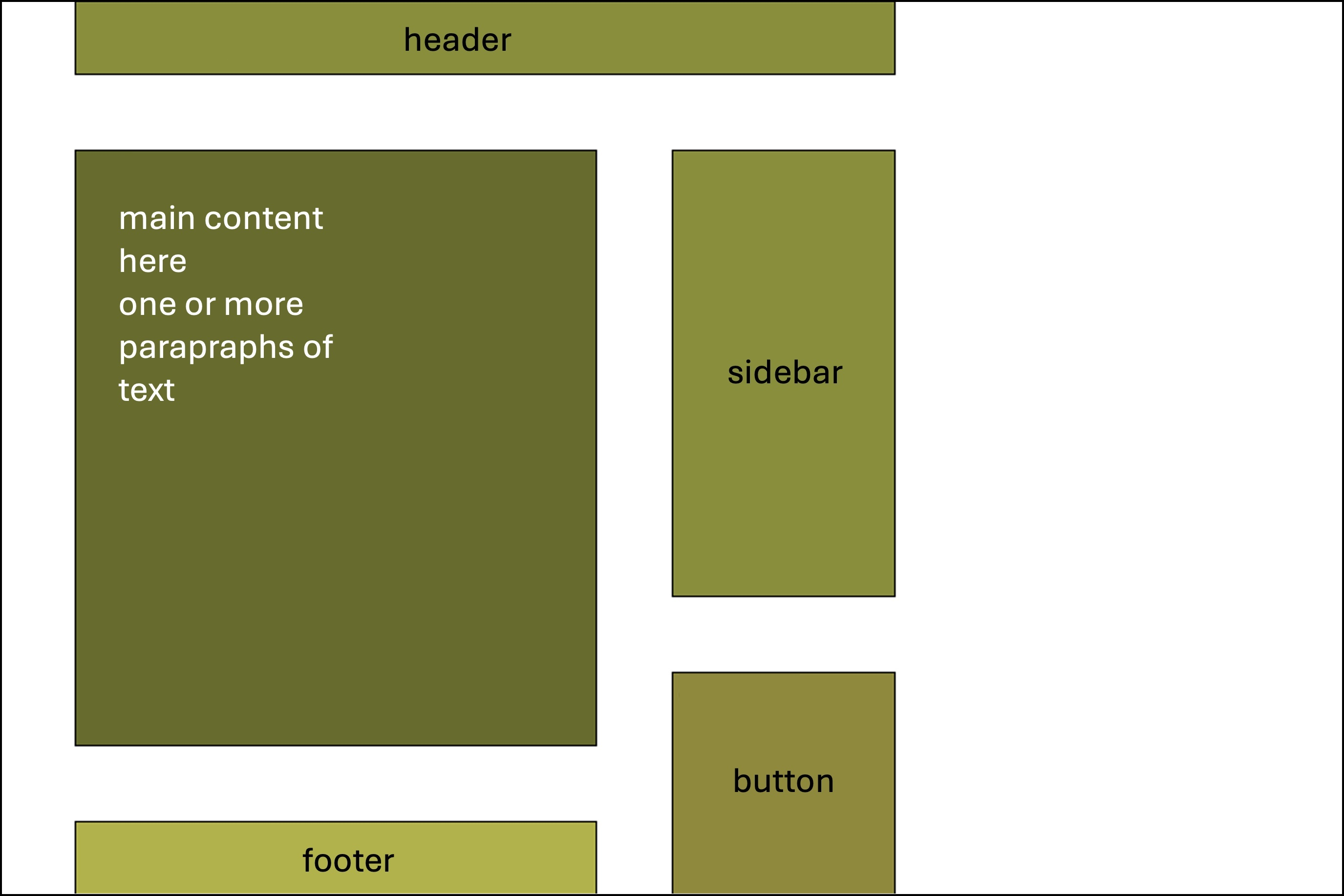}
    {Full--Simple stimulus 3 used in the end-user evaluation.}
    {fig:exp1_full_simple_3}
\expstimulus
    {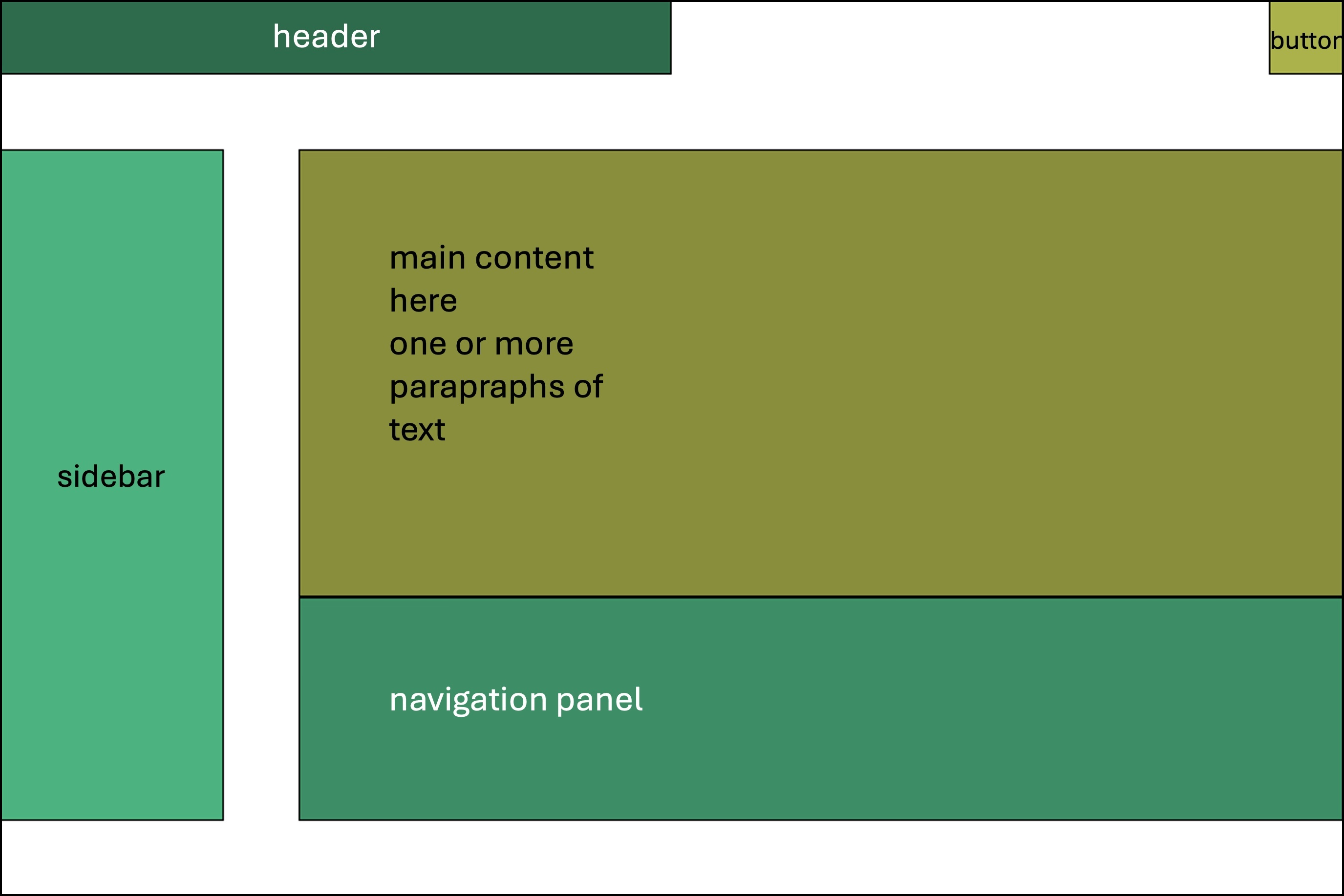}
    {Full--Simple stimulus 4 used in the end-user evaluation. Used also in the validation experiment.}
    {fig:exp1_full_simple_4}

\expstimulus
    {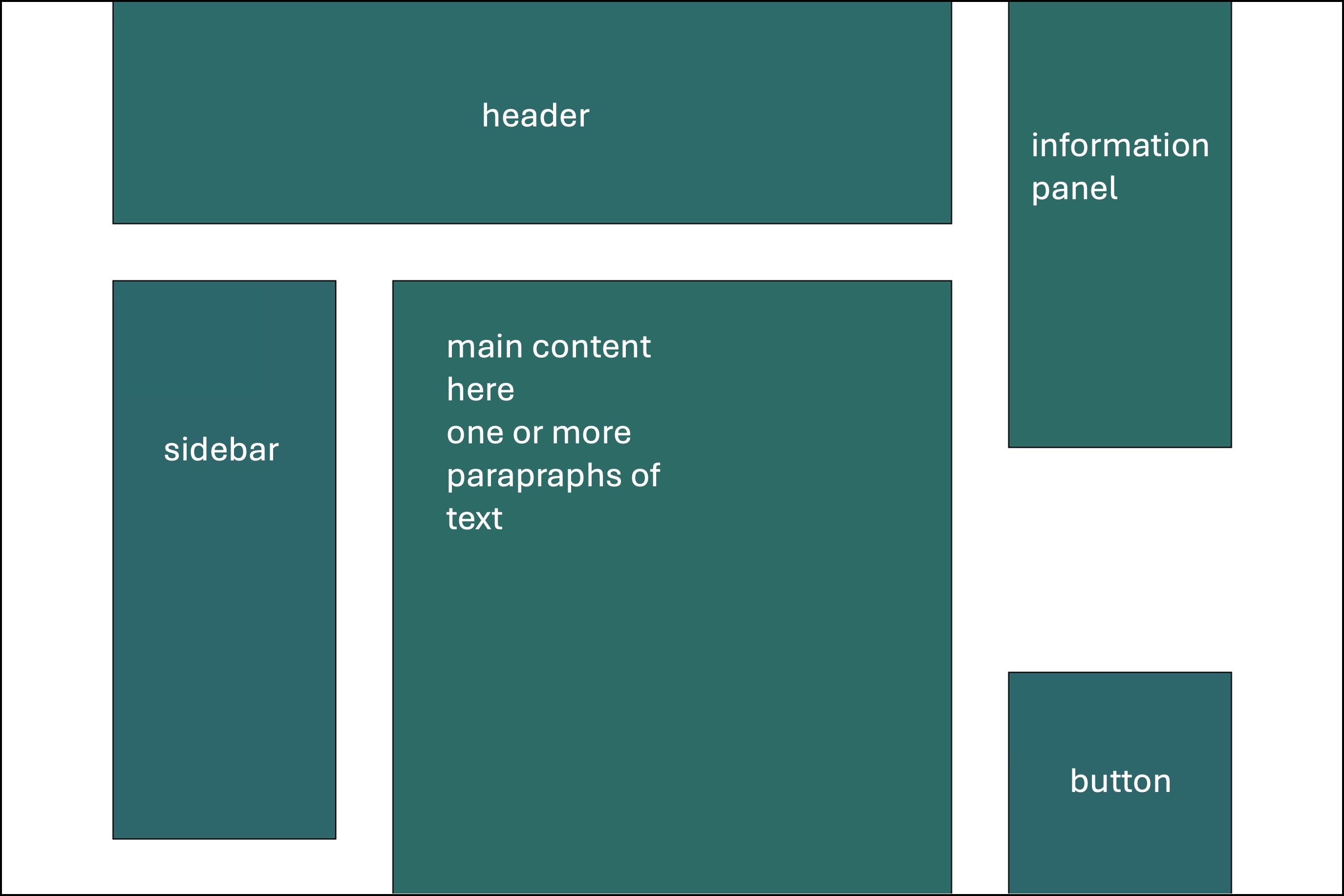}
    {Full--Simple stimulus 5 used in the end-user evaluation.}
    {fig:exp1_full_simple_5}
\expstimulus
    {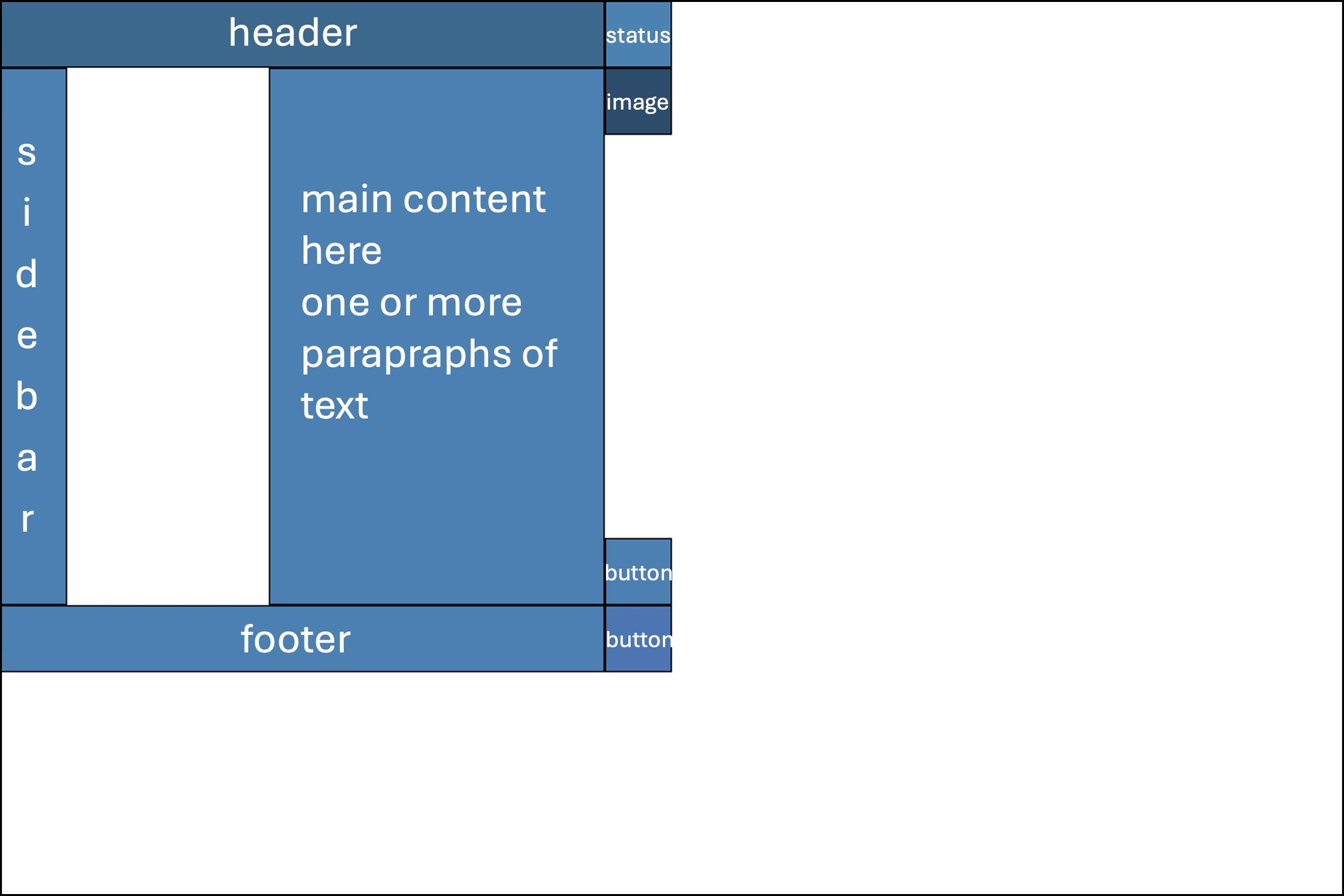}
    {Grid--Complex stimulus 1 used in the end-user evaluation.}
    {fig:exp1_reduced_complex_1}
\expstimulus
    {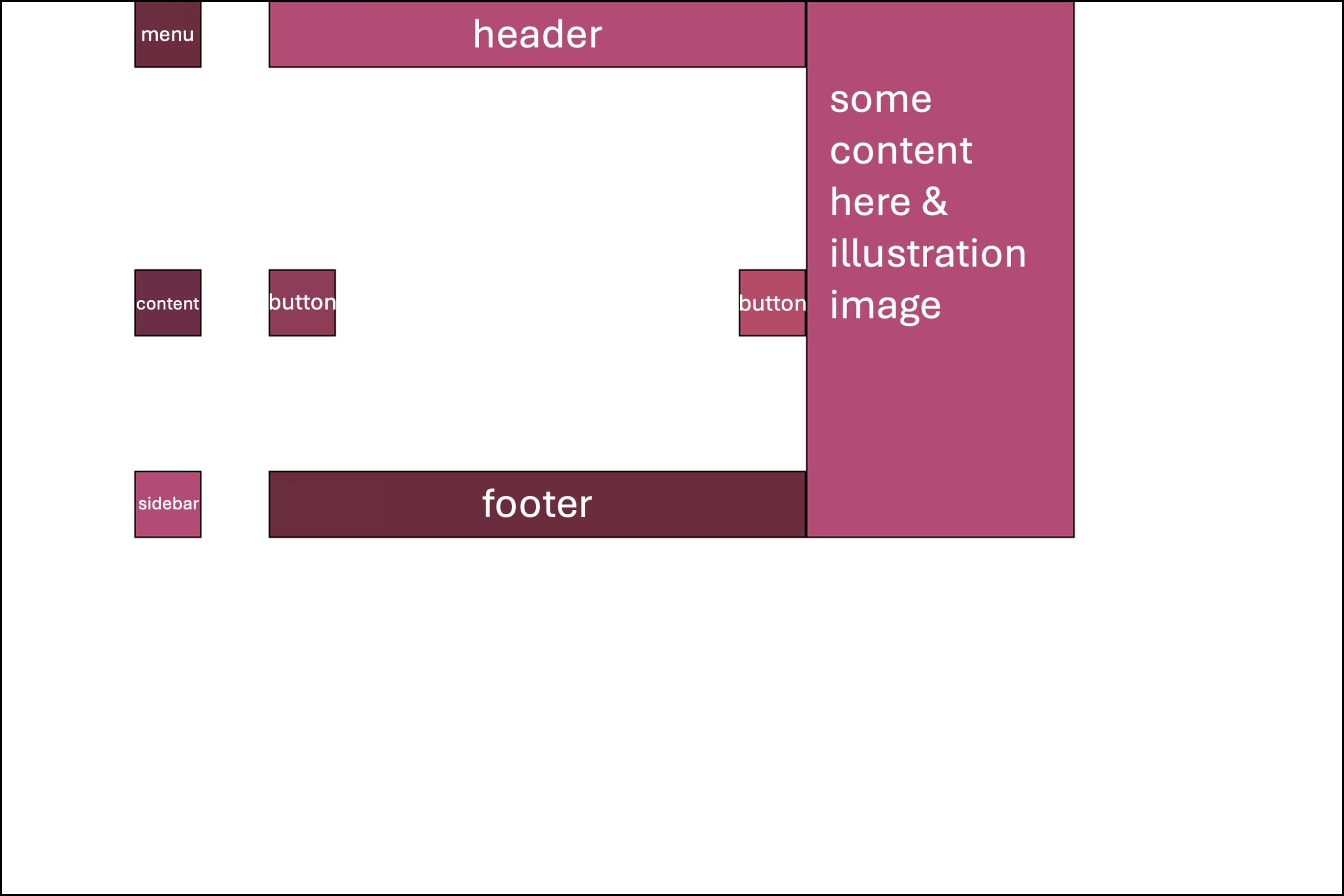}
    {Grid--Complex stimulus 2 used in the end-user evaluation.}
    {fig:exp1_reduced_complex_2}

\expstimulus
    {fig_exp1_Reduced_Complex_3.png}
    {Grid--Complex stimulus 3 used in the end-user evaluation.}
    {fig:exp1_reduced_complex_3}
\expstimulus
    {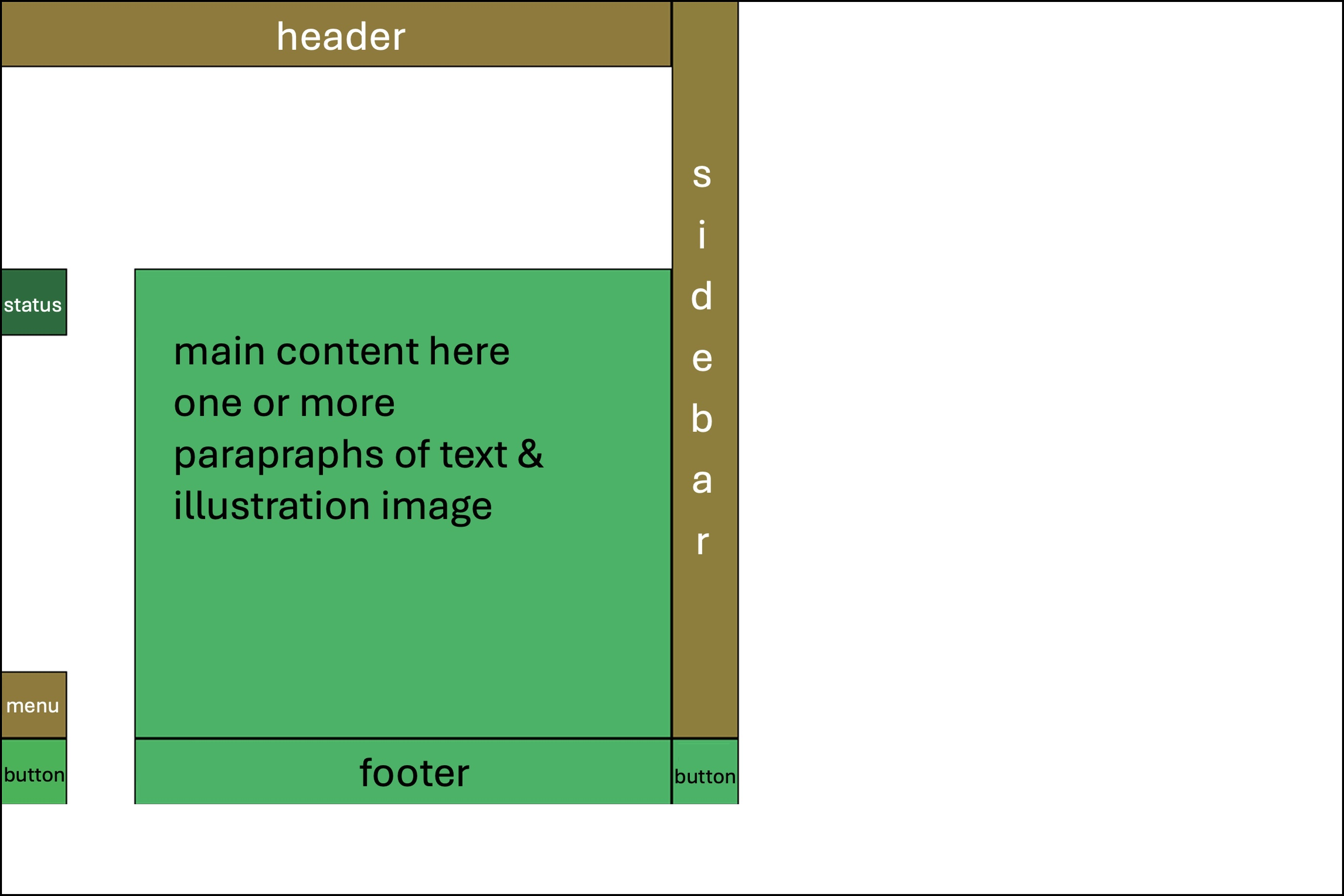}
    {Grid--Complex stimulus 4 used in the end-user evaluation. Used also in the validation experiment.}
    {fig:exp1_reduced_complex_4}
\expstimulus
    {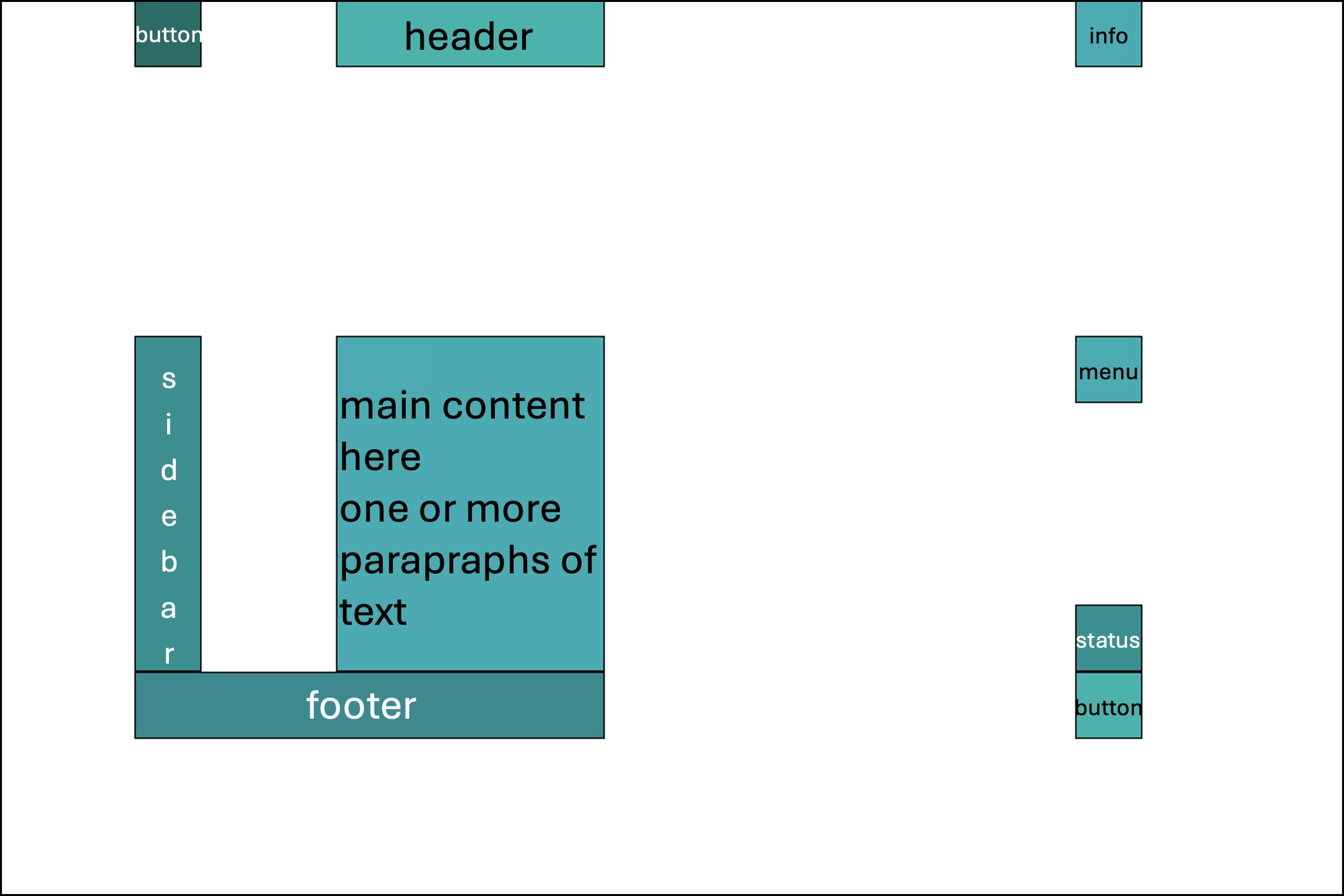}
    {Grid--Complex stimulus 5 used in the end-user evaluation. Used also in the validation experiment.}
    {fig:exp1_reduced_complex_5}

\expstimulus
    {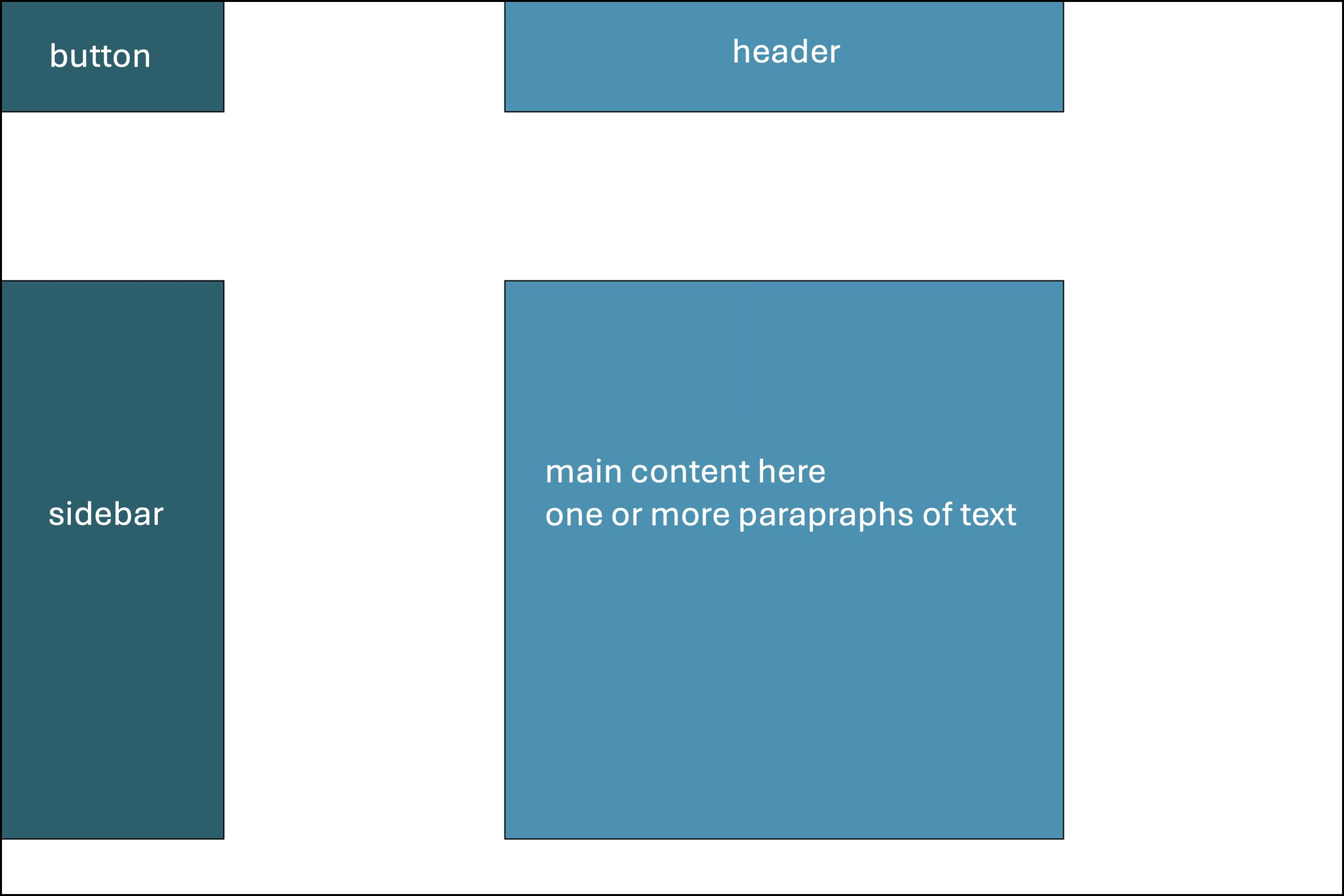}
    {Grid--Simple stimulus 1 used in the end-user evaluation. Used also in the validation experiment.}
    {fig:exp1_reduced_simple_1}
\expstimulus
    {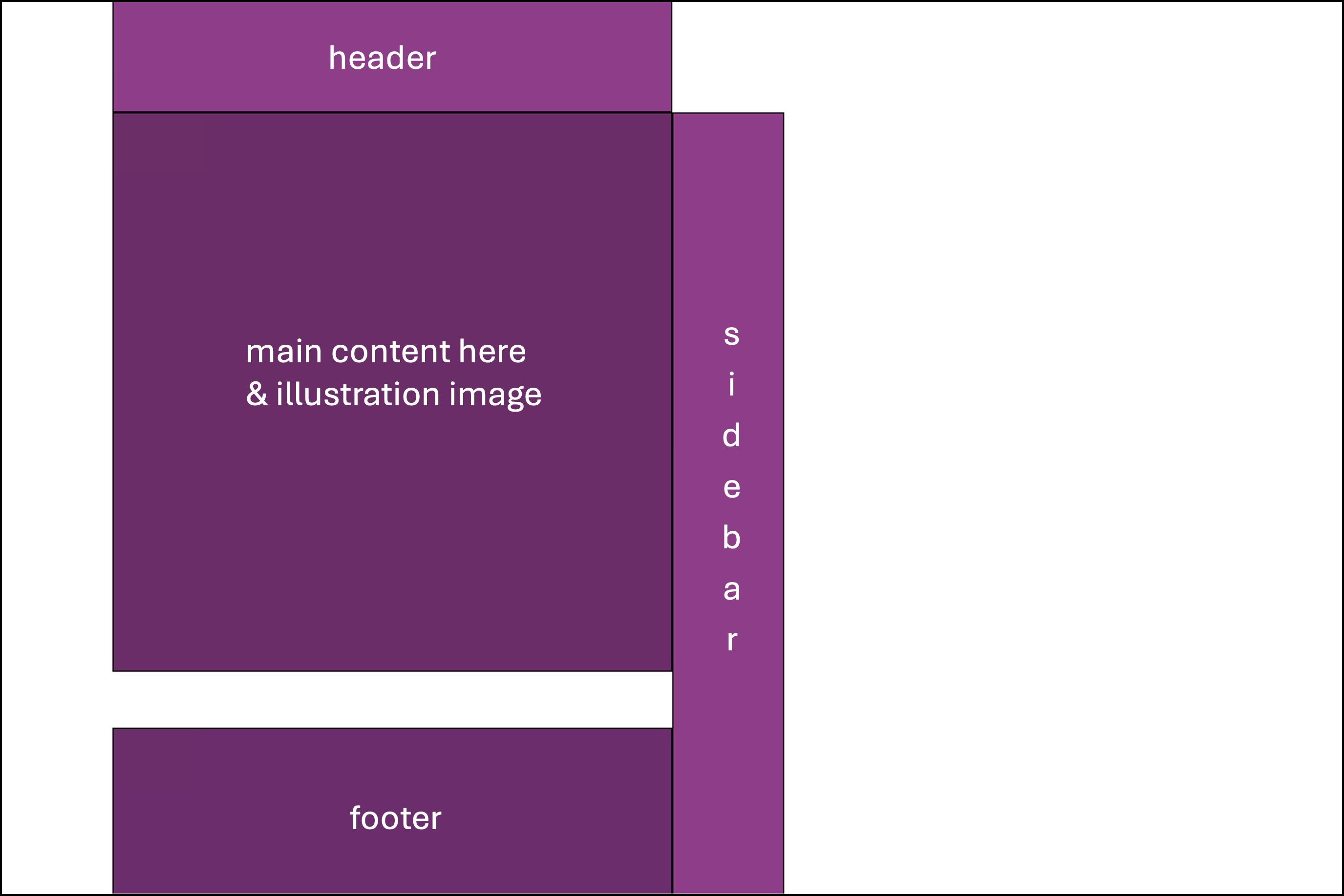}
    {Grid--Simple stimulus 2 used in the end-user evaluation.}
    {fig:exp1_reduced_simple_2}
\expstimulus
    {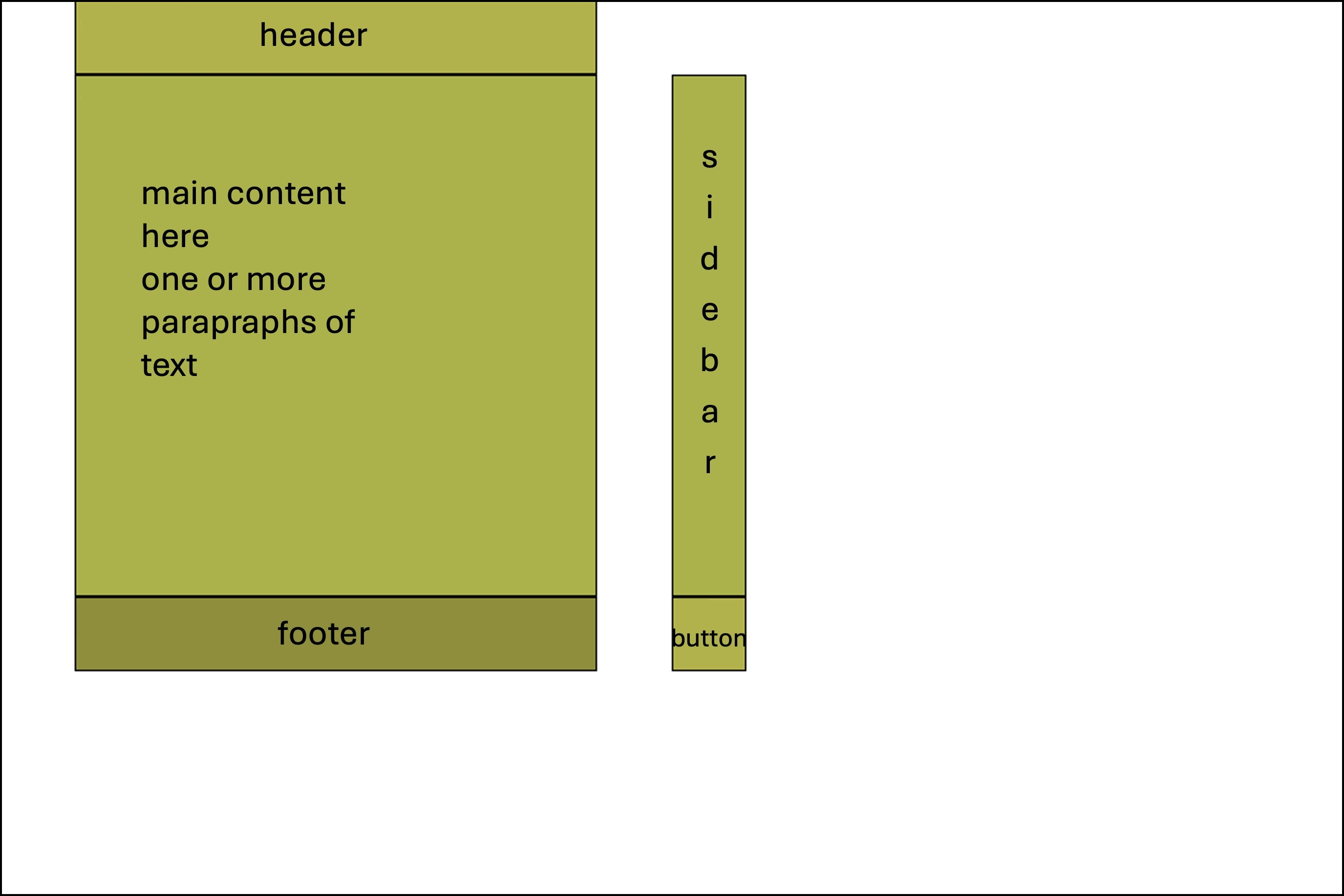}
    {Grid--Simple stimulus 3 used in the end-user evaluation.}
    {fig:exp1_reduced_simple_3}

\expstimulus
    {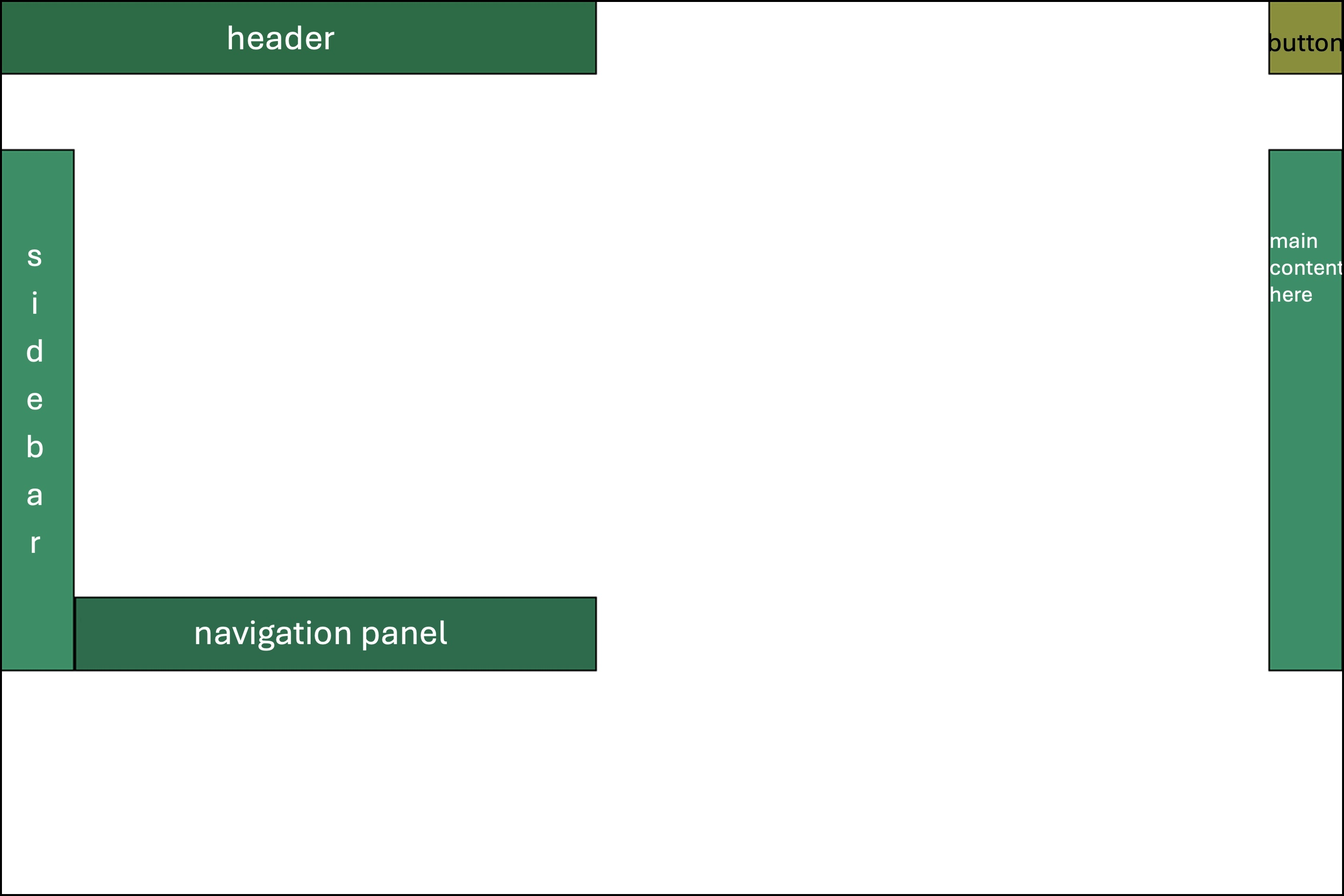}
    {Grid--Simple stimulus 4 used in the end-user evaluation. Used also in the validation experiment.}
    {fig:exp1_reduced_simple_4}
\expstimulus
    {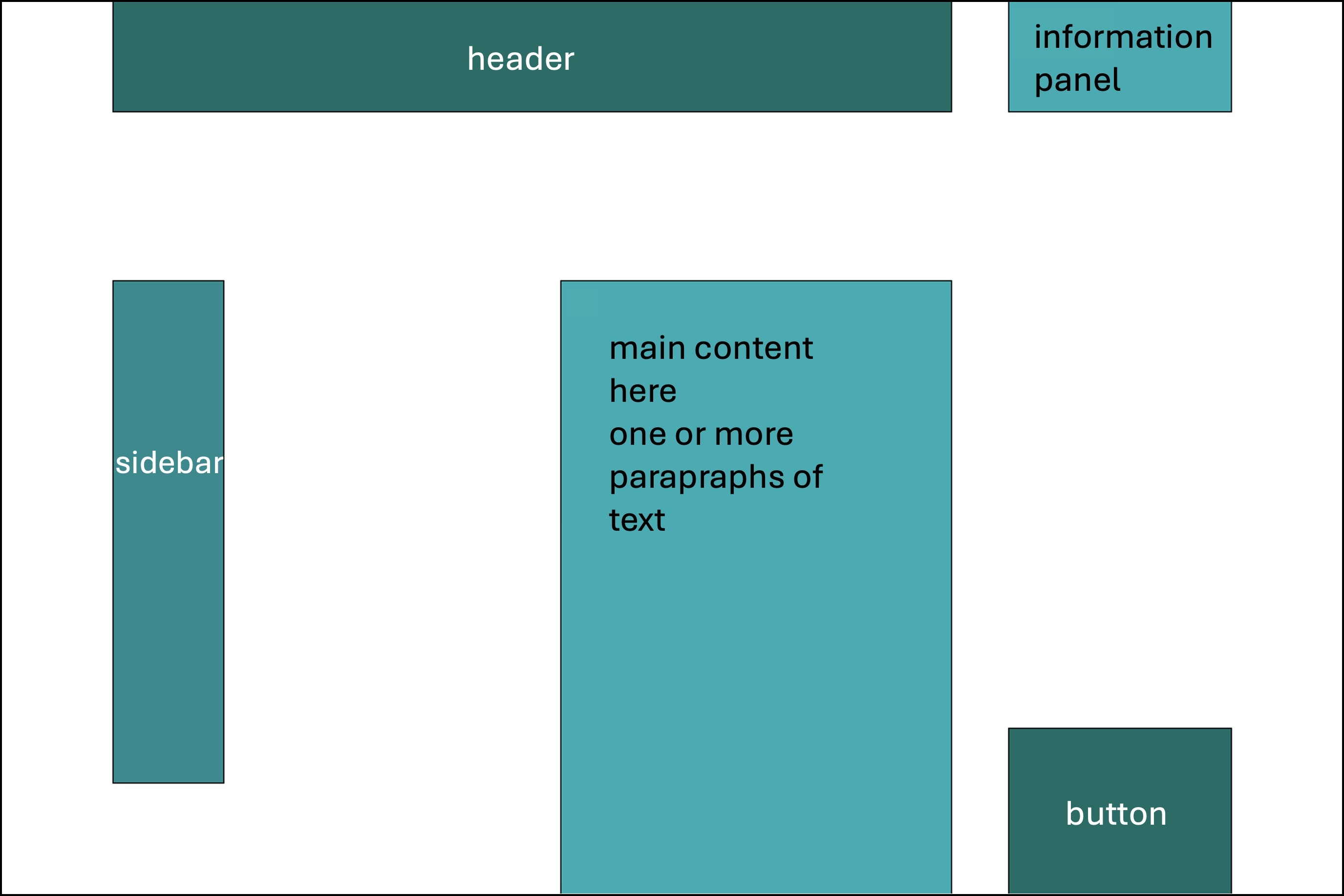}
    {Grid--Simple stimulus 5 used in the end-user evaluation. Used also in the validation experiment.}
    {fig:exp1_reduced_simple_5}
\expstimulus
    {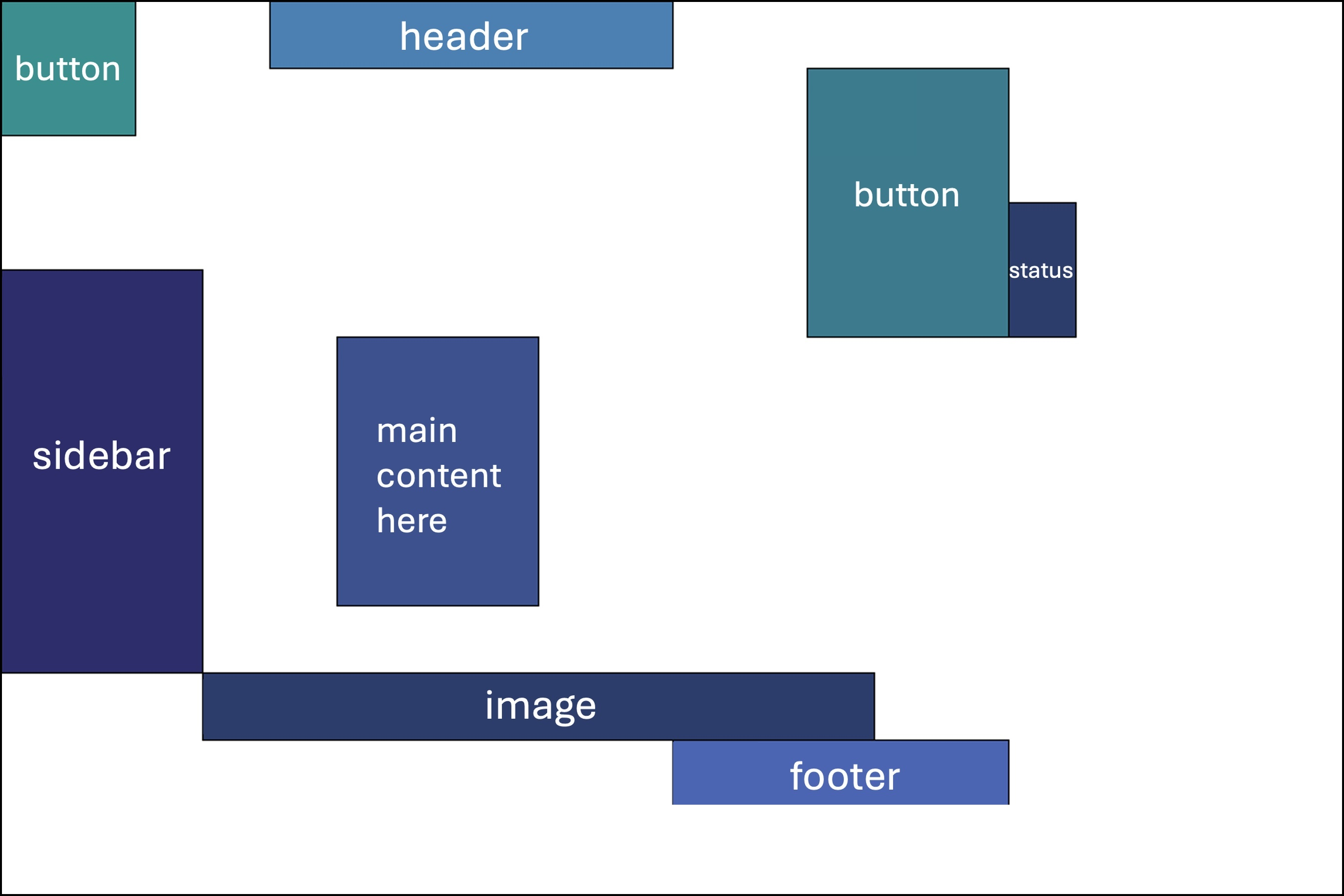}
    {Random--Complex stimulus 1 used in the end-user evaluation.}
    {fig:exp1_random_complex_1}

\expstimulus
    {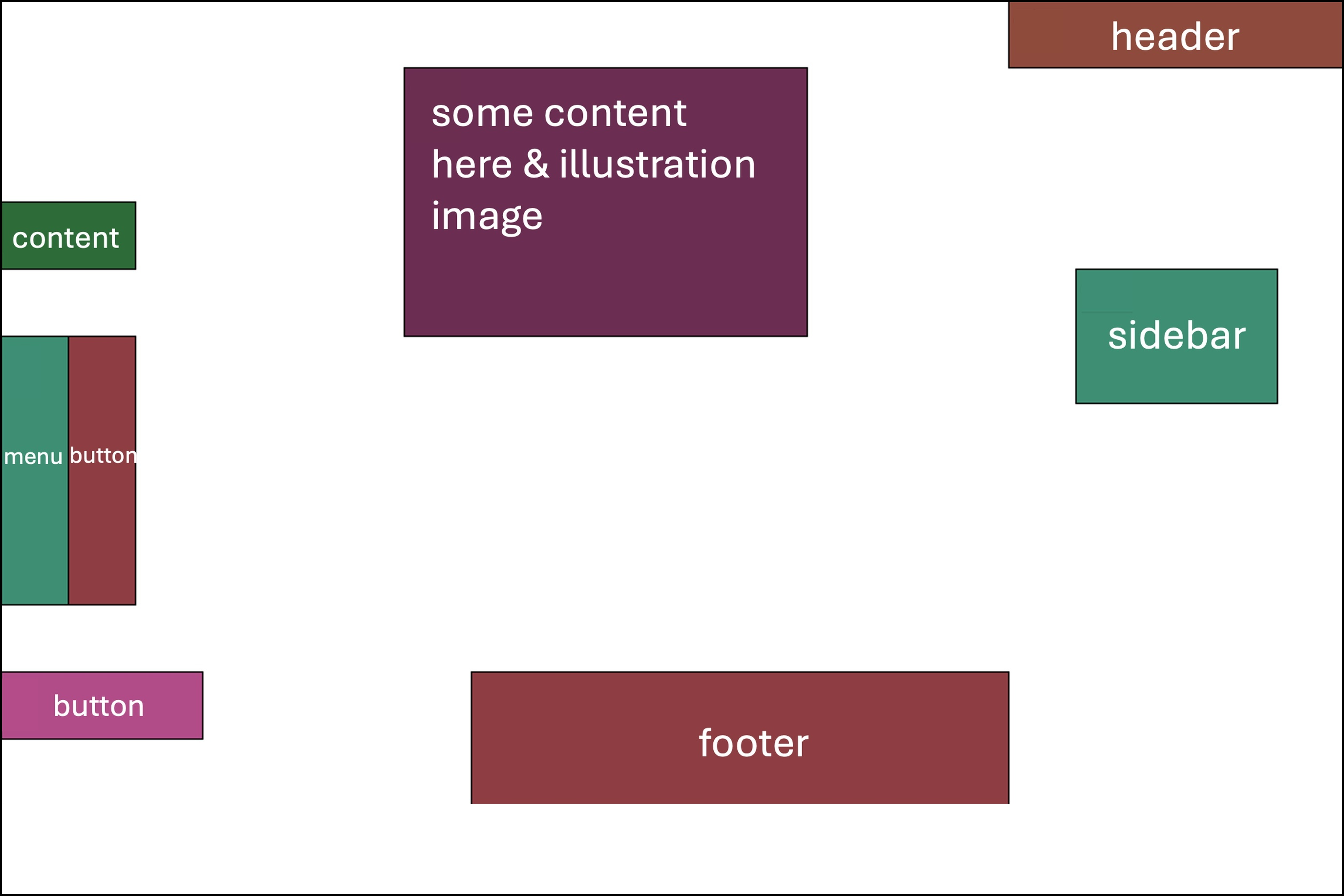}
    {Random--Complex stimulus 2 used in the end-user evaluation. Used also in the validation experiment.}
    {fig:exp1_random_complex_2}
\expstimulus
    {fig_exp1_Random_Complex_3.png}
    {Random--Complex stimulus 3 used in the end-user evaluation.}
    {fig:exp1_random_complex_3}
\expstimulus
    {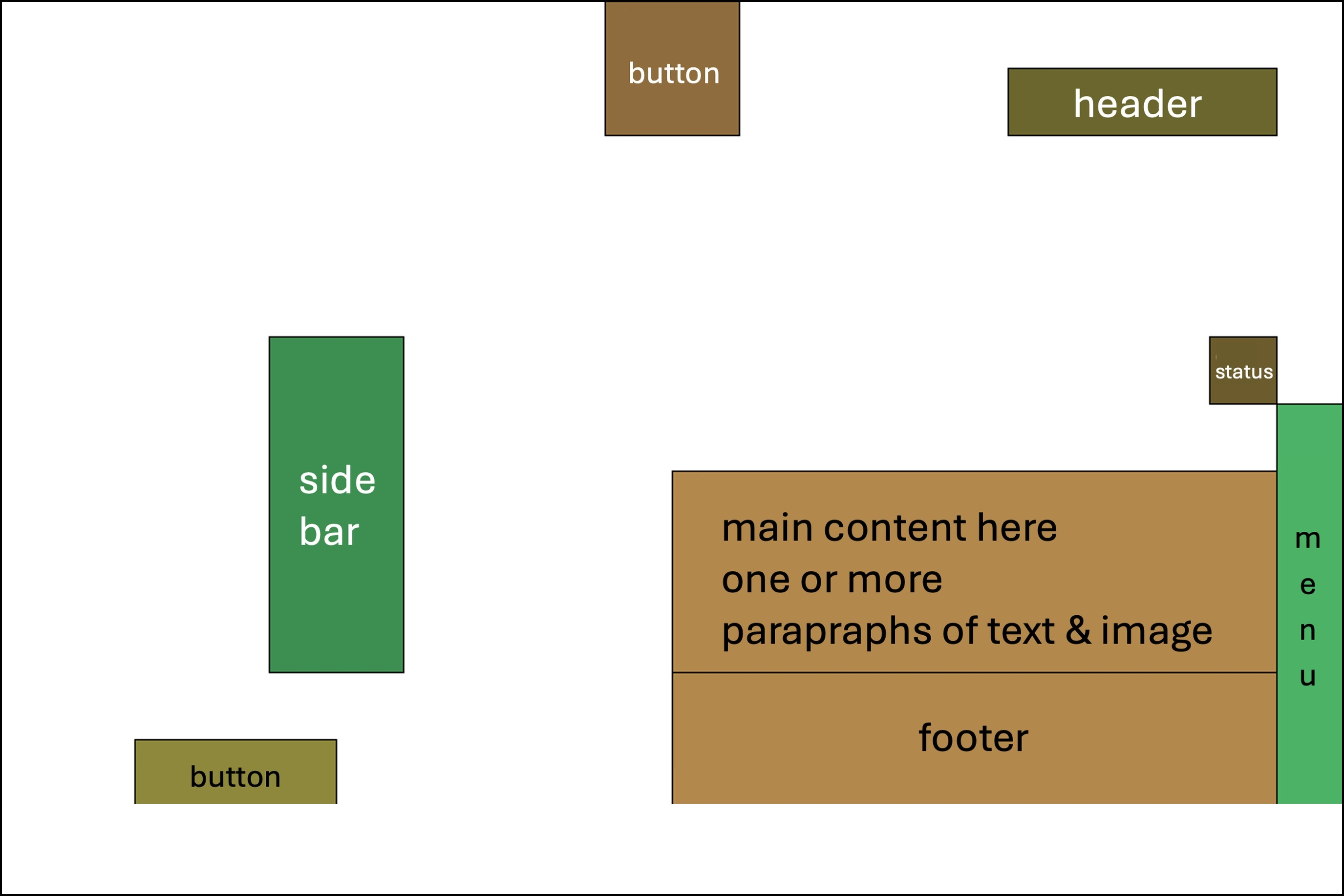}
    {Random--Complex stimulus 4 used in the end-user evaluation. Used also in the validation experiment.}
    {fig:exp1_random_complex_4}

\expstimulus
    {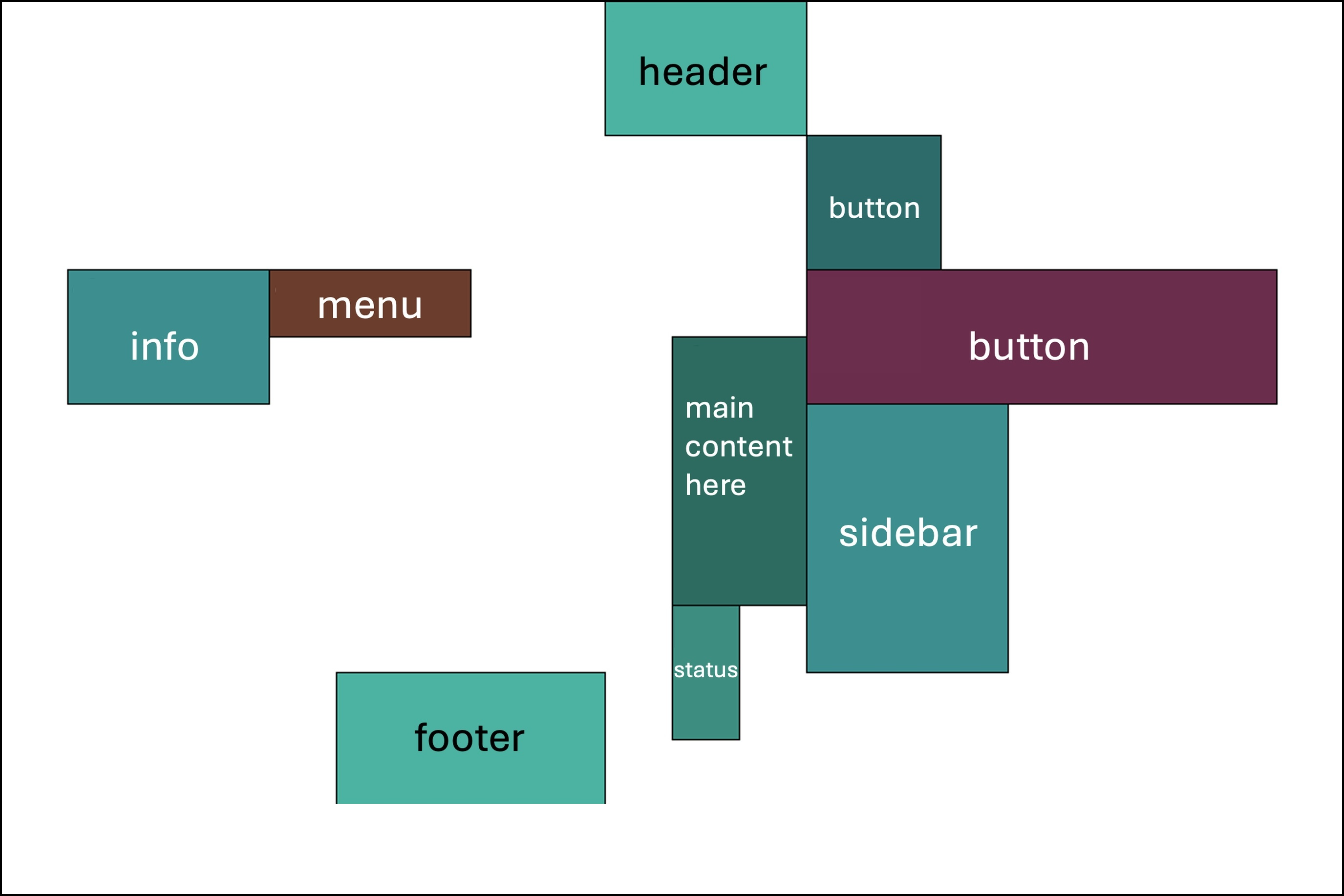}
    {Random--Complex stimulus 5 used in the end-user evaluation. Used also in the validation experiment.}
    {fig:exp1_random_complex_5}
\expstimulus
    {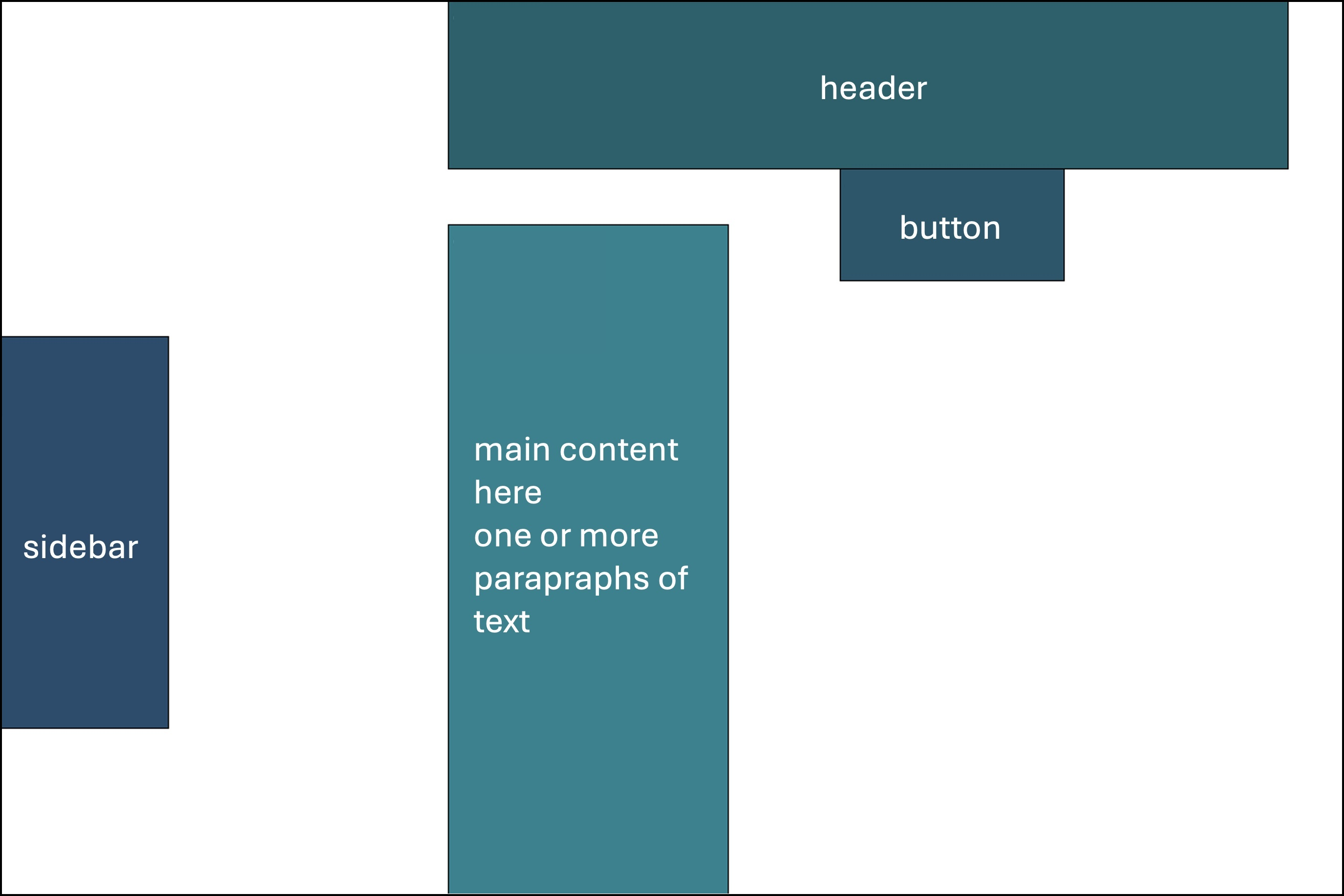}
    {Random--Simple stimulus 1 used in the end-user evaluation. Used also in the validation experiment.}
    {fig:exp1_random_simple_1}
\expstimulus
    {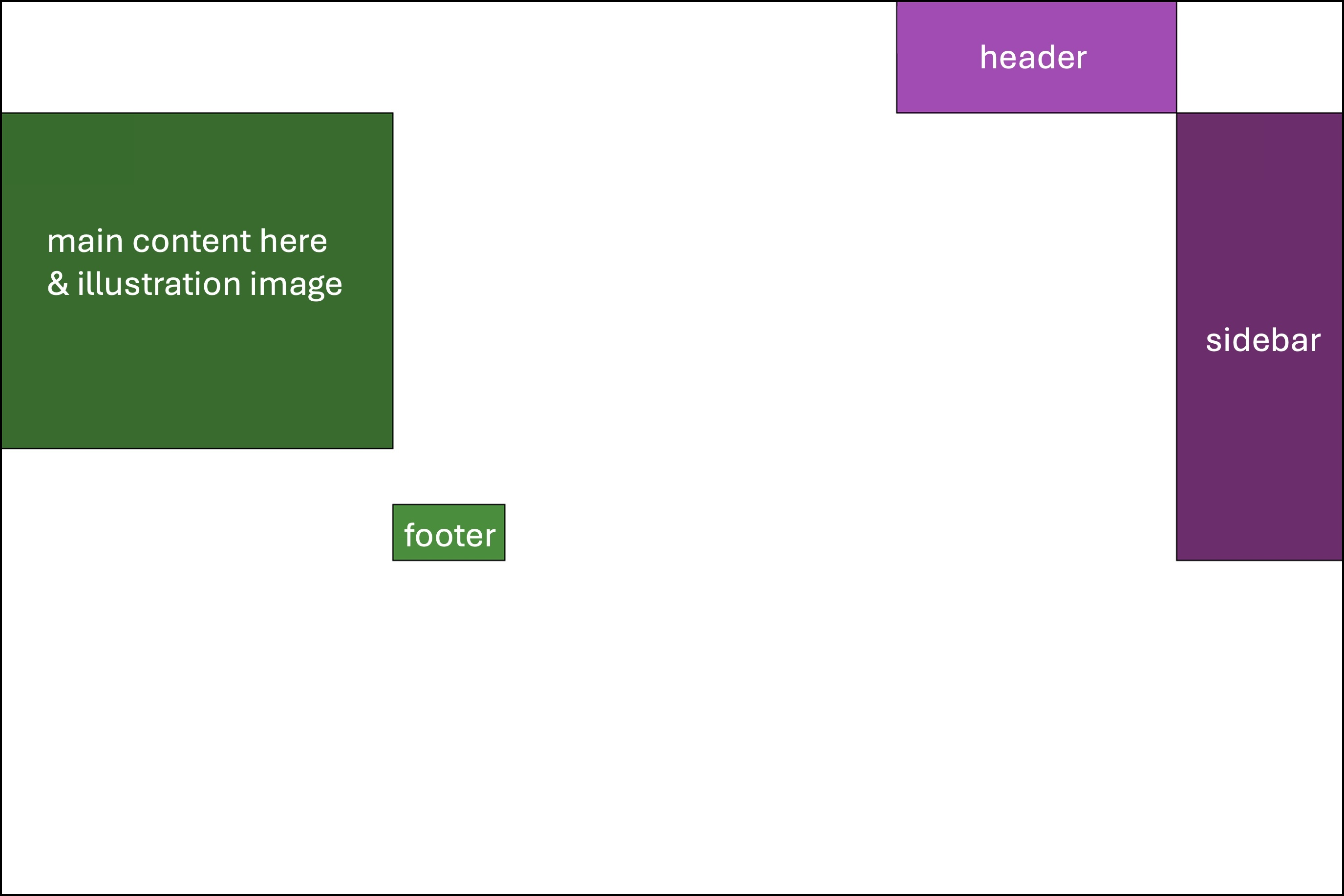}
    {Random--Simple stimulus 2 used in the end-user evaluation.}
    {fig:exp1_random_simple_2}

\expstimulus
    {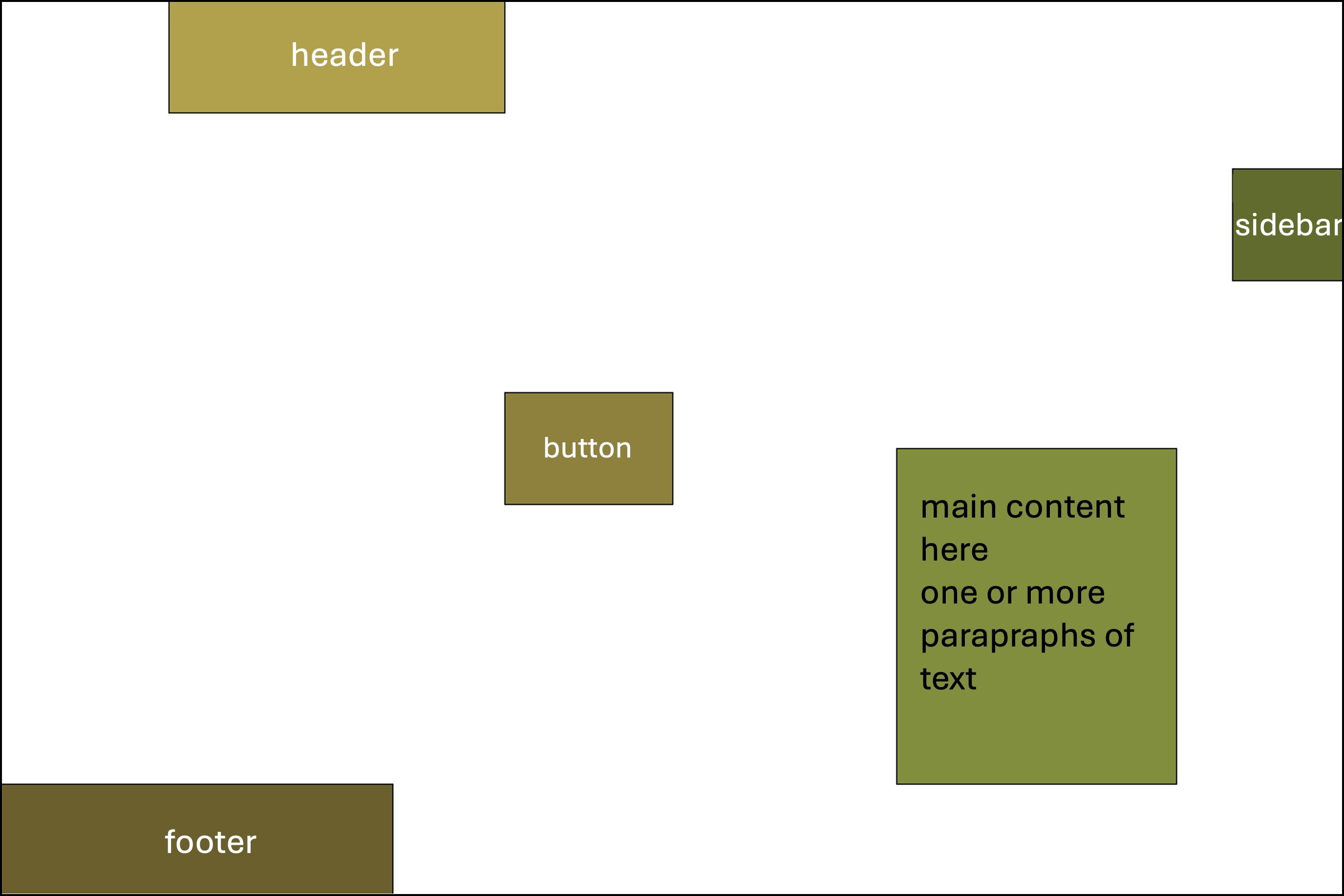}
    {Random--Simple stimulus 3 used in the end-user evaluation.}
    {fig:exp1_random_simple_3}
\expstimulus
    {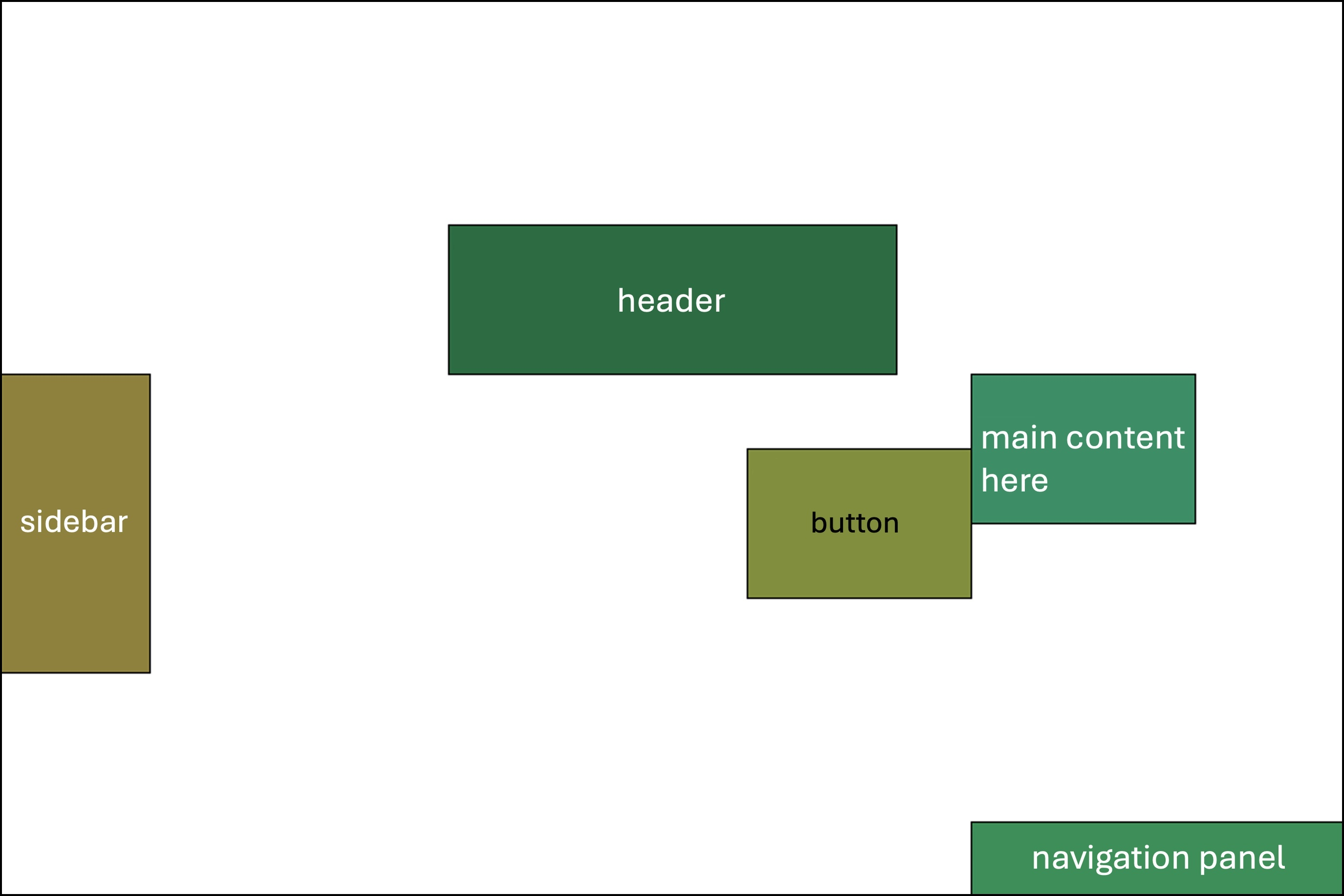}
    {Random--Simple stimulus 4 used in the end-user evaluation.}
    {fig:exp1_random_simple_4}
\expstimulus
    {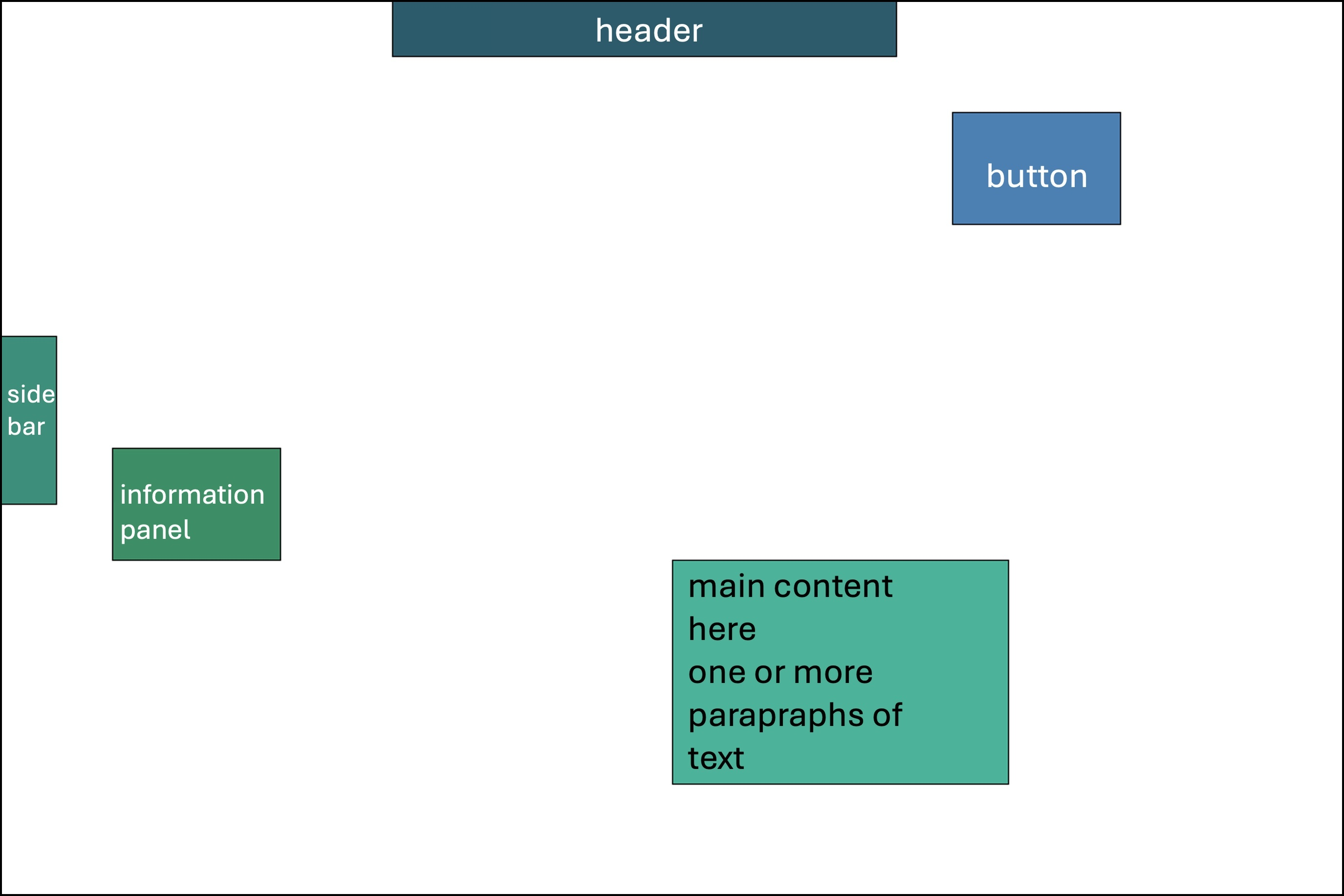}
    {Random--Simple stimulus 5 used in the end-user evaluation. Used also in the validation experiment.}
    {fig:exp1_random_simple_5}

\subsection{Screenshots of the End-User Experiments}
\label{appendix:exp1_screens}
\newcommand{\expscreen}[3]{%
    \begin{figure}[H]
        \centering
        \includegraphics[
            width=0.9\linewidth,
            height=0.4\textheight,
            keepaspectratio
        ]{figures/#1}
        \caption{#2}
        \label{#3}
    \end{figure}
}
\expscreen
    {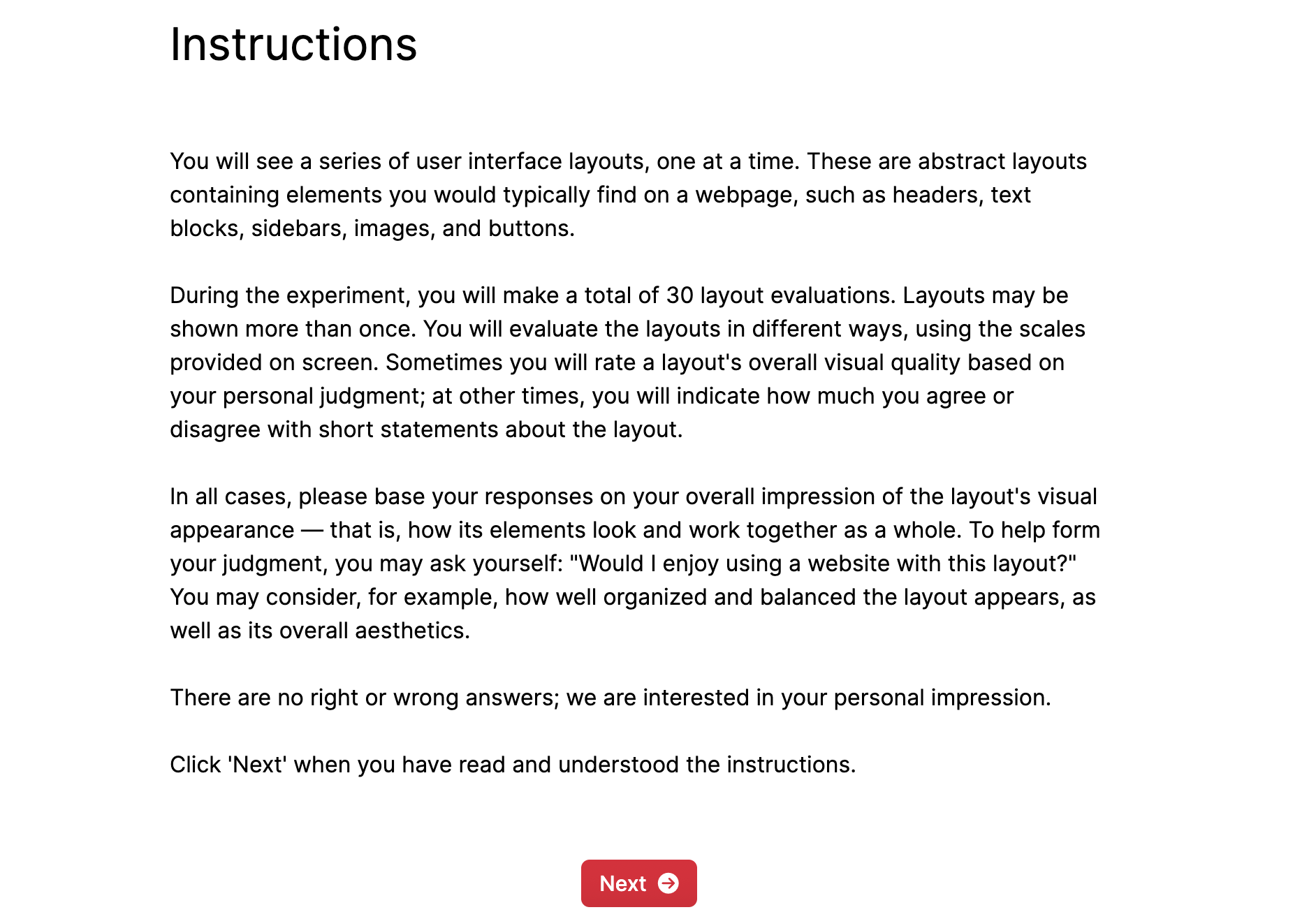}
    {Instructions presented at the beginning of the end-user experiment and its validation experiment.}
    {fig:exp1_screen1}
\expscreen
    {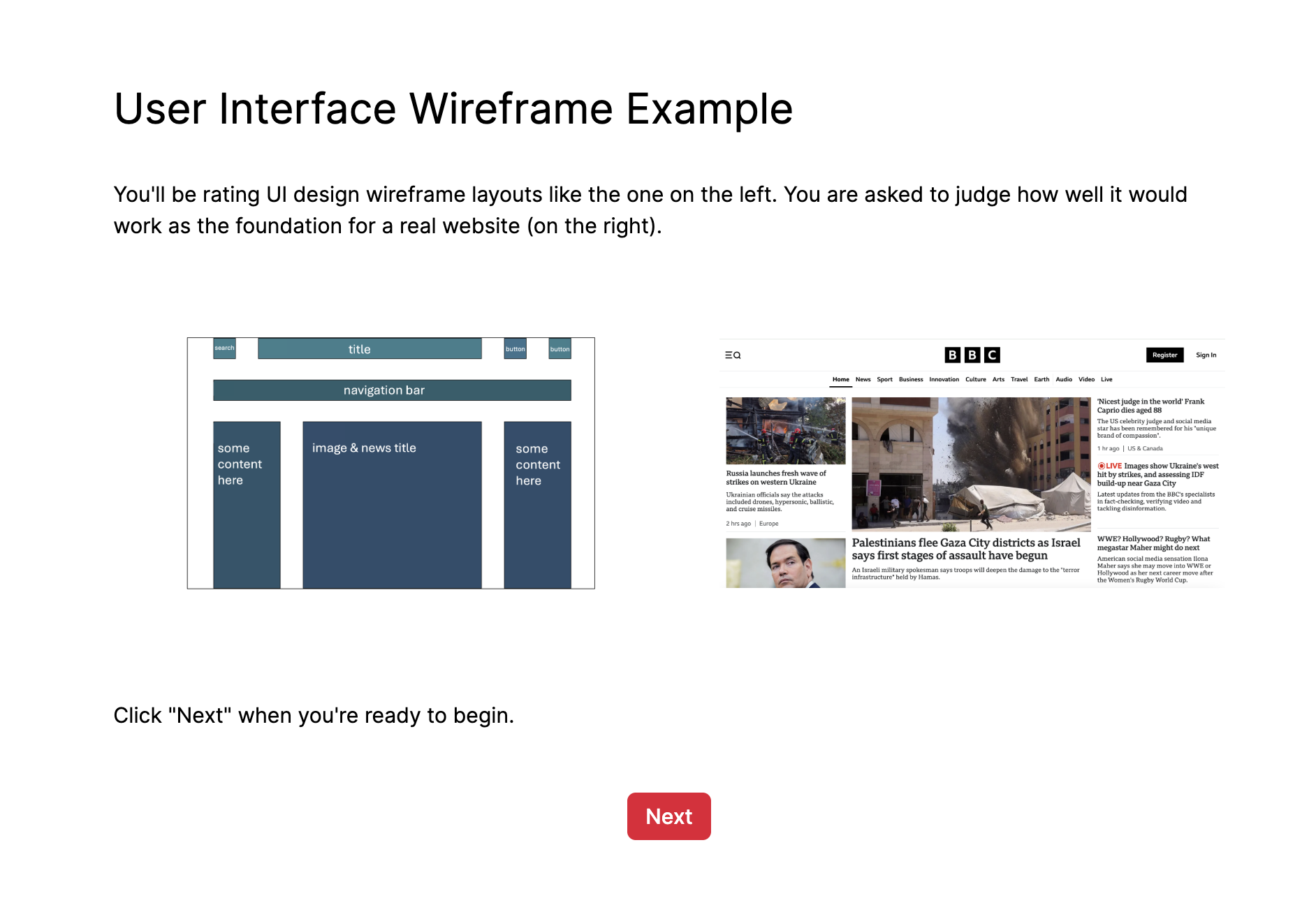}
    {Example illustrating the relationship between an abstract wireframe and a real website. Presented at the beginning of the end-user experiment and its validation experiment.}
    {fig:exp1_screen2}
\clearpage
\expscreen
    {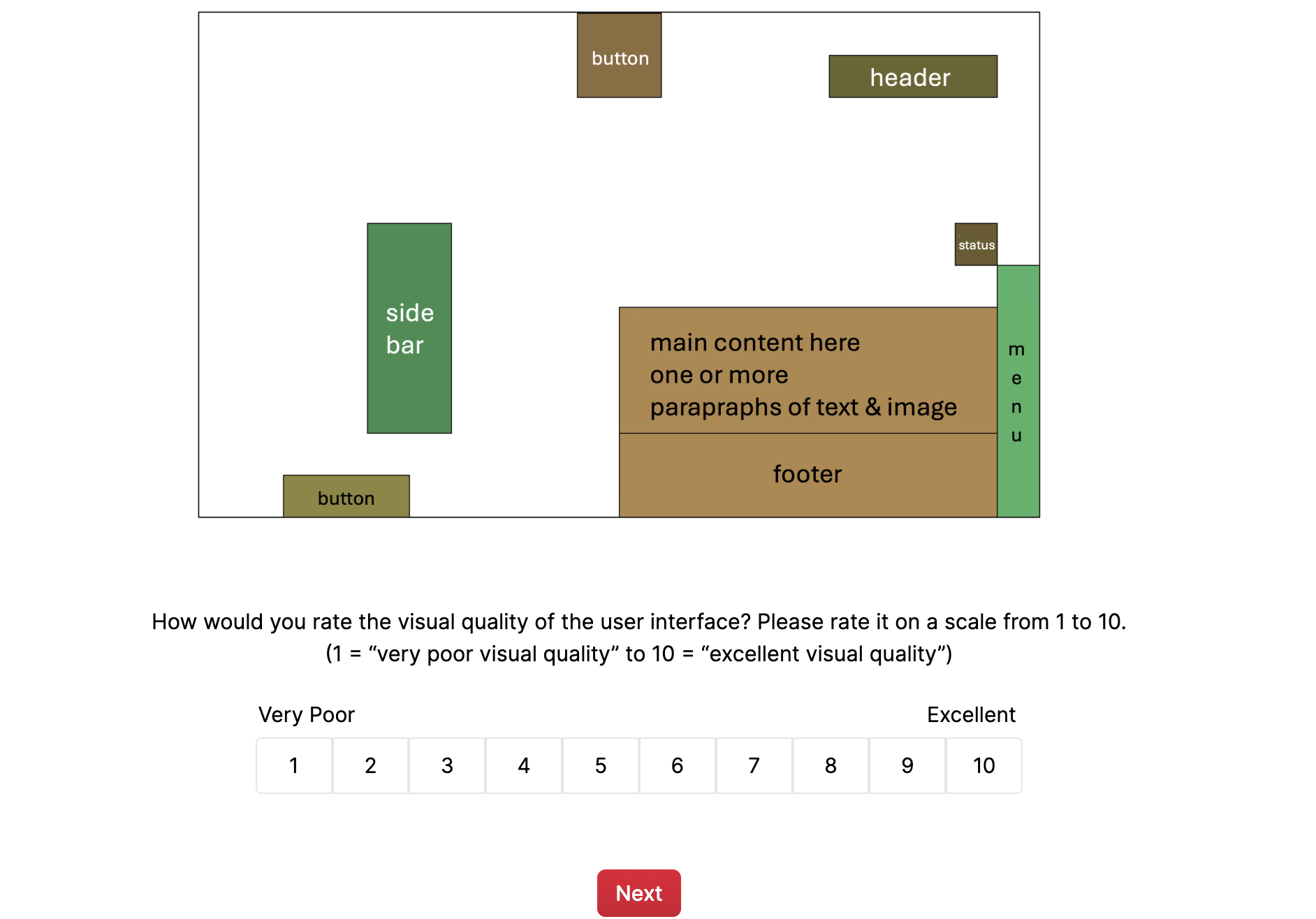}
    {Visual-quality rating screen used in the end-user experiment and its validation experiment.}
    {fig:exp1_screen4}
    
\expscreen
    {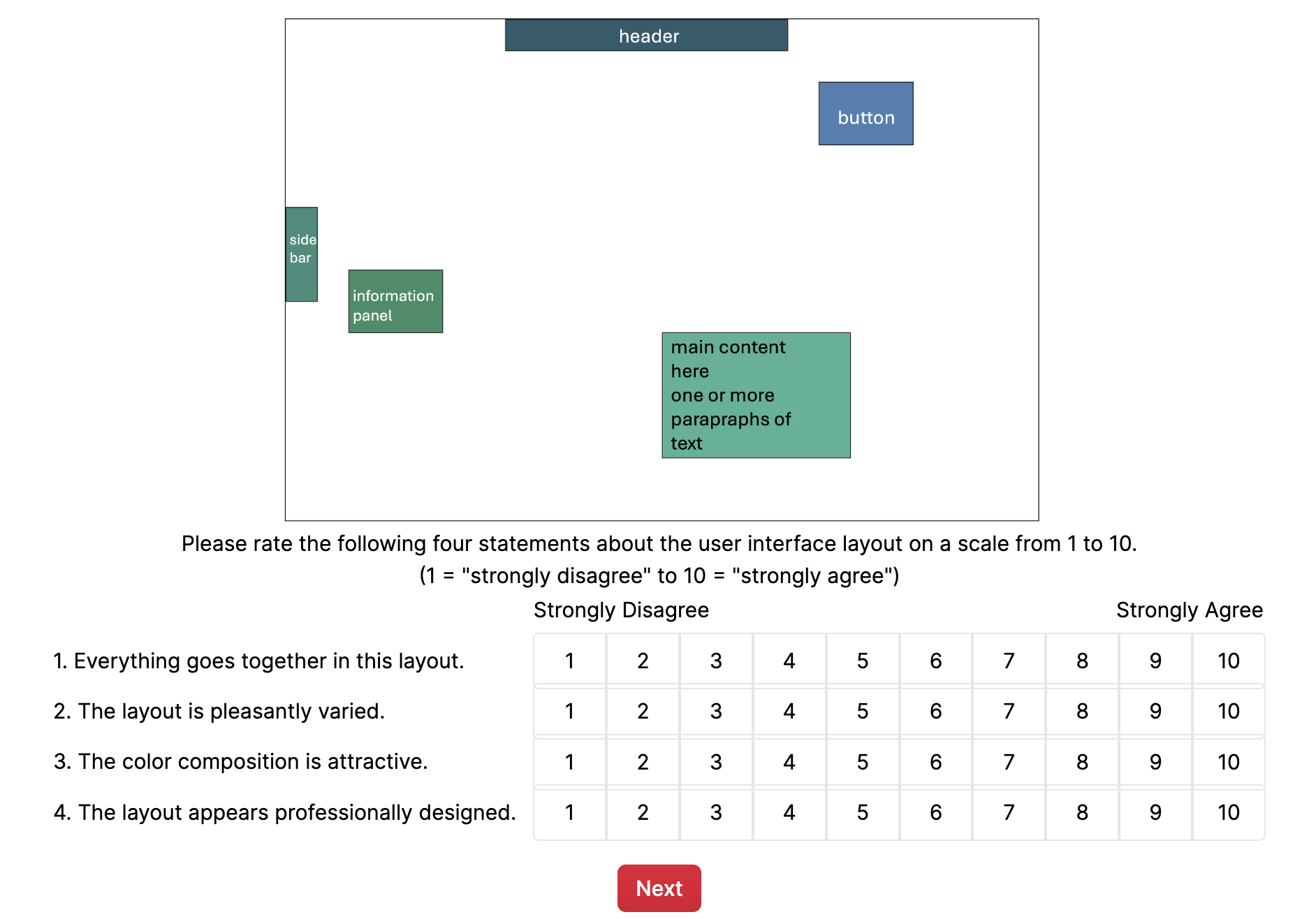}
    {Multi-item layout-rating screen used in the validation experiment.}
    {fig:exp1_screen3}

\clearpage
\section{Designer Evaluation Protocol and Interview Guide}
\label{appendix:exp2}

\subsection{Participant Orientation}

Before the main tasks, the researcher explained the purpose of the study and demonstrated how to use the GUI. Participants were informed that they could express design specifications through controls for element size, saliency, spatial relations, grouping, and color selection. The GUI translated these inputs into constraints for the ASP optimizer; participants were not shown or required to write ASP code.

Participants were instructed to read each design scenario, enter the specifications they considered appropriate, run the optimization, inspect the generated layout, and revise their specifications if desired. They then completed a three-element practice task to become familiar with the GUI before beginning the three main tasks.

\subsection{Experimental Procedure}

The researcher followed the procedure below:

\begin{enumerate}
\item Explain the study, obtain informed consent, and introduce the experimental procedure.

\item Demonstrate the GUI and allow the participant to complete the practice task. Procedural questions were answered before the main tasks began.

\item Present the three main design tasks in counterbalanced order. For each task, the participant reviewed the scenario and element list, entered design specifications through the GUI, and invoked the optimizer.

\item Display one generated layout after each optimization request. If the solver established optimality before the 30-second cutoff, the final optimal layout was displayed immediately; otherwise, the best valid layout found by the cutoff was displayed. Participants could revise their specifications and run the optimizer again until they indicated that they had finished the task.

\item Ask the post-task questions after each task. After all three tasks, administer the overall satisfaction rating and final interview. The researcher recorded the responses in notes during the interviews.

\end{enumerate}

During the tasks, the researcher provided procedural clarification when requested but did not provide design suggestions.

\subsection{Interview Guide}

The same post-task questions were asked after each of the three main tasks and are therefore listed only once. Neutral follow-up prompts, such as ``Why?'' or ``Could you explain further?'', were used when clarification was needed.

\subsubsection{Post-Task Questions}

\begin{enumerate}
\item What is your initial reaction to the generated layout? Does it reflect what you had in mind when entering your specifications? Why or why not?

\item Which aspects of the layout do you find most successful?

\item Did you notice any unexpected arrangements or relationships? Was anything inconsistent with the specifications you entered?

\end{enumerate}

\subsubsection{Overall Rating and Final Interview}

The following rating was administered once, after participants had completed all three main tasks:
Overall, how satisfied are you with the layouts generated by the tool, on a scale from 1 to 5? Why?

The researcher then asked the following open-ended questions:

\begin{enumerate}
\item Overall, how was your experience using the GUI to create layouts? What aspects of the interface were easy or difficult to use?

\item Did you encounter any difficulties or confusion when entering design specifications through the GUI?

\item Were there any design requirements that you wanted to express but could not specify using the available options? What additional capabilities would make the tool more useful?

\item Which specification categories did you find particularly useful or not useful?

\item Could you see this tool being integrated into your design workflow? Why or why not?

\end{enumerate}

\clearpage
\section{Results of Model Efficiency Evaluation}
\label{appendix:model-efficiency}
\setcounter{figure}{0}

\begin{figure*}[htbp]
    \centering
    \includegraphics[width=1\linewidth]{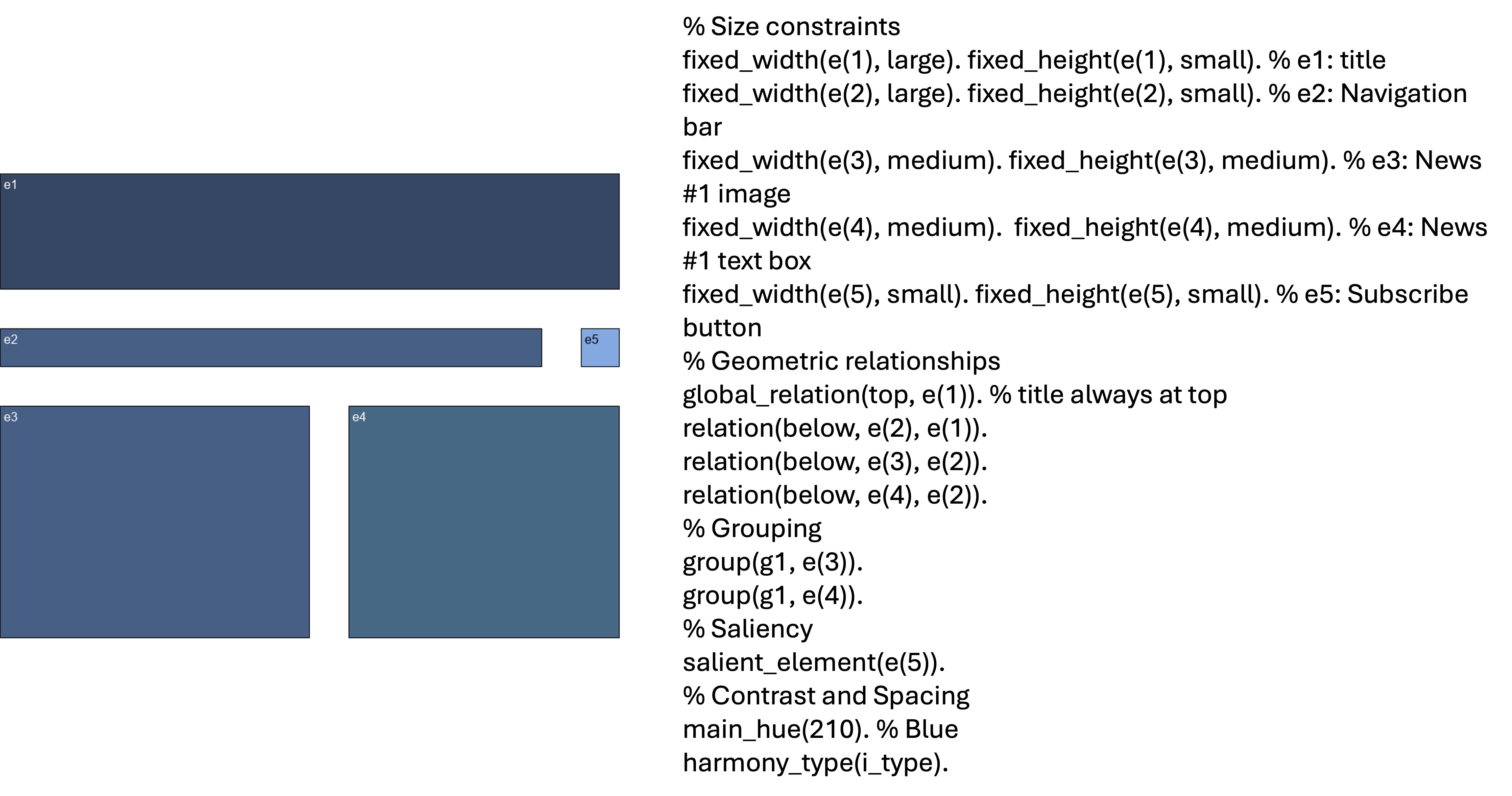}
    \caption{ Optimized layout and corresponding ASP constraints for 5-element news website design task.}
    \label{fig:appendix_t1_1}
\end{figure*}

\begin{figure*}[htbp]
    \centering
    \includegraphics[width=1\linewidth]{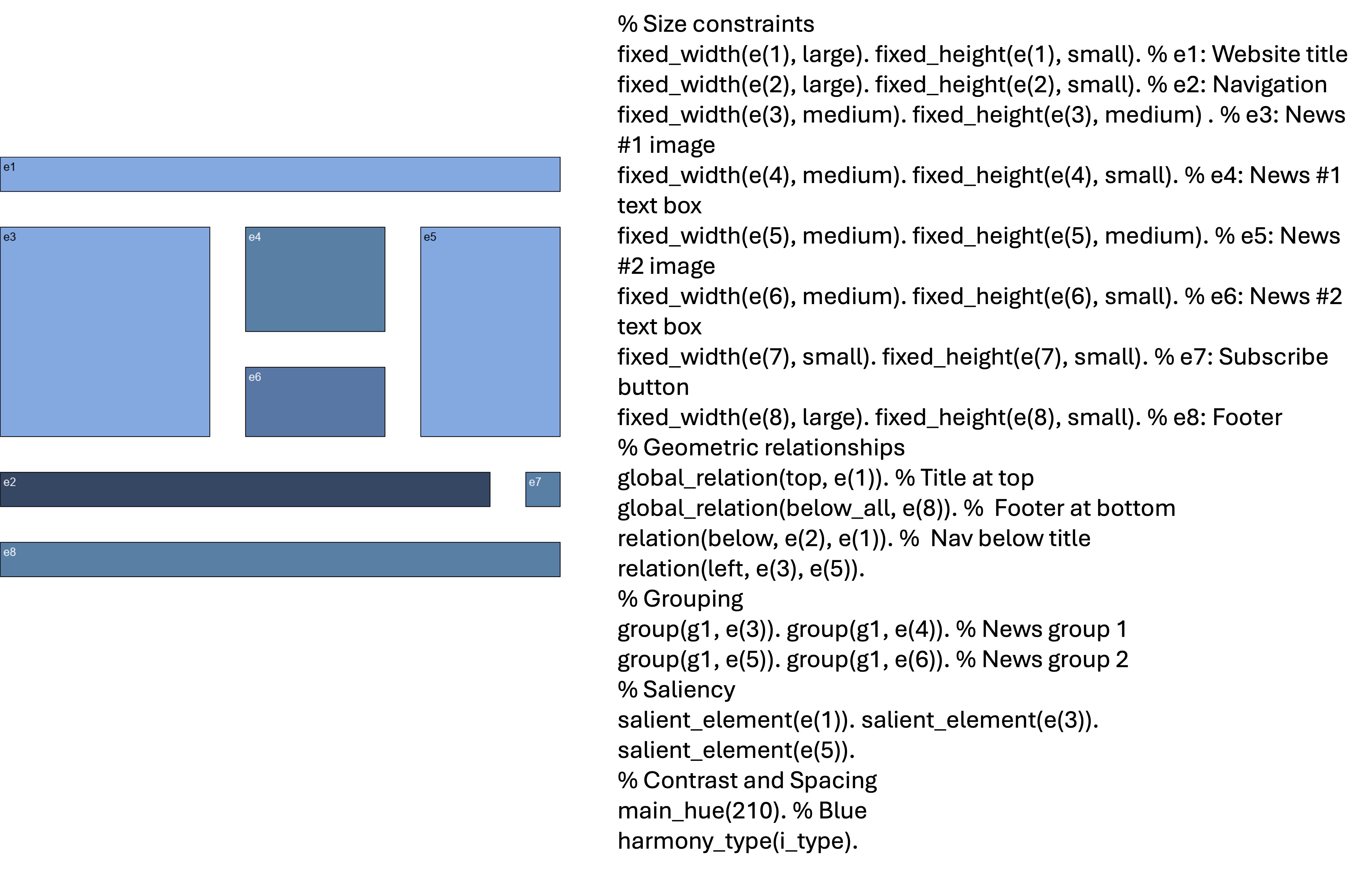}
    \caption{ Optimized layout and corresponding ASP constraints for 8-element news website design task.}
    \label{fig:appendix_t1_2}
\end{figure*}

\begin{figure*}[htbp]
    \centering
    \includegraphics[width=1\linewidth]{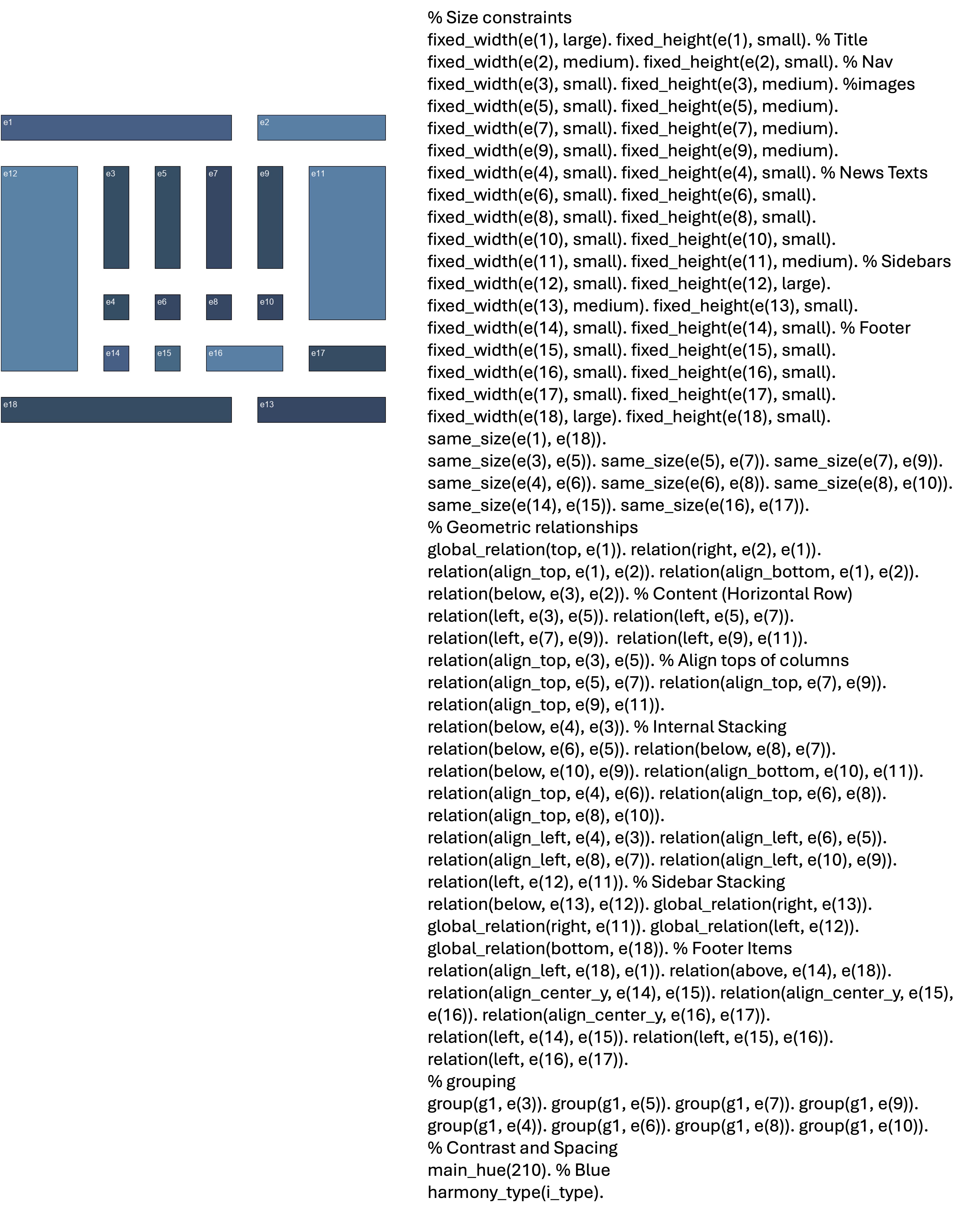}
    \caption{ Optimized layout and corresponding ASP constraints for 18-element news website design task.}
    \label{fig:appendix_t1_3}
\end{figure*}

\begin{figure*}[htbp]
    \centering
    \includegraphics[width=1\linewidth]{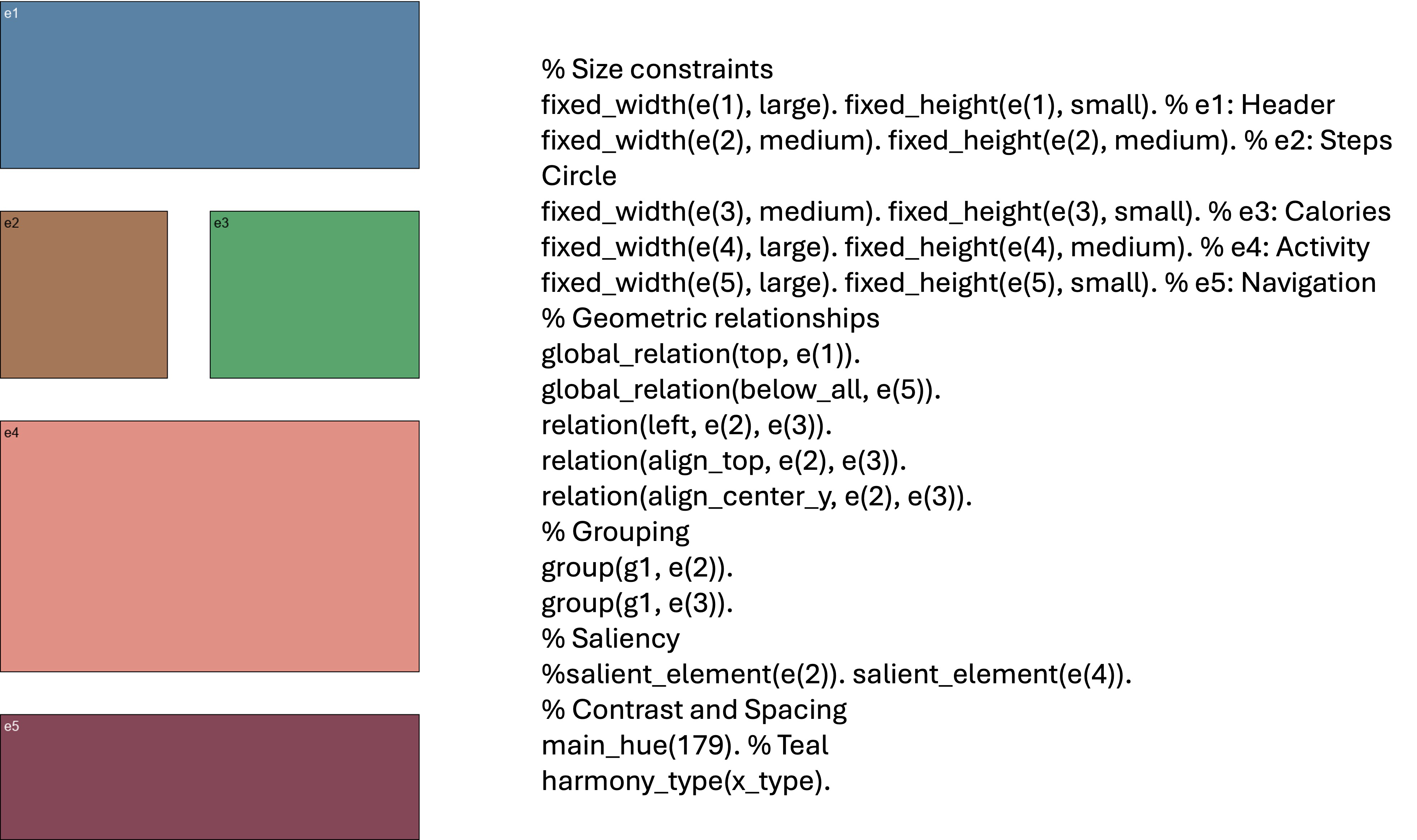}
    \caption{ Optimized layout and corresponding ASP constraints for 5-element fitness app design task.}
    \label{fig:appendix_t2_1}
\end{figure*}

\begin{figure*}[htbp]
    \centering
    \includegraphics[width=1\linewidth]{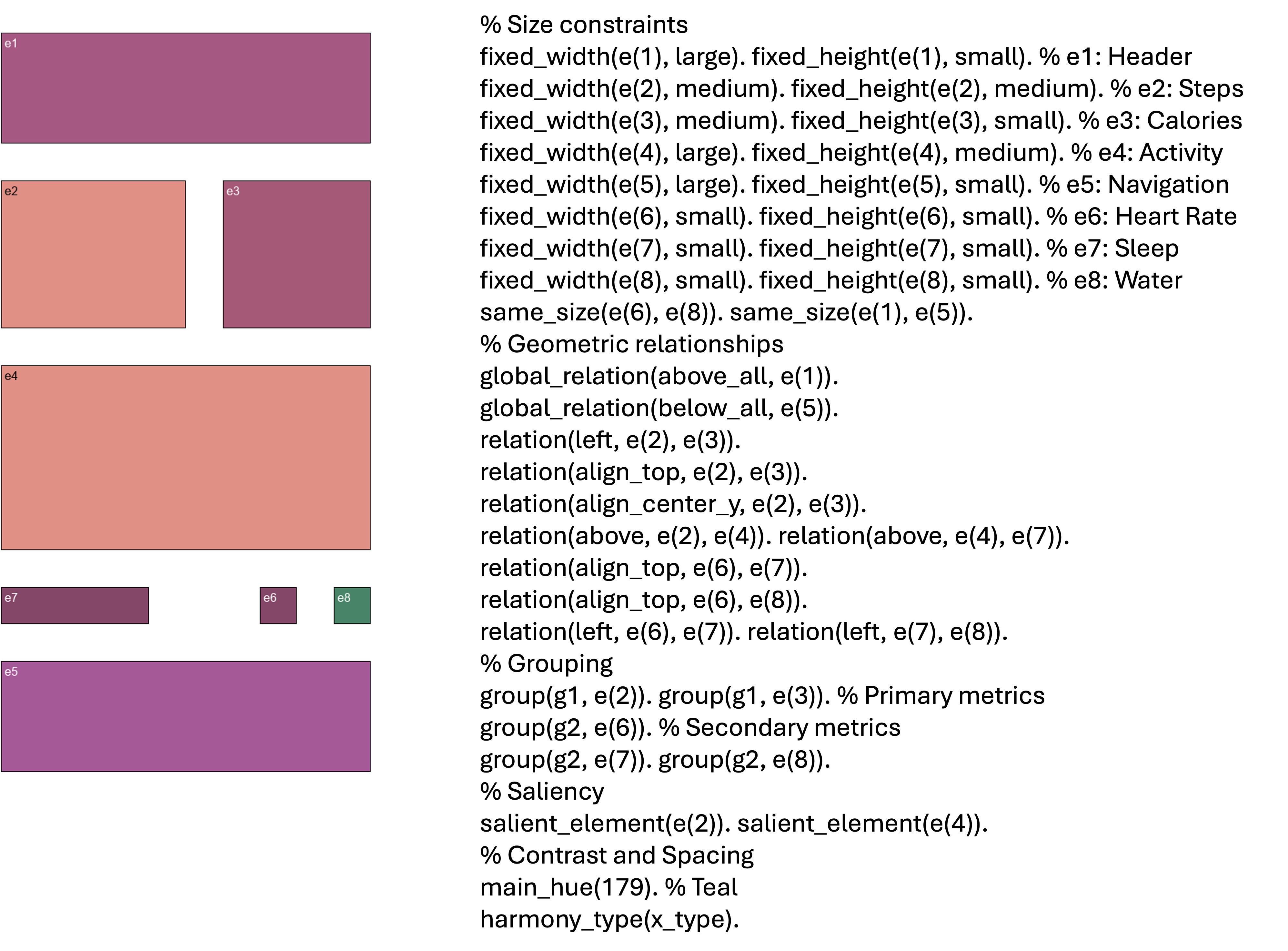}
    \caption{ Optimized layout and corresponding ASP constraints for 8-element fitness app design task.}
    \label{fig:appendix_t2_2}
\end{figure*}

\begin{figure*}[htbp]
    \centering
    \includegraphics[width=1\linewidth]{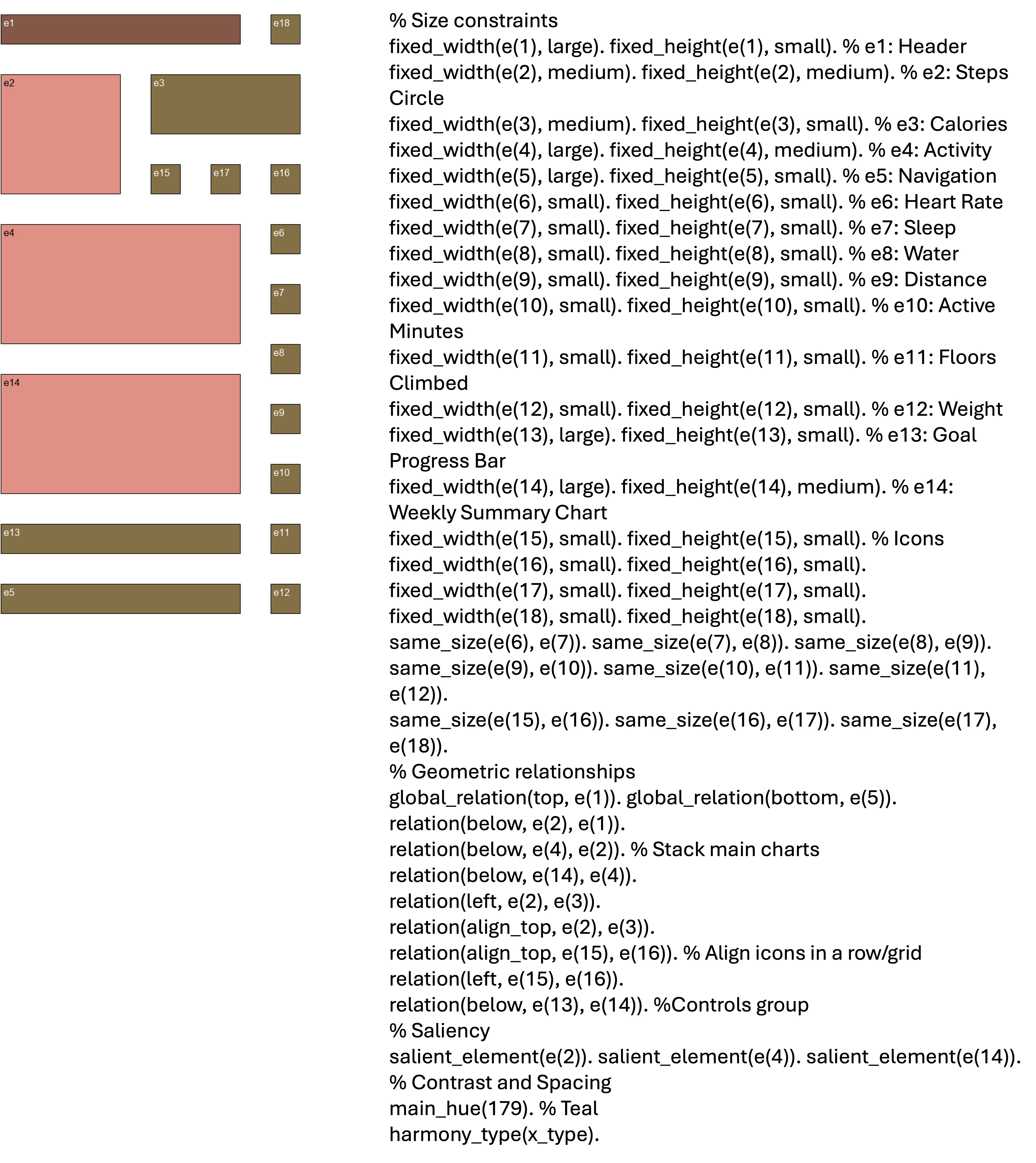}
    \caption{ Optimized layout and corresponding ASP constraints for 18-element fitness app design task.}
    \label{fig:appendix_t2_3}
\end{figure*}

\begin{figure*}[htbp]
    \centering
    \includegraphics[width=1\linewidth]{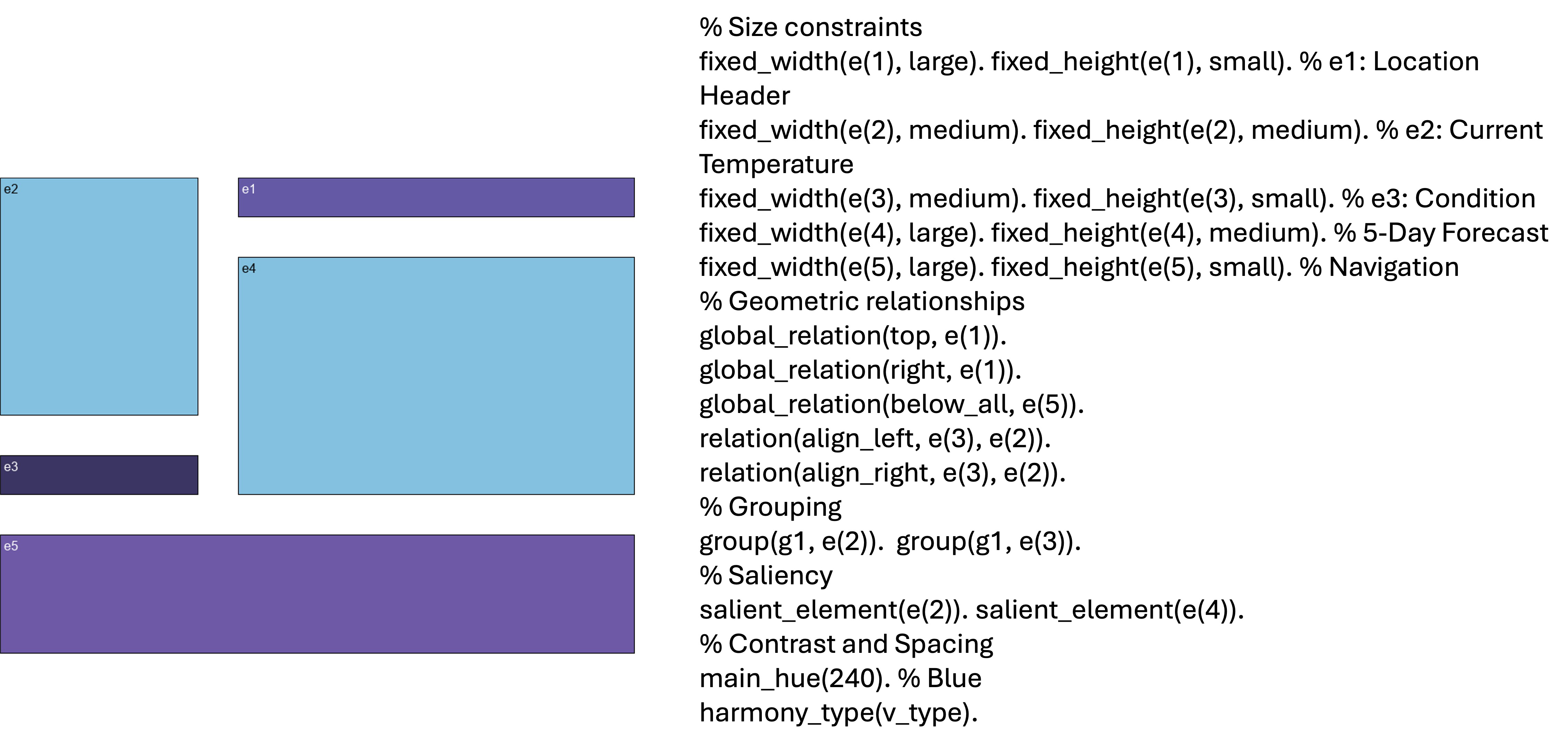}
    \caption{ Optimized layout and corresponding ASP constraints for 5-element weather forecast website design task.}
    \label{fig:appendix_t3_1}
\end{figure*}

\begin{figure*}[htbp]
    \centering
    \includegraphics[width=1\linewidth]{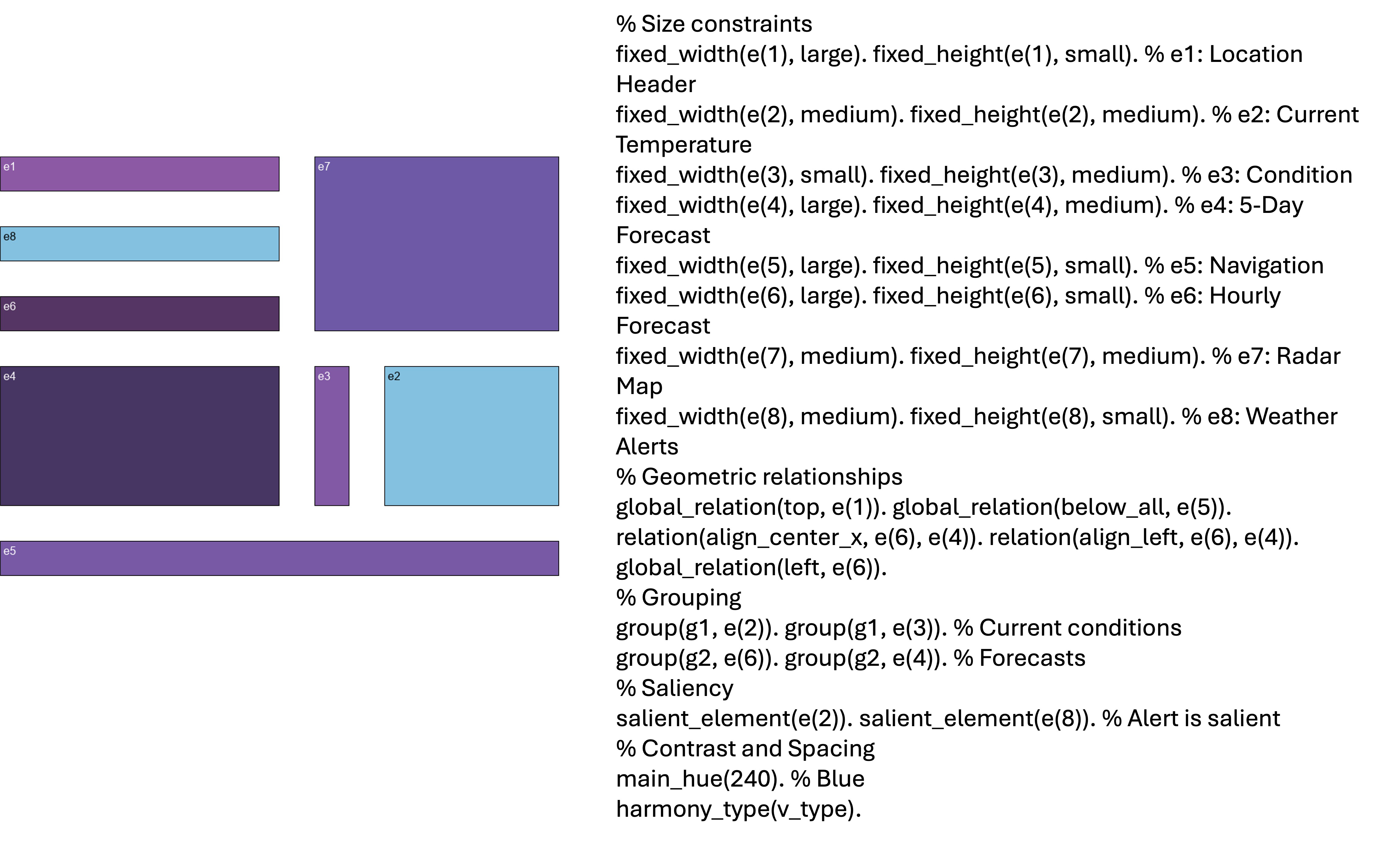}
    \caption{ Optimized layout and corresponding ASP constraints for 8-element weather forecast website design task.}
    \label{fig:appendix_t3_2}
\end{figure*}

\begin{figure*}[htbp]
    \centering
    \includegraphics[width=1\linewidth]{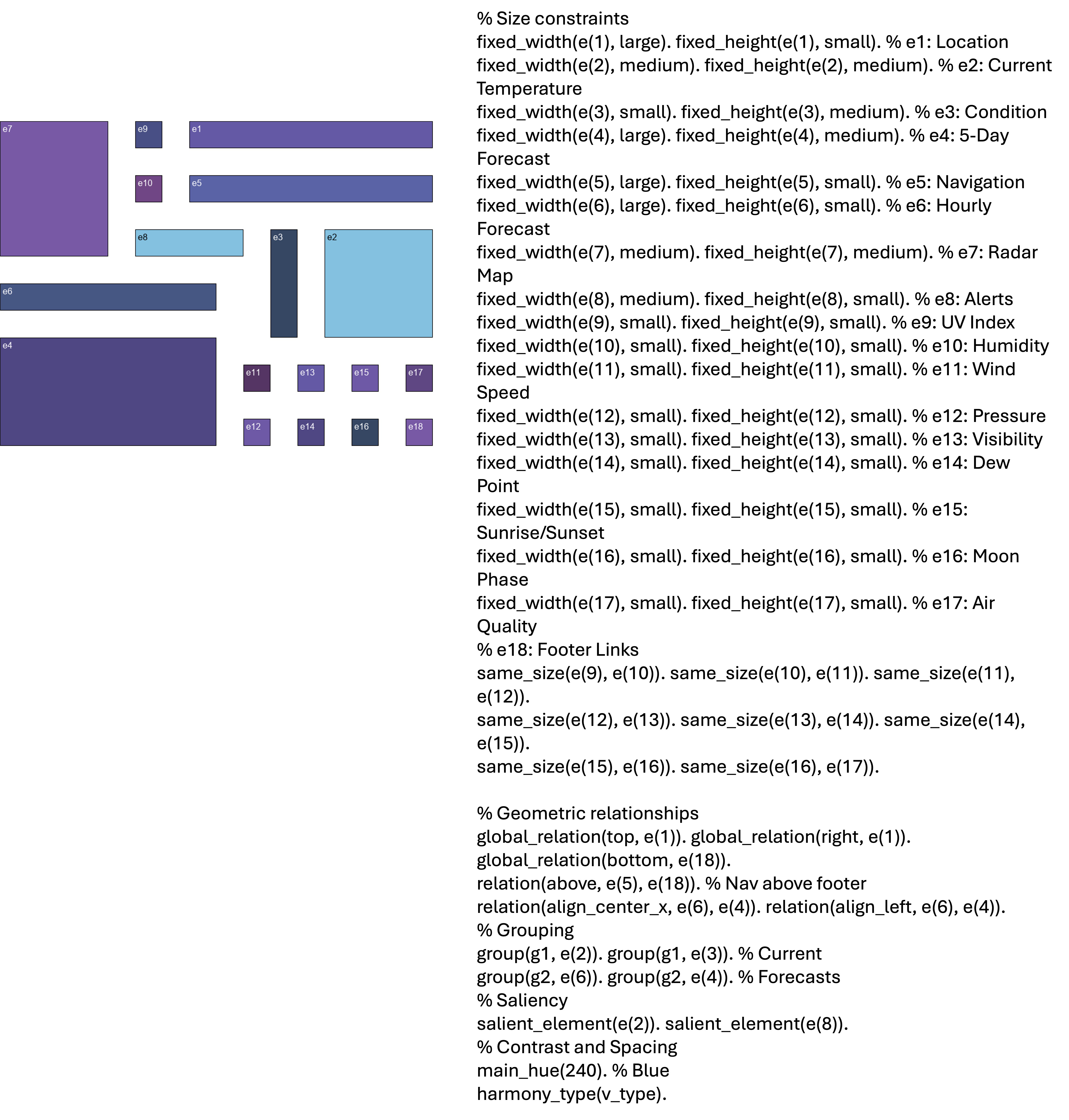}
    \caption{ Optimized layout and corresponding ASP constraint for 18-element weather forecast website design task.}
    \label{fig:appendix_t3_3}
\end{figure*}











\end{document}

%% file: 01-introduction.tex
Graphical user interface (GUI) design is an iterative, detail-oriented process that demands significant time and resources \citep{cross2006designerly}.
Designers must balance functional requirements, aesthetic choices, and constraints across platforms and screen sizes \citep{watzman2002visual}.
Even small layout changes can affect usability and perception, making evaluation both costly and time-intensive.
Current practice relies heavily on repeated prototyping and usability testing.
Existing computational tools are often narrow in scope, or inflexible, leaving much of the work manual \citep{oulasvirta2020combinatorial}.
There is therefore a need for more comprehensive automated support in design.

GUI design is challenging because interfaces must align with two domains: the external task environment and the internal cognitive environment of users \citep{simon2019sciences}.
Effective design requires minimizing discrepancies in both, meeting task demands while accommodating perceptual and cognitive capacities as well as individual needs and preferences.
Psychological research can inform design, but its application often remains abstract and difficult to implement directly.
To become actionable, such knowledge must be expressed in computational form, enabling systematic evaluation and adaptation of layouts \citep{oulasvirta2019s,oulasvirta2020combinatorial}.
Embedding cognitive models into optimization procedures allows design tools to move beyond surface heuristics toward principled and interpretable improvements \citep{howes2023towards}.

While computational tools for creating interface layouts have advanced, each existing paradigm leaves important gaps.
Integer programming provides provable guarantees and real-time exploration but is presently limited to grid-based formulations and linear objectives, making it difficult to capture richer design semantics~\citep{dayama2020grids,dayama2021interactive}.
Heuristic and evolutionary algorithms expand the design space and handle nonlinear objectives, yet they lack feasibility guarantees and often require substantial computation to avoid invalid layouts~\citep{shiripour2021grid}.
Deep learning methods, including generative and graph-based models, capture complex spatial regularities and produce realistic layouts, but they depend on large annotated datasets and perform poorly on out-of-distribution designs or tasks requiring explicit interpretability~\citep{jiang2022coarse,jiang2024graph4gui}.
Novel large language model (LLM) based approaches are promising, but lack in reliability and explainability \citep{kolthoff2025guide}.
Motivated by these limitations, we investigate \emph{Answer Set Programming} \citep[ASP;][]{lifschitz2008asp} as a declarative solver-based paradigm for GUI layout optimization.
ASP combines the guarantees of constraint solving with non-monotonic reasoning capabilities. This allows us to model psychological design principles as "defaults with exceptions" (e.g., elements use uniform sizing unless designated as salient), enabling both generation and evaluation of layouts under explicitly defined aesthetic and functional objectives.

\begin{figure*}[ht]
    \centering
    \includegraphics[width=1\linewidth]{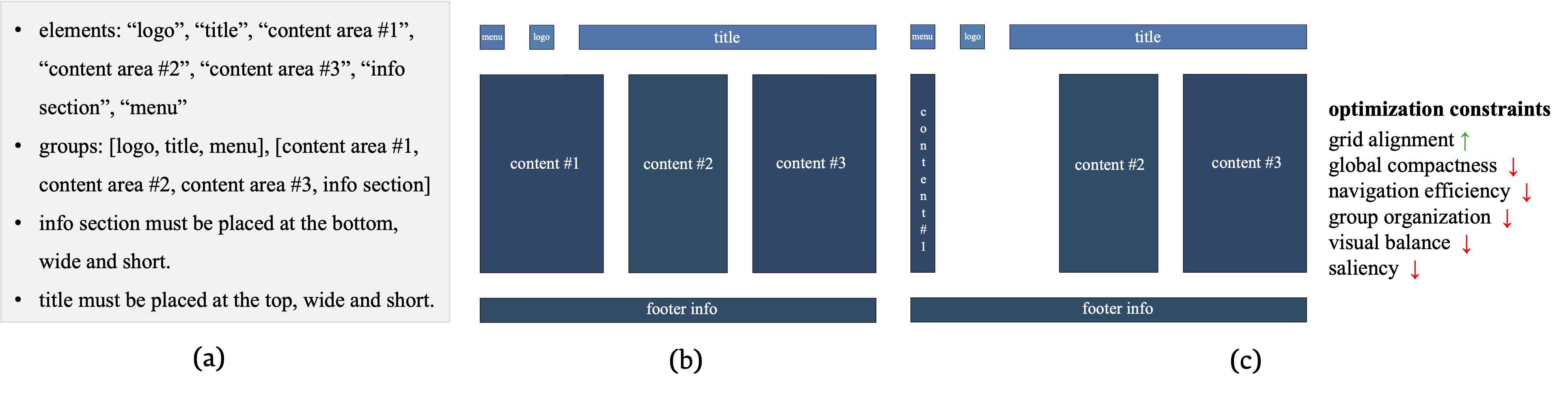}
    \caption{Overview of our ASP-based UI layout optimization workflow. (a) Designers begin with a sketch, partial layout, or set size, with declarative statements (e.g., alignment, grouping, contrast, screen size). (b) The solver generates an optimized layout using constraint categories grounded in psychological and visual design principles (grid alignment, global compactness, interaction efficiency, group organization, positional balance, saliency). (c) Designers can examine why a layout was generated or how an adjustment would affect the applied rules.}
    \label{fig:fig1}
\end{figure*}

Figure~\ref{fig:fig1} illustrates our approach to computational UI design with ASP.
Designers begin with sketches, partial layouts, or merely the desired number of elements, and encode principles such as alignment, grouping, and contrast into declarative statements.
The solver then generates optimized layouts, enabling quick exploration of alternatives and adaptation across devices.
Designers can also inspect the implemented \rvreplace{aesthetic rules}{design rules}, providing explainability.
This process supports a principled and transparent workflow in which constraints are directly interpretable.

In validation with online users, we show that ratings of visual quality clearly favor layouts generated with our ASP-based model over those produced with simple heuristics: encoding a broad set of aesthetic and cognitive constraints yields more appealing layouts.
However, our tool is not intended to replace expert intuition, which reflects years of tacit knowledge about aesthetics, usability, and cognitive principles of design.
Rather, it complements expertise by making these principles explicit, computationally testable, and actionable through optimization.
We present evidence that designers with intermediate experience rate the generated layouts highly and report that the constraint-based specification process encourages them to articulate design relationships more clearly.
Participants emphasize geometric relations, grouping, and automated color harmony as especially useful features, and note that the tool provides reference layouts that accelerate iteration and concept exploration.
We envision future integration of the tool into early-stage ideation and rapid prototyping workflows.
Beyond ideation, the system supports exploration of perceptual and cognitive design principles \emph{in silico}, enabling experts to evaluate trade-offs and adapt layouts efficiently for different devices or user needs.

In sum, the contributions of this paper are:
\begin{enumerate}
\item the first application of \emph{Answer Set Programming} (ASP) to GUI layout optimization, demonstrating its rich representation capabilities;
\item a formal set of aesthetic and cognitive design constraints operationalized in a declarative optimization pipeline; and
\item \rvreplace{two}{three} empirical studies with designers and end users that demonstrate\rvdel{the usability of our tool and} the perceptual benefits of ASP-optimized layouts \rv{and provide formative evidence for the design tool in early-stage ideation}.
\end{enumerate}
All code and experimental materials reported in this paper will be published open access in a Git repository upon publication.

%% file: 02-relatedwork.tex
\subsection{Psychological Grounding of UI Design Principles}
\label{sec:psyc_principles}

The idea that technological design choices should be psychologically justified has been articulated since at least the 1960s~\citep{cardnewell,engelbart1962program,licklider1960man,norman1988design}.
Here, we focus on the psychological grounding of GUI design, which requires integrating theories of perception, cognition, and motor control.
From a perceptual perspective, establishing a clear visual structure is essential for communicating information and preventing overload~\citep{watzman2002visual}.
Grids provide such structure by distributing visual elements along consistent alignment points~\citep{elam2004grid,darejeh2024cognitive,cowan2010magical}.
When elements share common horizontal and vertical alignment lines~\citep{balinsky2009aesthetic}, gaze can scan and process content more efficiently, helping manage perceptual complexity within the limits of visual working memory through systematic scanning patterns.

In addition to visual structure, GUI design must balance density and clarity: overly dense layouts feel cluttered and overwhelming, while sparse layouts can appear disconnected~\citep{watzman2002visual,white2011elements}.
Efficient use of white space communicates visual structure, such as grouping related information.
The center of mass principle, derived from physics but applied to visual perception, shows that users perceive the weight distribution of elements, shaping impressions of layout stability and balance~\citep{cholewiak2015perception,arnheim1954art}.
Eye-tracking studies further demonstrate that spatial relationships affect how users parse interfaces and navigate complex layouts~\citep{pernice2017fshaped,soegaard2021visual}.
In other words, the perceptual system organizes elements into coherent clusters based on spatial proximity and intervening empty space.
The Gestalt principle of proximity reflects this: elements near each other are perceived as a group, reducing the cognitive load of processing them individually~\citep{wagemans2012gestalt,graver2012best}.
Recent research using computational cognitive modeling has started to uncover the visual working memory structures that could explain this principle~\citep{sourulahti2026modeling}.

Human vision is \emph{foveated}, meaning it can accurately encode only a small portion of the visual field at a time~\citep{findlay1997saccade}.
During search, the visual system prioritizes fixating on elements with unique local features, making them \emph{salient} or likely to ``pop out''~\citep{wolfe2017five}.
Designers can exploit this effect to direct attention to important elements~\citep{leiva2020saliency,waugh2025standout}.
Common features that influence saliency include color, orientation, shape, and size~\citep{jokinen2020adaptive,wolfe2017five}.
Designers can also use feature familiarity; for example, if a function is strongly associated with a particular color, retaining that color in a new design helps guide users toward it~\citep{jokinen2020adaptive}.

Principles of color perception extend beyond saliency or subjective preference.
Certain color sets are perceived as more harmonious depending on their relative positions in color space~\citep{tokumaru2002color}.
Harmonious combinations can be computed using geometric templates applied to hue wheel sectors~\citep{cohenor2006color}, which provide criteria for assigning colors in layouts.
This does not imply that specific colors should be preferred, but rather that particular combinations are perceived as more pleasing.
Research in cultural psychology further shows that color perception varies across populations, emphasizing the need to adapt interface color design to each cultural context~\citep{elliot2014color}.

A common way for users to access GUI elements is through pointing with a mouse or touch.  
Efficient (fast) and effective (error-free) access plays a major role in overall usability.  
Thus, the design of visual elements must account for the variability of the human motor system.  
Fitts’ law is the classic formulation of the speed–accuracy tradeoff in pointing and has been validated across many interactive environments and modalities~\citep{mackenzie1992extending}.  
It captures psychomotor constraints by expressing pointing time as a function of movement distance and target size.  
Importantly, Fitts’ law can be parameterized for different contexts and individual differences~\citep{jastrzembski2007model}, enabling predictions of how element positioning and sizing affect usability across users and tasks.  

\subsection{Computational Optimization of UIs}

Early computational aids for graphical layout trace back to \emph{Sketchpad}~\citep{sutherland1964sketch}, which introduced geometric constraints for interactive graphics editing.
Sketchpad pioneered constraint solving for layout relationships and established the foundation for later constraint-based systems.
DeltaBlue~\citep{freeman1990incremental} extended this approach with incremental constraint hierarchies, maintaining solutions, as constraints were added or removed.
Cassowary~\citep{badros2001cassowary} advanced real-time layout solving for linear equality and inequality constraints, leading to adoption in layout engines such as Apple's Auto Layout.
Chorus~\citep{hosobe2001modular} generalized constraint solving to nonlinear geometric and graph-layout constraints through a modular architecture with soft constraint hierarchies.
These early systems demonstrated the potential of formal solvers for interface layout but largely emphasized numerical optimization, often requiring manual prioritization of constraints and offering limited flexibility for modeling abstract design rules.

Grammar-based and model-driven layout approaches emerged as alternatives.
Weitzman and Wittenburg’s Relational Grammars~\citep{weitzman1994automatic} encoded layout knowledge as grammar rules that map structured content to visual arrangements, enabling adaptive multimedia presentations.
DesignScape~\citep{o2015designscape} introduced an energy-based model to generate refinement and brainstorming suggestions for novice designers, prioritizing creativity support over optimality guarantees.
Sketchplore~\citep{todi2016sketchplore} combined sketching interfaces with optimization-based exploration, showing how solver feedback can aid design ideation.
These systems demonstrated the value of integrating computational reasoning with interactive exploration, but mostly relied on heuristic scoring or sampling, limiting both explainability and extensibility.

More recent work has advanced toward fully solver-driven optimization.
GRIDS~\citep{dayama2020grids} represents a key contribution, formulating layout design as a mixed-integer linear program that delivers real-time interactive performance, guarantees geometric constraint satisfaction, and supports diverse layout generation.
Interactive Layout Transfer~\citep{dayama2021interactive} extended this approach to design retargeting, validating layouts against style guides and templates.
However, these methods focus mainly on grid-based layouts and linear objectives; encoding richer design rules often requires major reformulation of the solver model, as their monotonic nature struggles to efficiently represent the defeasible "rule-of-thumb" logic inherent in design.
~\citet{shiripour2021grid} applied genetic algorithms for multi-objective optimization, incorporating nonlinear objectives such as clutter reduction and color harmony, but their stochastic approach lacks guarantees and frequently encounters infeasible solutions.
While solver-driven methods achieve high precision, they remain difficult to adapt to evolving constraints and new design semantics.

Machine learning-based methods take a complementary direction.
\citet{jiang2024graph4gui} introduced Graph4GUI, a graph neural network that learns element relationships and predicts layouts with contextual awareness, showing strong performance in design autocompletion tasks. 
\citet{jiang2022coarse} proposed a coarse-to-fine variational autoencoder that decodes layouts hierarchically, capturing both global and local structure and improving generation quality over flat generative models.
These methods, however, require large labeled datasets, provide limited interpretability, and struggle to enforce strict geometric or semantic constraints, limiting their applicability in high-stakes design settings.
FrameKit by \citet{wu2024framekit} shifted focus toward authoring workflows, offering programming-by-example keyframe interpolation for adaptive UI design.
While FrameKit provides flexibility for practitioners, it is not a solver and does not directly address optimization.

Research has also addressed the computational operationalization of visual aesthetics for evaluation.
\citet{miniukovich2015computation} proposed eight foundational metrics, including grid quality, contour congestion, and white space, showing that quantifiable properties correlate with perceived aesthetics.
The Aalto Interface Metrics (AIM) platform~\citep{oulasvirta2018aalto} aggregated 17 empirically validated measures, providing an online service and open-source codebase for image-based GUI evaluation. 
\citet{wang2025interpretable} advanced interpretability with the Visual Grouping Distribution metric, which combines DOM tree structure with pixel-based analysis to identify design issues and suggest improvements.
These approaches focus on assessment rather than optimization, relying on image-based pipelines that produce aggregate scores or visual overlays.

\rv{
Finally, our approach also shares a declarative foundation with a system called \emph{Clinguin}, which enables developers to construct domain-specific interactive interfaces directly in ASP by representing interface elements, attributes, and user-triggered operations as facts and rules ~\citep{beiser2025asp}.
It is principally an application framework for exposing and manipulating the evolving state of an ASP system through a graphical interface: its UI encoding specifies how domain states are rendered and how user events are communicated back to the solver.
Our work, in contrast, uses ASP as a computational design method that searches over alternative GUI layouts and selects among them according to explicit usability and aesthetic objectives grounded in HCI and psychology.
Although both approaches use ASP to describe interfaces, Clinguin concerns the declarative construction and operation of interactive frontends, while our contribution concerns layout generation, evaluation, and optimization based on metrics and simulations defined via computational interaction.
These roles are complementary: a Clinguin application could employ the proposed optimization model to generate its layout, or Clinguin could serve as an interactive frontend for exploring layouts produced by our model.}

Our work advances this state-of-the-art by adopting \emph{Answer Set Programming} (ASP) as a declarative solver framework for GUI layout optimization.
ASP combines the guarantees of integer programming with the modeling flexibility of logic programming, offering high elaboration tolerance ~\citep{mccarthy1998elaboration}. This allows us to encode qualitative, compositional, and conditional constraints that can be incrementally refined without restructuring the entire optimization pipeline. Unlike purely numerical approaches, ASP supports non-monotonic rule hierarchies, explicit reasoning, and enumeration of multiple optimal or near-optimal layouts, offering interpretable solutions that can be systematically adapted to new requirements.
Compared to data-driven methods, ASP requires no training data and guarantees that all generated layouts satisfy the encoded constraints, providing robustness in out-of-distribution design tasks.
We propose that this combination of expressivity, explainability, and solver guarantees positions ASP as a next-generation framework for computational GUI layout design.
In addition, as our approach explicitly implements aesthetic metrics such as grid alignment and color harmony, it serves both generative and evaluative purposes.

%% file: 03-definition.tex
\subsection{Layout Optimization with ASP}

\citet{lifschitz2008asp} introduced Answer Set Programming (ASP) as an approach to declarative problem solving
with many applications in academia and in industry \citep{erdem2016aspapplications}.
Rather than solving a problem by telling a computer how to solve it, the idea is to describe what the problem is and compute the solution.
This is carried out with a combination of a simple but rich modeling language
and high-performance solving capacities.
Modeling in ASP has its roots in the fields of knowledge representation and logic programming \citep{gebser2016modeling},
while solving is based in methods from deductive databases and boolean constraint solving \citep{kaufmann2016solving}.

Modern ASP systems,
such as \texttt{clingo}~\citep{gekakasc17a},
provide solver guarantees comparable to those of integer programming systems,
but offer a more flexible modeling language. 
%
GUI design involves a variety of constraints and preferences, 
such as non-overlap, alignment, grouping, whitespace, and color harmony.
Many of them are qualitative (e.g., ``the header must appear at the top''), compositional (e.g., nested groups of elements), or conditional (e.g., ``if a button is marked as primary, no other button in the same container may be primary'').
While numerical optimization techniques like integer programming excel at representing linear constraints,
extending them to handle such complex concepts can be cumbersome.
ASP offers a natural way to encode them declaratively,
without linearization or hand-crafted heuristics,
while still providing a \emph{complete} search algorithm. 
%
Moreover, ASP supports multi-objective optimization, 
enabling the solver to handle multiple competing design criteria simultaneously.


In ASP, a problem is represented by a \emph{logic program}, which is a set of rules.  
Such a program may have zero, one, or many \emph{answer sets}, each being a set of (logical) atoms.  
The basic idea of ASP is to model a problem by a logic program whose answer sets
correspond to the solutions of the problem,  
and then to use an ASP system to compute these answer sets.
When formalizing GUI design with ASP, the input provided by the designer can be represented by a set of \emph{facts}, 
the simplest type of rules:
\begin{lstlisting}
element(logo). element(title). ...
width(16). height(9). ...
\end{lstlisting}
%
These facts use the predicate \lstinline|element| to declare elements
such as \lstinline|title| or \lstinline|logo|,
as well as properties of the canvas, such as its relative
\lstinline|width| (\lstinline|16|) and \lstinline[breaklines=false]|height| (\lstinline[breaklines=false]|9|).
%
%
A program consisting only of these facts has a unique answer set:
\begin{lstlisting}
{ element(logo), element(title), ...,
  width(16), height(9), ... }
\end{lstlisting}
%

The following \emph{choice rule}
can be used to select the x-coordinate of the top-left corner of each element:
\begin{lstlisting}
{ x(E,X) : 1 <= X <= W } = 1 :-
  element(E), width(W).
\end{lstlisting}
For every element \lstinline|E| given width \lstinline|W|,
it chooses exactly one atom from the set
\lstinline|{ x(E,X) : 1 <= X <= W }|.
%
%
Similar rules can be specified for 
the y-coordinate, 
the width, and 
the height of each element. 
With this, the whole program selects one rectangle per element and, 
correspondingly, generates one answer set for each combination of rectangles. 

However, these rectangles may overlap.
To avoid this, a \emph{normal rule} is used to define which points \lstinline[breaklines=false]|(A,B)| are occupied by each element:\footnote{The code snippets presented in this section are illustrative 
and do not correspond exactly to the implementation used in the experiments. 
The actual system has been optimized for maintainability and performance.
Full model code will be made publicly available in a Git repository upon publication.}
\begin{lstlisting}
occupies(E,A,B) :-
  x(E,X), w(E,W), X <= A, A <= X + W,
  y(E,Y), h(E,H), Y <= B, B <= Y + H.
\end{lstlisting}
This rule says that element \lstinline|E| occupies point \lstinline[breaklines=false]|(A,B)|
if 
\lstinline|A| is between its x-coordinate \lstinline|X| and \lstinline|X + W|, 
and 
\lstinline|B| is between its y-coordinate \lstinline|Y| and \lstinline|Y + H|, 
where \lstinline|W| and \lstinline|H| are the width and height of \lstinline|E|, respectively.
The rule does not generate new answer sets, 
but extends each of them with the appropriate \lstinline|occupies(E,A,B)| atoms. 

Next, answer sets with overlapping rectangles can be eliminated using a \emph{constraint rule}:
\begin{lstlisting}
:- occupies(E1,A,B),
   occupies(E2,A,B), E1 != E2.
\end{lstlisting}
%
This constraint removes the answer sets 
where two different elements, \lstinline|E1| and \lstinline|E2|,
occupy the same point \lstinline[breaklines=false]|(A,B)|.



\subsection{Optimization Objectives}
\label{sec:objectives}
The optimization framework translates psychological principles into quantifiable objectives that guide layout generation towards both aesthetic and functional goals. 
Each objective encodes a specific design principle as a \emph{minimization statement} in ASP, 
possibly supported by auxiliary rules. 
These statements can be balanced against each other,
either by grouping them or by assigning them different priorities. 
This flexibility allows the modeler to navigate trade-offs between competing design goals. While a pre-defined set of priorities are provided as a configurable default, designers with differing preferences can easily rearrange the hierarchy to align with their specific design or project needs.
Modern ASP systems include specialized techniques for computing optimal answer sets of programs with minimization statements, 
allowing the solver to efficiently handle multiple objectives simultaneously.

\rv{
The pre-defined hierarchy ranks the objectives, from highest to lowest priority, as layout fundamentals (alignment and global compactness), interaction efficiency, group organization, positional balance, and size saliency. 
This ordering is motivated by both psychological and computational considerations.
Psychologically, it follows the coarse-to-fine trajectory of visual processing, in which viewers extract global structure before local detail \citep{wagemans2012gestalt,navon1977forest}. Structure-defining objectives therefore outrank refinements such as balance and saliency, which shape aesthetic impressions at later processing stages \citep{leder2004model}. 
Computationally, the solver resolves priorities from the top down, so placing structural objectives first prunes the largest portion of the search space (e.g., the grid alignment between elements) early and speeds up convergence to optimal layouts. 
We do not claim this ordering is universal. 
Instead, we treat it as a configurable default.
Designers can inspect and reorder the priorities directly, and Section 6 shows how the objectives and their priorities can be adapted to individual users and different design scenarios. 
This default hierarchy was used in both user studies reported later in this paper, providing preliminary evidence of its practical utility. 
The following subsections describe each objective within the hierarchy, including its underlying principle, intended contribution to GUI quality, and implementation in ASP.
}

\textbf{Grid Alignment}.
Grid alignment objectives translate the spatial organization principles from Subsection \ref{sec:psyc_principles} into computational form by reducing visual fragmentation through shared positioning. 
\rv{When element edges share a small set of alignment lines, the eye can scan content along consistent paths, keeping perceptual complexity within the limits of visual working memory~\citep{elam2004grid,balinsky2009aesthetic,cowan2010magical}. 
Fewer distinct lines thus make a layout appear more structured and easier to parse.}
To this end, 
the logic program is extended by a minimization statement that
counts distinct alignment lines across all elements, 
and minimizes their number: 
\begin{lstlisting}
#minimize {
  1,X : x(E,X);
  1,X : x(E,A), w(E,W), X = A + W
}.
\end{lstlisting}
For every answer set, this statement collects the 
pairs \lstinline[breaklines=false]|(1,X)| such that either 
\lstinline|X| is the x-coordinate of an element \lstinline|E|, 
or \lstinline|X| is the sum of that coordinate 
and the width \lstinline|W| of \lstinline|E|.
%
It then minimizes the sum of the first element of each such pair, 
thereby reducing the number of distinct alignment lines. 
The full program includes an analogous statement for the y-coordinate, 
and also captures local alignment within groups.




\textbf{Grouping}.
The grouping objective implements Gestalt proximity principles 
by encouraging related elements to cluster together visually. 
\rv{Proximity is a well-established Gestalt principle: elements placed near one another are perceived as belonging together, allowing users to parse functional regions rather than process each element separately ~\citep{wagemans2012gestalt}. 
Subsection \ref{sec:visualsearch} offers additional non-heuristic support, showing that grouping can emerge when layouts are optimized with a visual search simulator.}
To achieve this, 
the program defines for each element \lstinline|E| 
its center 
\lstinline[breaklines=false]|(X,Y)| and 
the centroid \lstinline[breaklines=false]|(GCX,GCY)| of its group \lstinline|G|, 
and minimizes the distance \lstinline|D| between both:
\begin{lstlisting}
#minimize {
  D,E :
    element_center(E,X,Y),
    element_group(E,G),
    group_centroid(G,GCX,GCY),
    D = |X - GCX| + |Y - GCY|
}.
\end{lstlisting}
%
The proximity principles also extend to hierarchical grouping, allowing subgroups to nest within parent groups. Elements are simultaneously optimized to cluster around the centroids of both their immediate subgroup and the enclosing parent group, ensuring cohesive structure at multiple scales.

\textbf{Saliency}.
Saliency objectives implement both size-based and color-based visual prominence through contrast maximization mechanisms.
\rv{Because human vision can encode fine detail only within a small region of the visual field at a time, visual search prioritizes elements whose features contrast with their surroundings~\citep{findlay1997saccade,wolfe2017five}. 
Making important elements larger therefore draws attention to them, while similar sizes among the remaining elements keep them from competing for it.}
Size saliency maximizes the dimensional differences between important ("salient") and unimportant ("non-salient") elements while minimizing size variation within each category. 
Accordingly, the logic program uses a dual optimization approach that maximizes differences between categories,
while minimizing differences within categories. 
This ensures that key elements stand out clearly while maintaining visual consistency among less important elements. 
The following minimize statement 
encodes the size saliency between elements 
\lstinline|E1| and \lstinline|E2| 
of different categories. 
Using 
auxiliary definitions of their width and height differences \lstinline|WD| and \lstinline|HD|, respectively, 
it maximizes their sum \lstinline|D| by minimizing \lstinline|-D|:
\begin{lstlisting}
#minimize {
  -D,E1,E2 :
    salient_element(E1),
    not salient_element(E2),
    width_difference(E1,E2,WD),
    height_difference(E1,E2,HD),
    D = WD + HD
}.
\end{lstlisting}

\textbf{Compactness}.
Compactness objectives address the balance between information density and spatial efficiency 
by regulating whitespace at multiple levels. 
\rv{Dense layouts can appear cluttered, while sparse layouts can appear disconnected. 
Whitespace therefore communicates structure by separating elements while preserving their perceived relationships~\citep{watzman2002visual, white2011elements}.}
The program enforces minimum distances between adjacent elements within groups to ensure clear separation. 
Between different groups, additional spacing constraints create distinct functional regions within the layout. 
In addition, a global compactness objective minimizes empty space within the layout's bounding box, 
and targets an optimal ratio between occupied area and total canvas size, ensuring efficient use of available space.
This is represented by the following statement, 
that relies on the definition of the area 
\lstinline|BA| of the bounding box of the layout. 
\rvdel{and assumes an ideal ratio, which can be set by the designer, and is here set to be $80\%$}
\rv{The statement assumes a designer-specified ideal ratio, set here to $80\%$ as a pragmatic default rather than a universal optimum}:
\begin{lstlisting}
#minimize {
  D :
    width(W), height(H),
    bounding_box_area(BA),
    R = (BA * 100) / (W * H),
    D = |R - 80|
}.
\end{lstlisting}
%


\textbf{Positional Balance}.
This objective optimizes the spatial distribution of elements and groups relative to key 
positions on the canvas. 
\rv{People perceive the distribution of visual weight in a layout much like the mass of a physical object ~\citep{arnheim1954art,cholewiak2015perception}. 
When this weight is evenly distributed around a reference point, the layout appears stable and aesthetically pleasing ~\citep{bauerly2006computational}.}
For each type of layout alignment
(centered, left or right)
the program minimizes 
the distance \lstinline|D| 
from the centroid \lstinline[breaklines=false]|(GCX,GCY)| 
of each group \lstinline|G|
to the reference position \lstinline[breaklines=false]|(X,Y)| 
of the alignment, 
weighted by the size \lstinline|S| of the group:
\begin{lstlisting}
#minimize {
  D * S,G :
    layout_alignment(LA),
    alignment_position(LA,X,Y),
    group_size(G,S),
    group_centroid(G,GCX,GCY),
    D = |GCX - X| + |GCY - Y|
}.
\end{lstlisting}
%
%
%

\textbf{Interaction Efficiency}. 
Selection time objectives implement Fitts' Law
in order to 
minimize interaction effort between element pairs. 
\rv{Fitts’ law is a well-established model of the speed–accuracy tradeoff in aimed movement ~\citep{mackenzie1992extending}. Minimizing predicted movement time therefore directly serves interaction efficiency.}
The program minimizes the movement time \lstinline|MT|
from element \lstinline|E1| to \lstinline|E2|, 
that is defined
as a function of 
their center-to-center distance \lstinline|D| and \lstinline|E2|'s minimum dimension \lstinline|MD|: 
\begin{lstlisting}
#minimize {
  MT,E1,E2 :
    element(E1), element(E2),
    element_distance(E1,E2,D),
    minimum_dimension(E2,MD),
    MT = @fitts_law(D,MD)
}.
\end{lstlisting}
    
%
The function is represented by the external function 
\lstinline|@fitts_law|, 
which is implemented in Python and returns 
$a + b \cdot log_2(D/MD + 1)$,
where $a$ and $b$ are empirical constants that describe the user and the interaction modality \citep{mackenzie1992extending}.
%
%
This hybrid approach enables the solver to incorporate 
more involved mathematical computations,
while keeping the declarative specification that makes ASP particularly effective for design optimization.
Below, in Subsection \ref{sec:who}, we demonstrate how this objective can be extended to other computational models, such as for minimizing pointing errors.

\subsection{Design Specifications}
\textbf{Geometric Relations}. 
The program supports several optional constraints that provide additional design control. 
Geometric relations allow designers 
to specify spatial dependencies between elements, 
supporting 22 types of relationships, 
including 
pairwise positional constraints (``above'', ``below'', ``left of''), 
pairwise alignment constraints (``left-edge aligned'',``horizontally centered''), and 
global constraints (``top'', ``bottom'').
For example, the fact  
that the title should be at the top 
can be expressed by:
\begin{lstlisting}
relation(top,title).
\end{lstlisting}
and enforced by the constraint:
\begin{lstlisting}
:- relation(top,E1),
   element(E2),
   y(E1,YE1), y(E2,YE2),
   YE2 < YE1.
\end{lstlisting}
This eliminates answer sets in which \lstinline|E1| 
is required to be at the top
but there is an element \lstinline|E2| whose
y-coordinate is smaller than that of \lstinline|E1|.

\textbf{Element Dimensions}.
The program allows elements to be constrained to flexible size categories ("small", "medium", "large"). These categories map to percentage-based ranges tailored to the canvas dimensions (e.g., "large" allows widths between 50–100\% of the canvas). Designers can assign one of three size categories to each element. The solver is then free to optimize the exact pixel dimensions within these bounds to satisfy other objectives, such as aligning with a grid or minimizing whitespace.
For example, the requirement that the width of a title should be large can be expressed by:
\begin{lstlisting}
width(title,large).
\end{lstlisting}
and enforced by constraints that map the abstract size category to concrete pixel ranges based on the canvas dimensions: 
\begin{lstlisting}
option_size(large,50,100).

:- width(E,Size),
   option_size(Size,MIN,MAX),
   w(E,W),
   not MIN <= W <= MAX.
\end{lstlisting}

\textbf{Color Harmony}. Color harmony guides palette creation according to seven established harmonic templates (V-type, Y-type, etc.) proposed by~\citet{cohenor2006color}, ensuring aesthetically consistent color schemes, while maintaining sufficient contrast for functional distinction. These templates define geometric sector patterns on the color wheel that determine which colors work harmoniously together. For example, a V-type chooses adjacent colors from a single large sector of \(93.6^{\circ}\), forming an analogous color scheme. 
Designers specify a primary color that serves as the thematic foundation for the layout and select a template, and the program harmonizes all interface elements based on the selected template.
To achieve this, the program determines a single salient hue and a set of palette hues. It then assigns hues to elements based on their saliency status. The assignment of the salient hue to all salient elements can be expressed by:
\begin{lstlisting}
assign_hue(E,H) :-
  salient_element(E),
  salient_hue(H).
\end{lstlisting}
and for non-salient elements, the solver chooses one hue from the generated palette:
\begin{lstlisting}
{ assign_hue(E,H) : palette_hue(H) } = 1 :-
  element(E),
  not salient_element(E).
\end{lstlisting}

Beyond hue assignment, the program enforces color saliency through saturation and lightness constraints to ensure visual hierarchy. For example, ensuring that salient elements maintain high saturation for visibility can be enforced by:
\begin{lstlisting}
element_saturation(E,70) :-
  element(E),
  salient_element(E).
\end{lstlisting}

These optimization objectives and design specifications listed in Subsections 3.2 and 3.3 are combined within the logic program 
that is processed by the ASP solver, 
producing layouts that balance competing requirements according to designer-specified priorities and requirements. The declarative specification enables rapid experimentation with different objective combinations and weights. This supports iterative refinement of design criteria without requiring algorithmic re-implementation.

All metrics implemented by our approach are listed in Table \ref{tab:objectives}, which also analyzes what existing approaches implement, based on explicit statements in the respective papers.
The table also marks cases where a feature is implicit.
For instance, in GRIDS \citep{dayama2020grids}, geometric relations can be implied by fixing multiple elements in place, but there is no explicit element relation objective. 
\rv{The objective set is not intended to be exhaustive. It covers the psychologically grounded principles reviewed in Subsection 2.1 and several objectives used in prior computational design systems (Table 1). 
Because each objective is an independent module over a shared vocabulary, the set can be extended without restructuring the pipeline.}

\subsection{Advanced Simulator Integration}
\label{sec:simulators}
To move beyond static geometric heuristics, we facilitate integration of dynamic cognitive simulators directly into the optimization loop via a custom \texttt{clingo} propagator~\citep{kaminski2020howto}.
The propagator intervenes directly during the solving process, and monitors the assignment of simulator-relevant, layout-defining atoms.
Whenever the solver derives a candidate layout (either a complete answer set or a sufficiently grounded partial assignment), the propagator triggers the external cognitive simulator, which takes as its input the layout specification, and returns the metrics it is designed to evaluate.
Our implementation can be adapted to any simulator; here, we demonstrate it using a computational cognitive model of visual search~\citep{sourulahti2026modeling}.

This integration of ASP-based optimizer and external simulator fundamentally changes how design principles are handled.
In the example we present in this paper, we demonstrate how the Gestalt principle of grouping can emerge non-heuristically from optimizing a layout design by minimizing search time with a visual search model that accounts for element groupings.
Above, in Subsection \ref{sec:objectives}, grouping was enforced through rigid geometric constraints that require specific minimization of centroid distances and mandatory whitespace buffers.
With the visual search simulator used in Subsection \ref{sec:visualsearch}, these explicit rules can be relaxed or removed.
The simulator is based on computational principles, resulting in simulating human-like visual search that adapts to the presence of grouping.
It predicts faster search times for layouts where related elements are spatially proximate or share visual features (e.g., color), as these configurations reduce the load on visual working memory.
Consequently, when optimizing for search time, the system naturally converges on layouts that exhibit strong grouping properties without requiring explicit grouping constraints.


\begin{table*}[ht]
\centering
\small
\begin{tabular}{p{3.5cm}ccccccc}
\hline
\textbf{Objective / Constraint} &
\makecell{\textbf{GRIDS}\textsuperscript{1}} &
\makecell{\textbf{Transfer}\textsuperscript{2}} &
\makecell{\textbf{Genetic}\textsuperscript{3}} &
\makecell{\textbf{Graph4GUI}\textsuperscript{4}} &
\makecell{\textbf{VAE}\textsuperscript{5}} &
\makecell{\textbf{ASP}} \\
\hline
Grid Alignment         & \checkmark & \checkmark & \checkmark & \checkmark & \textbf{*} & \checkmark \\
Grouping (Gestalt)     & \textbf{*} & \textbf{*} & \checkmark & \checkmark & \textbf{*} & \checkmark \\
Visual Saliency &  \textbf{*} &  & \checkmark &  & & \checkmark \\
Compactness / Whitespace & \textbf{*} & \textbf{*} & \textbf{*} &  & & \checkmark \\
Positional Balance     &  &  &  &  &  & \checkmark \\
Selection (Fitts, etc.) &  &  & \checkmark  &  &  & \checkmark \\
Element Dimensions     & & \textbf{*} & \checkmark  & \textbf{*} & & \checkmark \\
Geometric Relations    & \textbf{*} & \textbf{*} &  & \textbf{*} & \textbf{*}  & \checkmark \\
Color Harmony          &  &  & \checkmark &  &  & \checkmark \\
Cognitive Simulators & & & & & & \checkmark \\
\hline

\end{tabular}
\caption{Comparison of optimization objectives across layout optimization methods and our ASP-based model. A checkmark (\checkmark) indicates that the method explicitly optimizes for the feature. An asterisk (\textbf{*}) indicates cases where a feature is implicit.
\textsuperscript{1}\citep{dayama2020grids}
\textsuperscript{2}\citep{dayama2021interactive}
\textsuperscript{3}\citep{shiripour2021grid}
\textsuperscript{4}\citep{jiang2024graph4gui}
\textsuperscript{5}\citep{jiang2022coarse}.}
\label{tab:objectives}
\end{table*}

%% file: 04-experiment1.tex

The goal of this experiment was to evaluate whether our ASP-based optimization method produces GUI layouts that are perceived as visually superior to those generated under simpler optimization conditions from the perspective of general internet users. \rv{In addition to the main experiment, we conducted a follow-up experiment that validated the rating question against an established multi-item scale and tested whether the effect of the optimization condition replicates (see Subsection \ref{sec:validation}).}

\subsection{Method}

\subsubsection{Participants}
We recruited $N=100$ participants (mean age = 32 years; 54 identified as female) through the Prolific platform.
All participants were required to have normal or corrected-to-normal vision, and be fluent in English.
No design background was required from the participants, but it also was not explicitly prohibited.
They were compensated at a rate consistent with Prolific guidelines.
\rvdel{Three participants failed to complete the tasks properly, resulting in the final analysis of 97 participants.} \rv{Three participants were excluded because the task had been launched twice within their session (e.g., in a second browser window), corrupting their trial logs; 97 participants remained for final analysis.} 

\subsubsection{\rv{Design and }Materials}

\rv{The experiment followed a one-factor within-subjects design. 
The independent variable was the layout optimization condition with three levels (\emph{Random}, \emph{Grid}, and \emph{Full}), defined by the set of optimization objectives enabled during stimulus generation. 
The stimuli covered ten layout specifications, five simple (4--5 elements) and five complex (8--9 elements), with each specification optimized once under every condition, yielding 30 wireframes in total. 
Layout specification and complexity provided stimulus variety and were not analyzed as design factors. Each participant rated all 30 wireframes in a newly randomized order. 
The dependent variable was a single item, ``How would you rate the visual quality of the user interface?'', answered on a scale from 1 (``Very Poor'') to 10 (``Excellent''). 
The primary analysis was a multilevel model with the optimization condition as the fixed effect.}

\rvdel{We generated 30 wireframe GUI layouts, divided equally into three groups:}\rv{The three optimization conditions were:}
\emph{Random}, which only satisfied color harmony, non-overlap, and semantic constraints (e.g., header placement);
\emph{Grid}, which additionally optimized grid alignment; and
\emph{Full}, which used our full optimization model\rv{, further optimizing grouping, visual saliency, compactness, positional balance, and interaction efficiency (see Subsection 3.2)}.
\rv{\emph{Random} and \emph{Grid} thus serve as ablations of our full model rather than external baselines: the three conditions share the same pipeline and input specifications, and differ only in the objectives enabled, isolating the contribution of the objectives from implementation differences. Each element was labelled with its semantic role (e.g., header, sidebar, main content, footer, button, status bar), with larger elements carrying a short description of their prospective content, so the wireframes presented the kinds of content found on a typical webpage.}
We maintained color harmony in all groups, as a fully random color palette would have confounded the results in the \emph{Random} group.
\rvdel{For additional variety, half of the stimuli in each group had 4 elements, and the other half 8 elements.}

\begin{figure*}[htbp]
    \centering
    \begin{minipage}{0.3\textwidth}
        \centering
        \includegraphics[width=\linewidth]{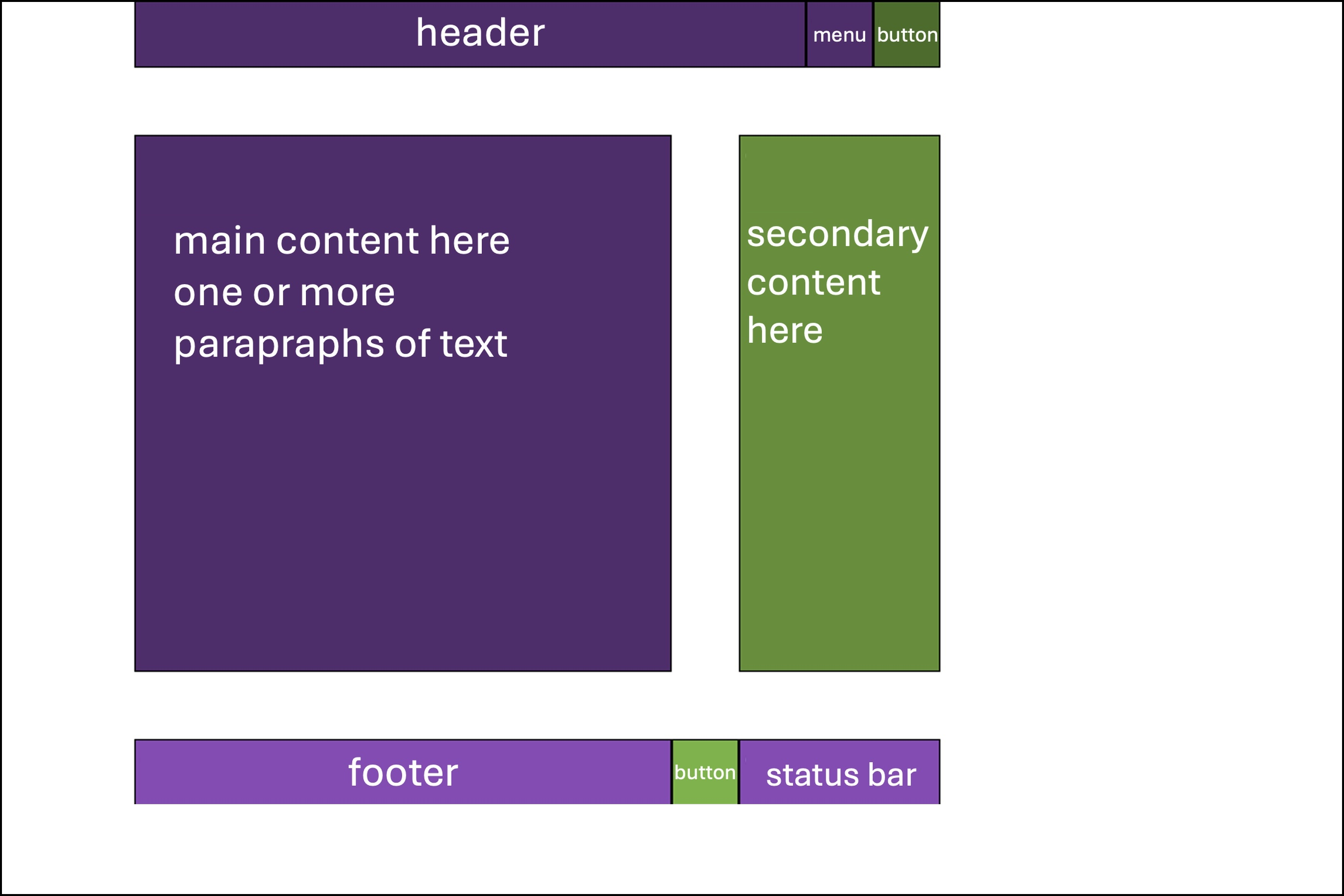}
        \vspace{3pt}
        {\small (a) Full}
    \end{minipage}
    \hfill
    \begin{minipage}{0.3\textwidth}
        \centering
        \includegraphics[width=\linewidth]{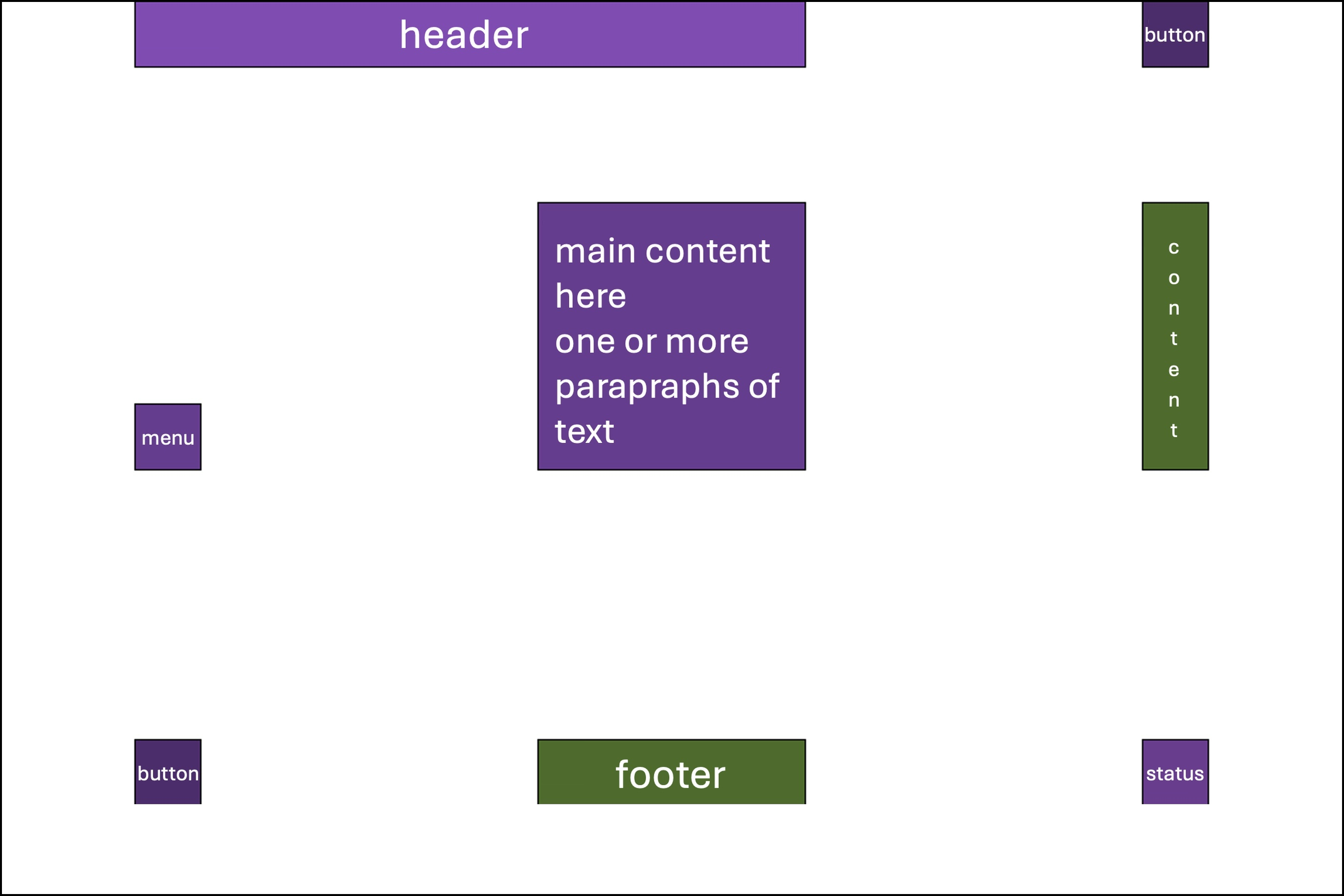}
        \vspace{3pt}
        {\small (b) Grid}
    \end{minipage}
    \hfill
    \begin{minipage}{0.3\textwidth}
        \centering
        \includegraphics[width=\linewidth]{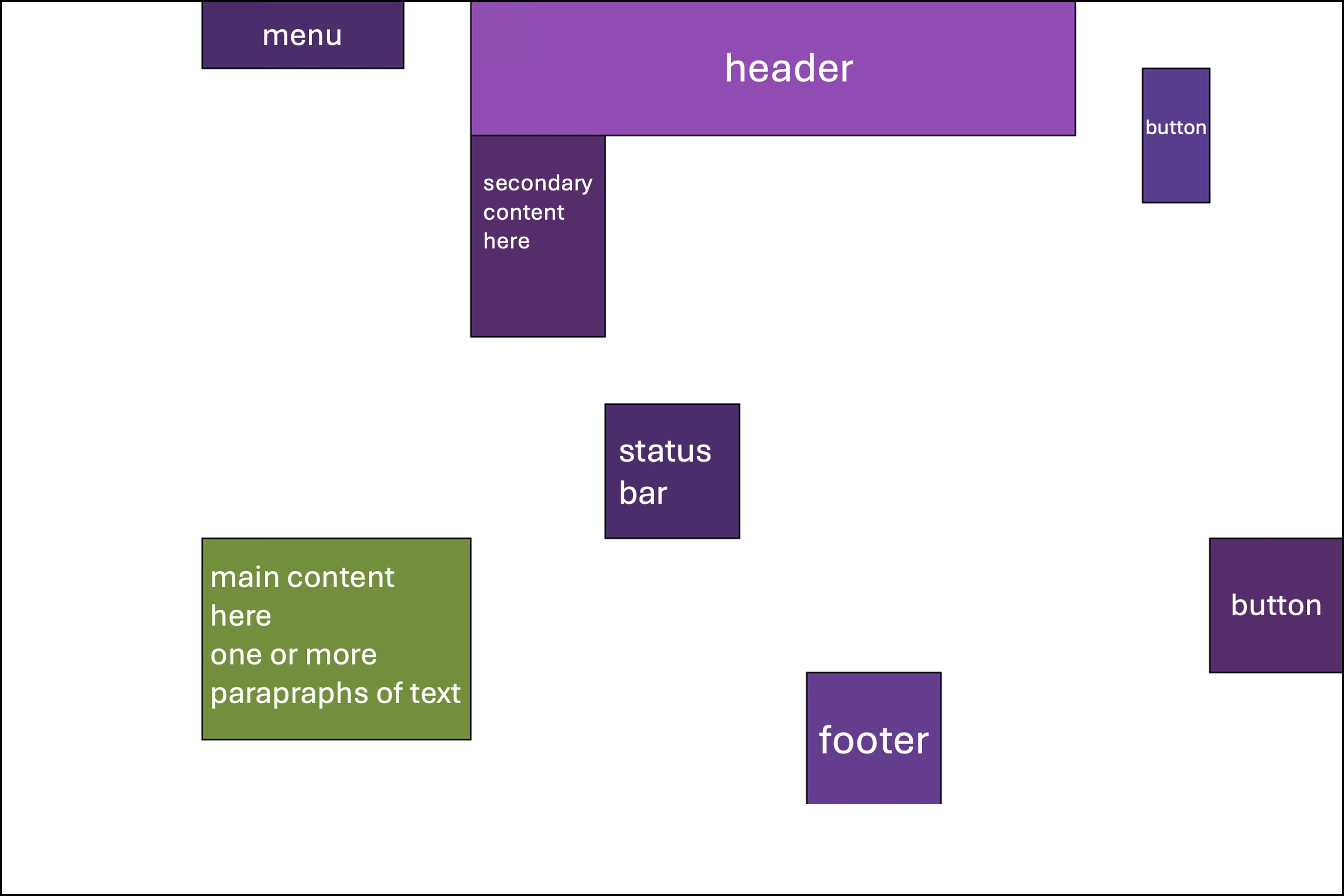}
        \vspace{3pt}
        {\small (c) Random}
    \end{minipage}
    \caption{One complex layout specification realized under the three experimental conditions: (a) Full, (b) Grid, and (c) Random.}
    \label{fig:exp1_stimuli}
\end{figure*}

\rv{All 30 stimuli were produced with the same pipeline. 
A layout specification fixed the canvas size, the elements with their semantic roles and size categories, and a small number of semantic positioning constraints. 
Each condition was defined in an objective file that enabled the corresponding optimization objectives. 
We optimized all ten specifications under all three objective files; the best layout from every run became a stimulus.
For a given specification and objective file, the optimizer always returns the same final layout, so the stimuli could not be selected among alternative outputs. 
The ten specifications themselves were authored iteratively: we inspected generated layouts and adjusted the specifications where needed, aiming for a diverse stimulus set. 
Since the final specifications were used identically in all three conditions, this process affects which layouts were tested, but not the comparison between conditions.
Figure~\ref{fig:exp1_stimuli} shows one complex specification realized under the three conditions; the full stimulus set is reproduced in Appendix~A.}

\subsubsection{Procedure}
To familiarize participants with wireframes, we displayed an example showing a news site screenshot alongside its corresponding wireframe abstraction.
\rv{This example served only to illustrate the wireframe abstraction before the task; the stimuli themselves were presented and rated as generic layouts, with no reference to a specific application domain.}
\rvdel{Participants rated layouts on a single-item scale from 1 (``Poor'') to 10 (``Excellent'').}
Each participant rated all 30 layouts, one at a time, in random order.
The on-screen instructions \rvdel{emphasized judging overall visual quality, organization, and usability.}
\rv{asked participants to rate the overall visual quality of each layout; to make this judgment easier, the instructions suggested factors to consider, such as organization, balance, and ease of use and understanding.}
The experiment was completed online in a single session.
\rv{Because the session was short and self-paced, we did not include scheduled breaks or attention checks.}

\subsection{Results}
\begin{figure}[ht]
    \centering
    \includegraphics[width=0.7\linewidth]{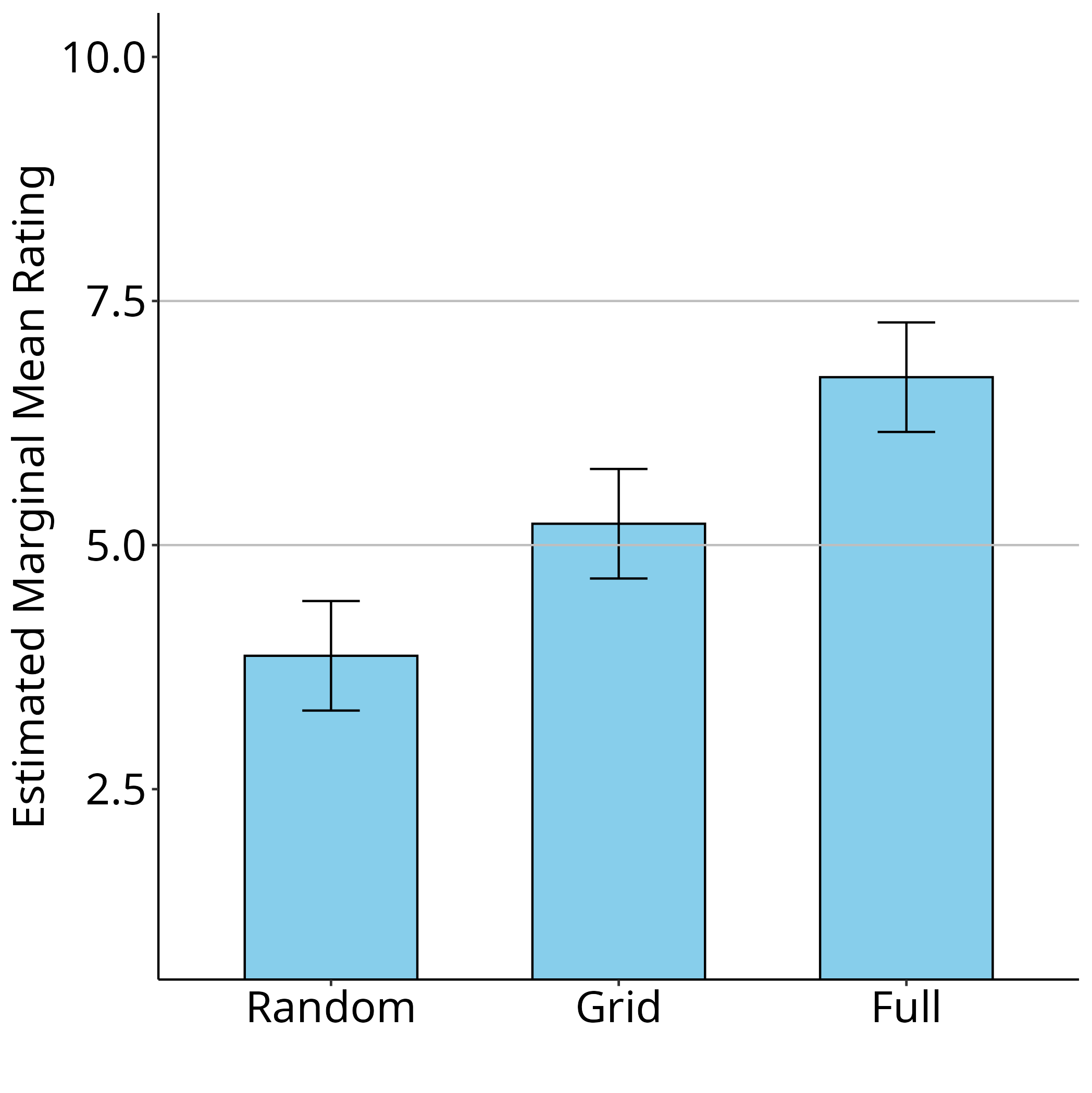}
    \caption{Estimated marginal mean ratings per group with 95\% CIs. Full optimization achieved the highest ratings, random layouts the lowest.}
    \label{fig:bargraph}
\end{figure}

\begin{figure}[ht]
    \centering
    \includegraphics[width=0.7\linewidth]{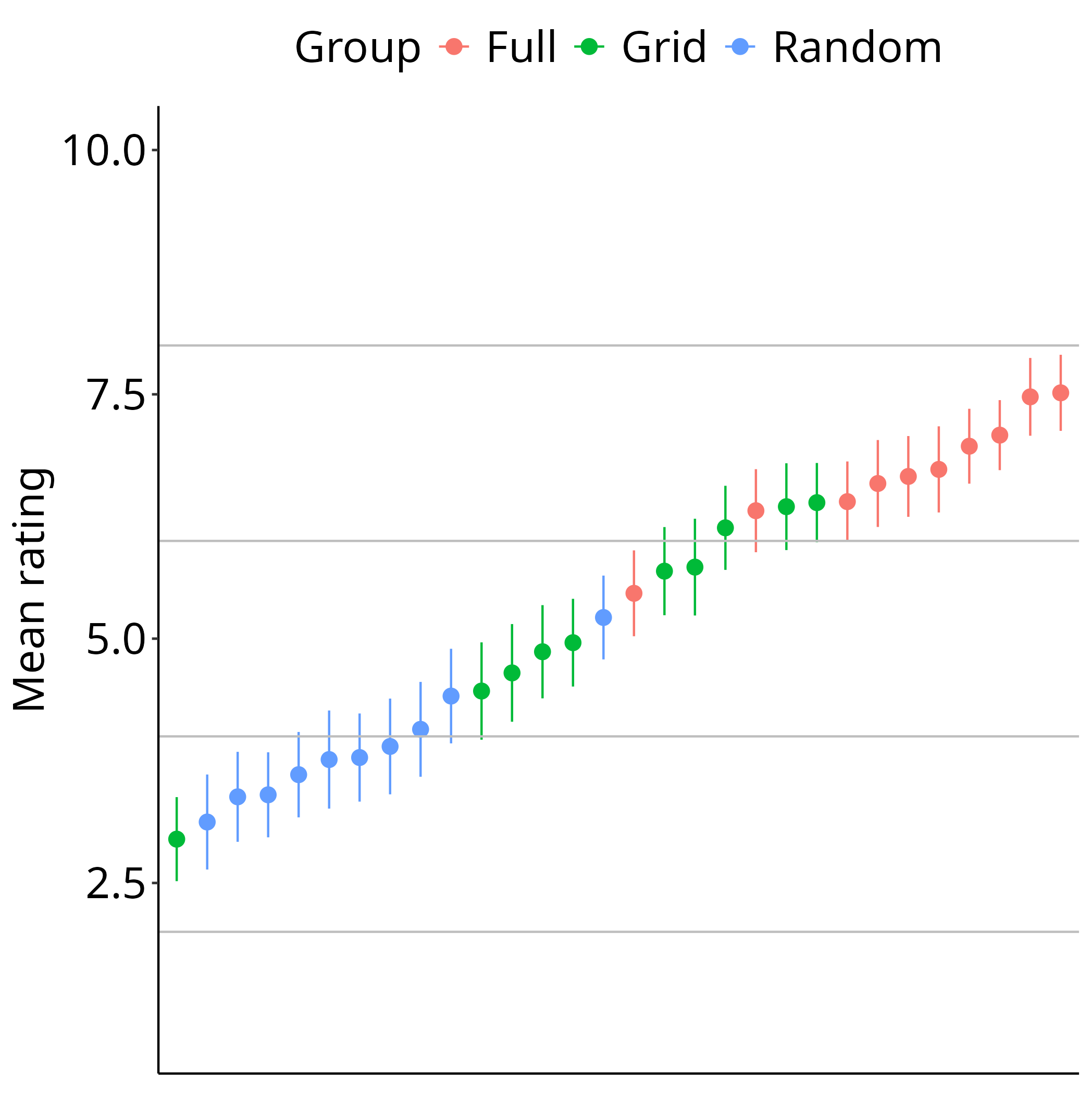}
    \caption{Mean rating of each individual stimulus, colored by condition, with 95\% CIs.}
    \label{fig:stimuli-means}
\end{figure}

We analyzed ratings using a multilevel model with fixed part for the optimization condition and random intercepts for participants and stimuli using \texttt{lme4}.
The intraclass correlation coefficients in the random part were 0.27 for the participant and 0.11 for stimulus.
The effect of condition on ratings was significant,
$F(2, 27)=32.9$, $p<.001$.
Figure~\ref{fig:bargraph} shows estimated marginal mean ratings with 95\% confidence intervals:
layouts from the \emph{Full} group were rated highest, followed by \emph{Grid}, with \emph{Random} rated lowest.
Pairwise comparisons confirm significant differences between \emph{Full} and both alternatives with large effect sizes
(Cohen’s $d = 0.81~\rv{[0.42, 1.20]}$ for \emph{Full} vs \emph{Grid}, and $d = 1.54~\rv{[1.15, 1.93]}$ for \emph{Full} vs \emph{Random}; \rv{95\% confidence intervals in brackets}).
Figure~\ref{fig:stimuli-means} shows per-stimulus means: \emph{Random} layouts cluster at the low end, \emph{Full} layouts at the high end, and \emph{Grid} layouts are intermediate. 
\rv{Participants completed the 30 rating trials in 3.59 min on average (SD = 1.60; median = 3.19 min), with a median of 5.0 s per trial. Trial completion times, analyzed with the same multilevel model on log-transformed times, did not differ between conditions, $F(2,27) = 1.31$, $p = .29$.}

\subsection{Discussion}
The results demonstrate that our optimization method reliably improves perceived visual quality of GUI layouts.
Layouts optimized with the full model were consistently rated higher than both the \emph{Grid} and \emph{Random} conditions, indicating that encoding a rich set of aesthetic and cognitive constraints yields meaningful improvements in perceived design quality.
The spread of per-stimulus means (Figure~\ref{fig:stimuli-means}) further supports this conclusion:
\emph{Full} layouts dominate the top of the scale, while \emph{Random} layouts cluster at the bottom.
\rv{Completion times also indicate that these judgments were made quickly and at a similar pace across conditions, suggesting that the rating advantage of \emph{Full} layouts was not accompanied by longer deliberation.} 
These findings validate our approach and justify its use as a foundation for further model-based layout generation.

\rv{The generalizability of our results to GUIs overall should be treated with caution.
We set an a priori recruitment target of 100 participants based on methodological judgment and the information provided by the repeated-measures design, in which every participant evaluated all 30 layouts, yielding approximately 3,000 ratings before exclusions.
No formal simulation-based power analysis was conducted before data collection.
Such an analysis would have required prior estimates of both participant- and stimulus-level variance that were not available for this novel task.
We therefore evaluated the precision of the resulting estimates using confidence intervals for the model contrasts and effect sizes.
However, due to the large variety of interfaces used in daily lives, the study does not show that our approach generalizes to all layouts; rather, it shows that given the wireframe designs we chose, the optimization reliably produces more highly rated layouts. 
These wireframes represent static, structured page layouts with clearly delineated elements. 
Whether the results extend to designs with less visible grids, such as large imagery or scroll-driven storytelling, or to interaction-heavy interfaces, remains to be tested.}

\rv{\subsection{Validation and Replication Experiment}
\label{sec:validation}
A single-item measure raises two questions: whether one item can capture perceived visual quality reliably, and what participants were in fact judging. 
We therefore conducted a follow-up experiment in which a subset of the original stimuli was rated again by a new sample, using both the original question and an adapted version of the VisAWI-S, a validated multi-item measure of perceived visual aesthetics \citep{moshagen2013short}. 
The scale includes one item per facet: simplicity, diversity, colorfulness, and craftsmanship. 
The experiment had two goals: estimating how closely ratings on the original question agree with the validated scale, 
and testing whether the effect of the optimization condition replicates.}

\rv{We recruited $N = 41$ participants (mean age = 33 years; 19 identified as female) through the Prolific platform under the same eligibility and compensation terms as the main experiment; 
participants of the main experiment were screened out, so that all raters were naive to the stimuli. 
No participants were excluded from the analysis. 
The sample size was determined a priori to provide at least 80\% power to detect a large correlation ($r = .50$) between the original single-item measure and each of the four VisAWI-S items, using a Bonferroni-adjusted significance level of $\alpha = .0125$ ($.05/4$). 
For the analyses reported below, the mean VisAWI-S score was used as the primary validation measure, while associations with the four individual items were examined as secondary analyses.}

\rv{The experimental design and procedure were identical to the main experiment (see Subsections 4.1.2 and 4.1.3), with two changes. 
First, we used 15 of the 30 wireframes. 
From each condition, we selected five wireframes: those with the lowest and highest mean rating in the main experiment, and three spaced evenly in between (marked in Appendix~A). 
The subset thus spans the observed rating range within each condition as well as overall.
Second, the experiment consisted of two rating sections using the same 15 wireframes. 
In one section, participants answered the original visual-quality question for each wireframe, whereas in the other section, they answered the four items of the VisAWI-S. 
The order of the two sections was counterbalanced across participants, and the order of the 15 wireframes was randomized separately within each section. 
Thus, each wireframe was presented twice, once with each rating measure. 
We retained the four VisAWI-S items but adapted their response scale from the original 7-point format to a 10-point agreement scale, matching the response range of the original visual-quality question.}

\rv{The four VisAWI-S items showed high internal consistency in the adapted 10-point format (Cronbach’s $\alpha = .94$). 
We analyzed agreement by predicting the single-item rating from the mean VisAWI-S score in a multilevel model with crossed random intercepts for participants and stimuli using \texttt{lme4}. 
The intraclass correlation coefficients in the random part were 0.15 for the participant and 0.16 for stimulus. 
The original single-item rating was associated with the mean VisAWI-S score after accounting for participant- and stimulus-level differences (standardized $\beta = .49$, $p < .001$).
The same pattern was observed for each individual VisAWI-S item, with standardized coefficients ranging from $\beta = .36$ to $.44$ (all $p < .001$). 
At the stimulus level, the mean single-item and VisAWI-S ratings in the follow-up experiment were highly correlated ($r = .98$ $[.95, .99]$, $p < .001$; 95\% confidence intervals in brackets).
The mean VisAWI-S ratings in the follow-up experiment were also strongly correlated with the single-item ratings from the main experiment ($r = .96$ $[.88, .99]$, $p < .001$). 
The single-item ratings were highly consistent across the two experiments ($r = .98$ $[.95, .99]$, $p < .001$).}

\rv{The effect of optimization condition also replicated for both measures, which we analyzed with the same multilevel model as in the main experiment. 
For the single-item ratings, the effect of condition was significant, $F(2,12) = 8.73$, $p = .005$, with \emph{Full} rated highest, followed by \emph{Grid} and \emph{Random}. 
\emph{Full} differed significantly from \emph{Random} with a large effect size (Cohen's $d = 1.52$ $[0.73, 2.32]$, Tukey-adjusted $p = .003$); the other pairwise comparisons were not significant. 
The VisAWI-S scores showed the same ordering, with a significant effect of condition, $F(2,12) = 8.03$, $p = .006$. 
\emph{Full} again differed significantly from \emph{Random} with a large effect size ($d = 1.43$ $[0.65, 2.21]$, Tukey-adjusted $p = .005$); the other pairwise comparisons were not significant. 
Participants completed the 30 rating trials in 7.03 min on average (SD = 3.49; median = 6.37 min), with median completion times of 5.47 s per single-item trial and 13.7 s per VisAWI-S trial.}

\rv{These results support the original single-item question as a measure of perceived visual quality. 
After accounting for participant- and stimulus-level differences, its association with the mean VisAWI-S score was strong, and the high stimulus-level correlation shows that the layouts were ordered very similarly by the two measures. 
This agreement was also observed across independent samples: the single-item ratings from the main experiment were strongly related to the VisAWI-S ratings from the new sample.
Together with consistent associations across all four facets, these results indicate that the single question captured a broad aesthetic judgment rather than primarily reflecting one aspect of the layouts. 
At the same time, because the stimuli were static and were not used interactively, these findings concern perceived visual quality and do not establish the question as a measure of usability. 
Beyond validating the measure, the similar ordering of optimization conditions across the single-item and VisAWI-S measures provides additional evidence that the advantage of \emph{Full} optimization is not specific to the original rating question.}

%% file: 05-experiment2.tex
The end-user evaluation assessed end-user perception of the quality of generated layouts.
The designer evaluation focused on the practical design utility of our ASP-based system: could the approach be integrated into designer practice?

\subsection{The Design Tool}
\begin{figure*}[ht]
    \centering
    \includegraphics[width=0.85\linewidth]{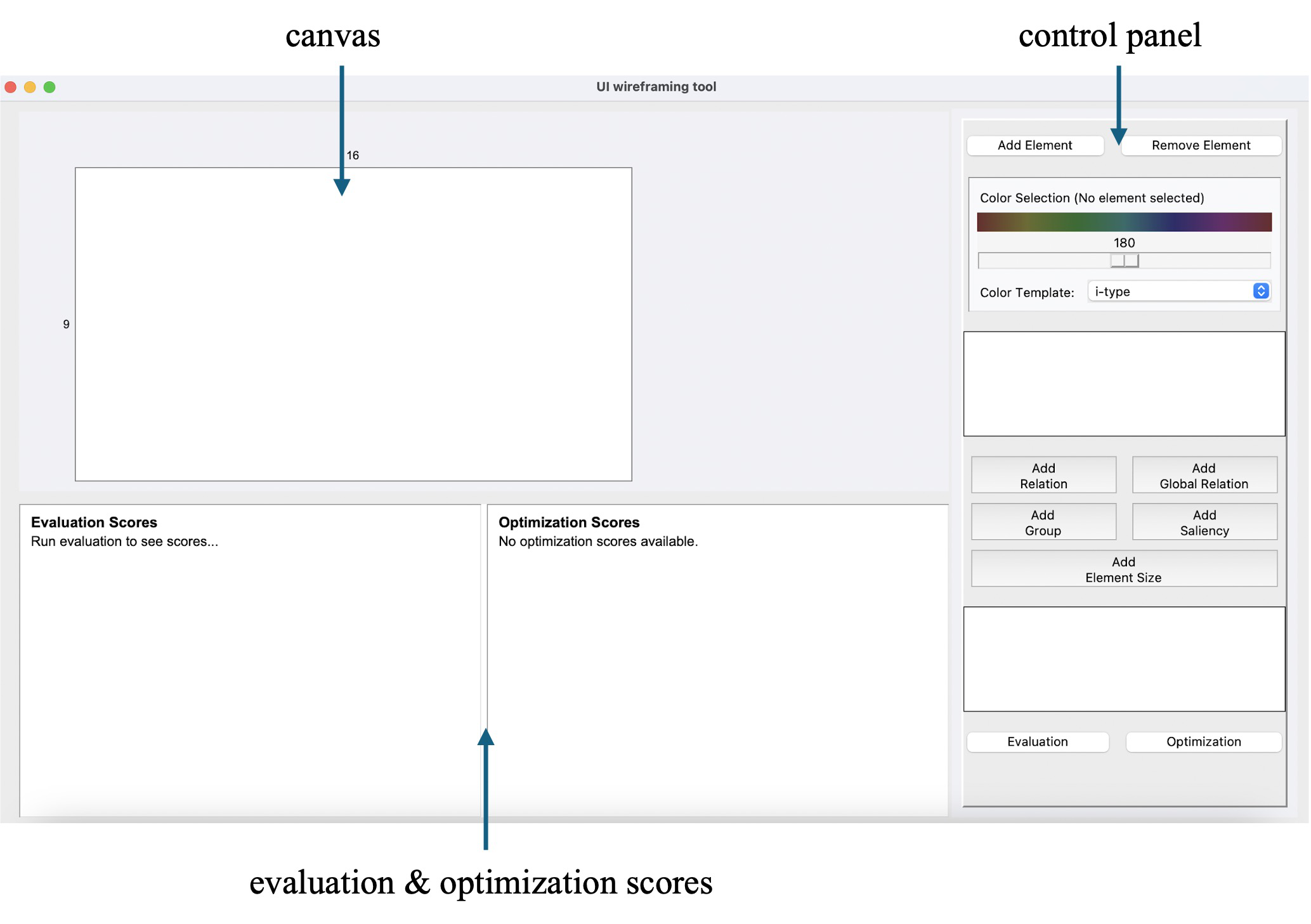}
    \caption{Prototype design tool interface for our ASP-based GUI optimizer. The canvas allows users to add elements and view optimized layouts. The control panel provides access to constraint specifications such as color harmony, grouping, and element sizes and relations. A scoreboard at the bottom displays results for all optimization objectives. Users can also adjust the optimized layout and re-evaluate it, with numerical differences in objective scores shown between the optimized and modified layouts.}
    \label{fig:GUI}
\end{figure*}

For the experiment, we developed a frontend GUI interface that allowed users to provide design specifications (see Figure~\ref{fig:GUI}). \rv{Pilot testing showed that participants struggled to imagine elements and their arrangement without seeing them on a canvas, and providing design specifications directly to the backend model would have required interaction with the ASP implementation. 
The frontend GUI therefore acts as a bridge between designers and the backend model, supporting an iterative workflow in which designers specify requirements, generate and inspect a layout, and revise their specifications~\citep{horvitz1999mixed}. 
The GUI was implemented as a research prototype that exposed the model’s current specification, optimization, and evaluation functions. 
Developing the breadth and level of refinement expected of a commercial design tool was outside the scope of this study.}

Users could define requirements across several constraint categories: element sizes (small, medium, large), element saliency (importance), geometric relations (spatial relationships between elements), grouping (related elements), and color selection (main color and template). 
\rv{Participants used design terminology to specify only task-specific requirements, such as element size, saliency, grouping, and geometric relations (e.g., fixing the header at the top or placing element A to the left of element B) rather than ASP syntax. 
The GUI translated these selections into ASP facts and constraints, while the optimization objectives described in Section 3 remained encoded in the backend.}
The interface included optimization and evaluation functions for generating and assessing layouts.
The optimization function called the backend ASP optimizer, producing layouts based on the specifications of users, while the evaluation function returned quantitative scores for the layout shown on the canvas. 

\rv{Each optimization request returned one layout. If the solver established optimality within 30 seconds, the final optimal layout was displayed; otherwise, the search ended at 30 seconds and the best valid layout found by that point was returned. 
Preliminary testing indicated that the solver generally produced a valid candidate layout within 30 seconds, even when the optimization process was not complete. 
We selected this cutoff to maintain a reasonable interaction pace when limited user specifications resulted in a large search space.}

\subsection{Method}

\subsubsection{Participants}
We recruited $N=8$ participants using snowball sampling, requiring prior experience with UI design. 
\rv{The experiment was designed as an initial, formative evaluation of the design tool rather than a summative usability evaluation. 
Given this focused aim and the relevant design experience of the participants, a sample of eight was considered suitable for obtaining initial feedback about the benefits and limitations of the tool~\citep{malterud2016sample,caine2016local}.}
Participants' ages ranged from 18 to 31 years, and their experience with digital design tools (e.g., Figma, Sketch, Canva) ranged from 1 to 7 years (mean 2.7).
Each participant received a 20~\euro{} gift card as compensation.

\subsubsection{Materials and Tasks}
The experiment was conducted on a MacBook Pro laptop running both the frontend tool and the backend ASP model. \rv{The laptop had an Apple M4 Pro processor (12 CPU cores and 16 GPU cores) and 24 GB of memory and ran macOS 15.7.3. The optimizer used Python 3.9.6 and clingo 5.8.0 with one solving thread, running on the CPU; the GPU was not used.} 
Participants completed three design tasks: a news website (landscape orientation, 8 elements), a weather website (landscape orientation, 9 elements), and a fitness app dashboard (portrait orientation, 7 elements).
Each task included a detailed scenario description and a list of elements for arrangement.
These tasks were designed to ensure variety for a thorough evaluation of the system.
A practice task was also included: designing a to-do list app in portrait orientation with 3 elements.

\subsubsection{Procedure}
Each session lasted about 2 hours.
Participants first received instructions and completed the practice task to familiarize themselves with the tool.
The three main design tasks were presented in counterbalanced order across participants.
For each task, participants followed an iterative process: they reviewed the description and requirements, specified design constraints in the GUI interface, and used the optimization function to generate a layout.
They then evaluated the generated layout and adjusted their specifications as needed. \rvreplace{After participants indicated that they had finished a task, they completed a structured interview about layout quality and the design process. After all tasks, a final interview was conducted to gather reflections on overall experience and to assess the tool's potential integration into participants' workflows.}{Participants completed a structured interview about layout quality and the design process when they indicated that they had finished each task. 
Once all three tasks were complete, they rated their overall satisfaction with the generated layouts on a 5-point scale. 
A final interview then gathered reflections on their overall experience and the potential integration of the tool into their workflows.} 
The interview questions were open-ended, encouraging both positive and critical observations. 
\rv{The satisfaction rating and the interview questions served as formative probes of participants' immediate experiences with the design tool; they were analyzed descriptively and not treated as validated measures of usability or another latent construct. 
The experiment procedure, rating item, and structured interview guide are provided in Appendix~B.}

\subsubsection{Data Analysis}
\rv{The overall satisfaction ratings were summarized descriptively using the mean and standard deviation. 
The researcher who conducted the interviews also analyzed the interview notes using qualitative content analysis \citep{hsieh2005three}. 
The responses were grouped into recurring categories concerning model capabilities, experiences with the frontend GUI, and workflow integration, and the number of participants contributing to each category was recorded. 
Because the analysis was conducted by one researcher, inter-rater reliability was not calculated \citep{mcdonald2019reliability}.}

\subsection{Results}
Across participants, the mean overall satisfaction rating for the generated layouts was 4.0 (SD = 0.76, scale $1$--$5$).
The interviews provided feedback on both the optimization model and participants’ experiences with the frontend. 
Participants valued the model’s comprehensive specification capabilities, with particular praise for geometric relations and grouping functions that supported organizing related elements.
Three participants highlighted the color harmony feature, noting its ability to generate coordinated palettes that reinforced a cohesive visual identity.

\rvdel{Regarding usability, layout outputs were displayed on the canvas within 30 seconds, and no participants reported being disturbed by the response time.} \rv{Regarding their experiences with the GUI, participants described the tool as simple and easy to use. None reported that the response time disrupted their work.}\rvdel{In our internal testing, the system can efficiently handle dozens of elements at the same time, provided that reasonable design constraints are specified.} 
Five participants reported that the tool provided valuable design guidance. 
One participant observed that ``one-click layout generation can serve as a reference for future design iterations.''
Two participants further noted that the constraint specification process encouraged them to articulate design relationships more explicitly compared to direct manipulation tools such as Figma or Canva.

Participants also expressed interest in expanded model capabilities.
A key concern for four participants was the lack of consistency and clarity in the element dimensions (small, medium, large), which mapped to ranges rather than exact values.
Several participants wanted more flexible control over grouping, such as automatic edge alignment within a group.
Some felt that the saliency feature had limited practical impact on their design process.
Two participants requested more element shape variety beyond rectangles, noting potential for creative or artistic applications.

In the final interview, participants were asked about the tool's potential integration into their workflow.
They most often identified value in the early design phase for concept development and brainstorming.
The tool was also seen as useful for rapid prototyping, particularly when content is fixed but layout design time is limited.
One participant noted that it improved their design confidence, commenting that the ``template-like approach provides structure for those with limited design experience.''

\subsection{Model Efficiency Evaluation}
To complement the qualitative feedback from designers, we conducted a supplementary performance experiment evaluating the \rvdel{ scalability and}computational efficiency of our ASP-based optimizer. This experiment aimed to determine whether the model could generate valid, optimized layouts within acceptable timeframes for interactive prototyping workflows across \rvreplace{varying interface complexity levels.}{different interface complexity levels (5, 8, and 18 elements). It examined whether the observed runtimes were compatible with interactive prototyping, rather than providing a comprehensive scalability analysis.}

\subsubsection{Method}
We replicated the three design tasks from the designer evaluation: a news website (landscape orientation), a fitness app dashboard (portrait orientation), and a weather forecast website (landscape orientation). Each task was tested under three complexity conditions, defined by element count: 5 elements (simple layouts), 8 elements (standard layouts), and 18 elements (complex information displays). \rv{The 5- and 8-element conditions fall near the lower and upper ends of the range of 4 to 9 elements used in the end-user stimuli and the three main designer-evaluation tasks. The 18-element condition doubles the upper end of this range to include a more information-dense interface.} These conditions thus range from minimal interfaces to information-dense applications, yielding 9 experimental configurations in total.

The experiment was conducted on the same MacBook Pro laptop \rv{and solver configuration} used in the main designer evaluation to ensure consistency. 
\rv{Unlike the designer evaluation, this experiment invoked the optimizer directly without the GUI frontend and involved no participants. 
The 30-second cutoff used to bound participant waiting time therefore did not apply; each run continued until the solver established the final optimal layout.} 
Each task specification incorporated various constraint combinations, including element sizing, grouping relationships, geometric constraints, saliency, and color schemes, reflecting the full range of specification types available in the design tool. 
\rv{The task themes themselves were not computational factors, since solver runtime depends on the structure of the encoded constraints rather than on their semantic content. 
At each element count, the themes provided three structurally different constraint specifications rather than a single benchmark case. 
Because the number and combination of constraints also changed as elements were added, the evaluation characterizes overall performance across representative configurations rather than isolating the effect of either element count or constraint structure.}
Time measurements were recorded from the moment the optimization function was invoked until the solver identified the final optimal layout. \rv{Each of the nine configurations was timed once.}

\subsubsection{Results}
\rv{Across the nine tested configurations, the optimizer achieved practical efficiency; all optimization runs reached a final optimal layout within 40 seconds (see Table~\ref{tab:task_times}).}
\rvdel{The optimizer achieves practical efficiency across varying layout complexities (see Table~\ref{tab:task_times}). While execution times varied with specific constraints and element counts, all optimization tasks completed successfully within 40 seconds.} For layouts with 5 elements, the mean generation time was 12.3\,s (SD = 12.02), increasing to 18.4\,s (SD = 15.57) for 8 elements, and 21.3\,s (SD = 8.25) for 18 elements. 
\rv{These standard deviations are calculated across the three tasks at each element count and describe differences among their constraint specifications, not variability of the solver on a fixed specification. Because each configuration was timed once, they do not reflect run-to-run variation.}
Complete results, including the final layouts for all nine design tasks and their corresponding constraint specifications, are presented in Appendix~C.

\begin{table}[t]
\centering
\setlength{\tabcolsep}{4pt} 
\begin{tabular}{@{}c c c c@{}}
\hline
\textbf{Element Count} & \textbf{Task 1 (s)} & \textbf{Task 2 (s)} & \textbf{Task 3 (s)} \\
\hline
5  & 25.0 & 1.1  & 10.8 \\
8  & 36.3 & 8.0  & 10.9 \\
18 & 13.1 & 21.2 & 29.6 \\
\hline
\end{tabular}
\caption{Layout optimization times for three design tasks (Task 1: News Website, Task 2: Fitness App Dashboard, Task 3: Weather Forecast) under three complexity conditions (5, 8, and 18 elements).}
\label{tab:task_times}
\end{table}

\subsection{Discussion}
The designer evaluation provides formative evidence that, even with a prototype-level frontend, our ASP-based optimization model can be made accessible to designers and may support design work.
Participants with different levels of experience found the tool simple to use, effective for brainstorming, and valuable for early-stage prototyping, describing it as a ``comforting'' source of objective design guidance.
The constraint-based specification approach helped designers articulate relationships between elements more explicitly than direct manipulation tools, and the solver’s one-click layout generation was viewed as a way to accelerate iteration and provide structured inspiration.
These findings suggest that our approach lowers the barrier for exploring computational layout optimization, demonstrating the feasibility of integrating ASP-based modeling into professional workflows despite the lightweight interface used in this study. 
\rv{Given the formative purpose and the snowball sample of eight designers, these findings characterize the experiences of this participant group rather than support population-level conclusions.}

Feedback on element size control highlights an area for refinement. 
We limited element specification to three coarse size categories (small, medium, large) to reduce complexity, but participants preferred finer-grained and more predictable control. 
This limitation is straightforward to address: ASP can accommodate both precise fixed sizes and flexible ranges, and future versions will expose these options in the interface. 
Participants also suggested improvements to grouping behaviors and element shape variety, both of which are tractable within a declarative modeling framework. 
Additionally, participants saw limited practical impact from the saliency feature, which is intended to make important elements visually distinctive and easier to locate. 
This feedback suggests that the instructions did not sufficiently explain the feature or its influence on generated layouts, but it raises an interesting question on how to use the tool to effectively communicate such cognitive design principles. 

\rv{The current interface also presented only one layout per optimization request. Future versions could present a small, diverse set of high-quality candidates to support comparison during early-stage ideation.} 
More broadly, this study shows that designers are receptive to computational tools that combine explicit reasoning with automated evaluation, and it offers clear guidance for evolving our system into a full-featured design companion that bridges exploratory ideation with rigorous optimization.

The model efficiency experiment supports the feasibility of ASP-based layout optimization across the representative design configurations tested.
With most configurations reaching final optimality within 30 seconds, the observed runtimes are consistent with designers’ reports that response times within this range did not disrupt their workflows.
\rvdel{Notably, increased task complexity did not yield strictly linear time increases. While the 8-element news website task required 36.3 seconds, its 18-element version of the same task completed in just 13.1 seconds. This result reveals that although additional elements typically increase computational complexity, certain constraint combinations can effectively prune the search space, accelerating convergence to optimal solutions.}
\rv{Runtimes varied substantially both within and across element-count conditions.
For example, the 5-element news and fitness tasks required 25.0 and 1.1 seconds, respectively, while the 18-element news task completed in 13.1 seconds, compared with 36.3 seconds for its 8-element version.
Such non-monotonic patterns can occur in solver-based search because runtime depends on how the active constraints shape the search space, rather than on element count alone. 
The present evaluation, however, does not isolate which constraint interactions produced these differences. 
These timings characterize the nine representative configurations tested and should not be interpreted as worst-case bounds or as a general scalability curve.}

These observations highlight a valuable direction for future research.
Future iterations of the design tool could incorporate predictive models to estimate optimization duration based on the active constraints and element definitions.
By identifying computationally expensive combinations in advance, the tool could provide estimated wait times and suggest specification adjustments that improve solver efficiency while preserving design intent.
This feature would enhance both transparency and interactivity, enabling designers to make informed trade-offs between constraint complexity and system performance before initiating optimization.
\rv{In addition, future benchmarking should independently vary element count and the number and type of constraints, repeat each configuration, and include underconstrained, tightly constrained, and unsatisfiable cases.}

%% file: 06-applications.tex
The following applications demonstrate how our ASP-based optimization approach can be used for a variety of design tasks.

\subsection{Menu Optimization}

Menu optimization is a recognized challenge in GUI design.
MenuOptimizer~\citep{bailly2013menuoptimizer} pioneered interactive optimization by combining predictive models of selection performance with ant-colony optimization to help designers place and group items, but it is limited to static scenarios and prespecified objectives.
More recently, integer programming has been combined with information foraging theory to generate efficient hierarchical menus by balancing known-item selection with exploratory search costs~\citep{dayama2021foraging}.
However, no single approach yet provides a flexible optimization pipeline that can adapt to diverse objectives such as memory, visual search, and grouping preferences.
Although our ASP-based optimizer is not specialized for menus, it can generate reasonable menu structures by treating menu items as graphical elements and adding menu-specific constraints, such as requiring horizontal alignment (stacking) within a menu.
Its declarative formulation also supports the addition of new objectives, including cognitive simulation models~\citep{chen2015emergence} or designer-specified constraints, without reformulating the optimization model.

\begin{figure}[!ht]
    \centering
    \includegraphics[width=0.34\linewidth]{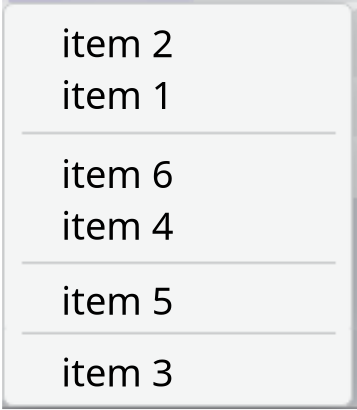}
    \caption{ASP-generated dropdown menu optimized for groups (items 1 and 2) and (items 4 and 6), a cross-group association between items 1 and 6, a negative association between items 2 and 3, and a fixed position for item 5.}
    \label{fig:app1}
\end{figure}

In a short demonstration, we optimized a 6-item dropdown menu by integrating multiple design constraints that reflect key principles of menu design, including frequency-based positioning, perceptual grouping, and spatial organization (see Figure~\ref{fig:app1}).
The menu contained two groups: items 1 and 2 formed one group, and 4 and 6 another.
Additionally, items 1 and 6 were associated and should be placed close together, while items 2 and 3 were specified to remain separate to avoid confusion.
Finally, item 5 had a fixed position (fifth from the top) that had to be preserved.

In future applications, the system could optimize multiple adjacent dropdown menus simultaneously by modeling both within- and between-menu hierarchies ~\citep{dayama2021foraging}.
Objectives such as item associations (from a lookup table or a cognitive model) and cognitive model based visual search could be incorporated to balance the number of items across menus while maintaining overall consistency.

\subsection{Dynamically Appearing Elements}

\begin{figure*}[!ht]
    \centering
    \begin{minipage}{0.4\textwidth}
        \centering
        \includegraphics[width=\linewidth]{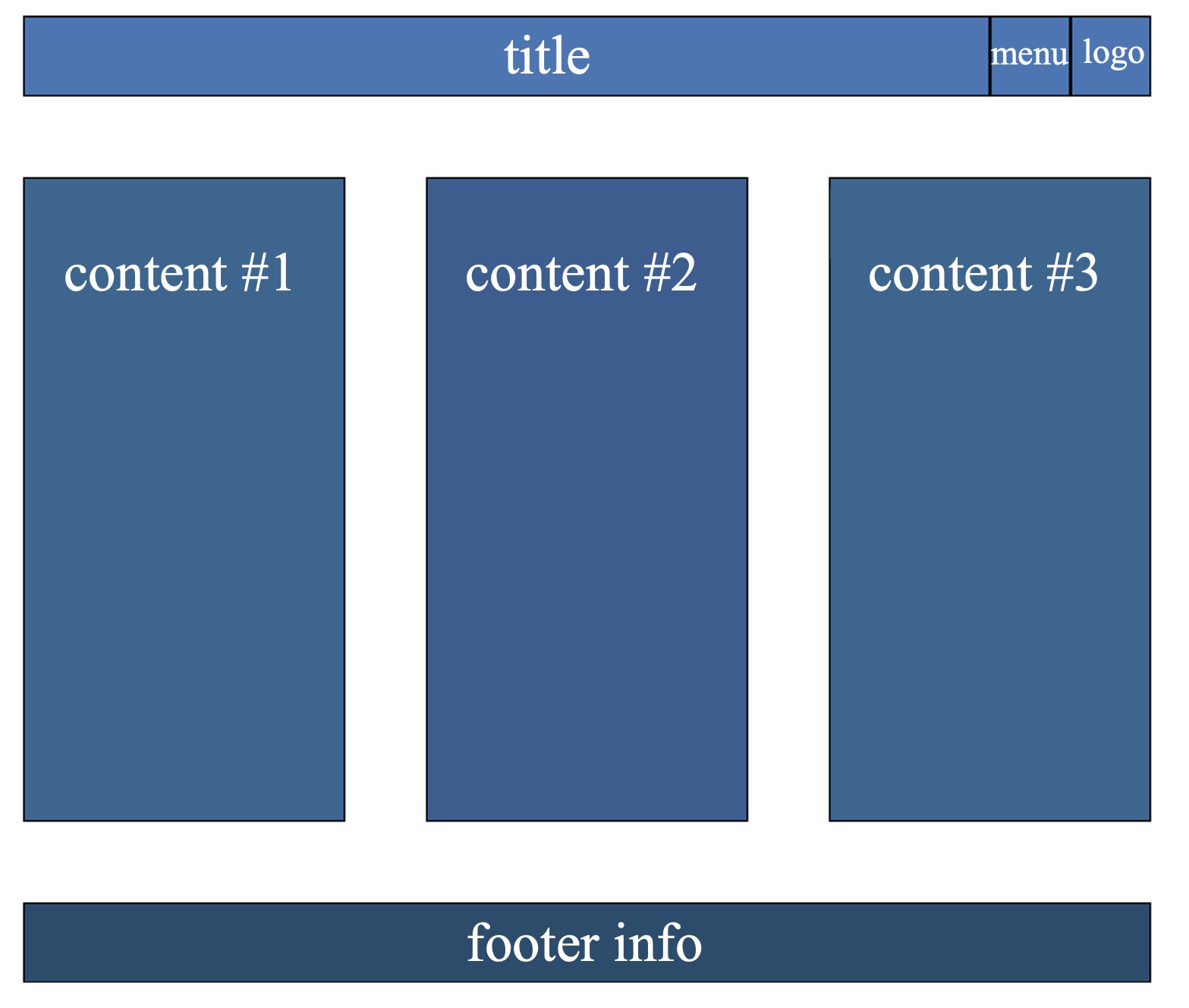}
        \vspace{3pt}
        {\small (a)}
    \end{minipage}\hspace{0.05\textwidth}
    \begin{minipage}{0.4\textwidth}
        \centering
        \includegraphics[width=\linewidth]{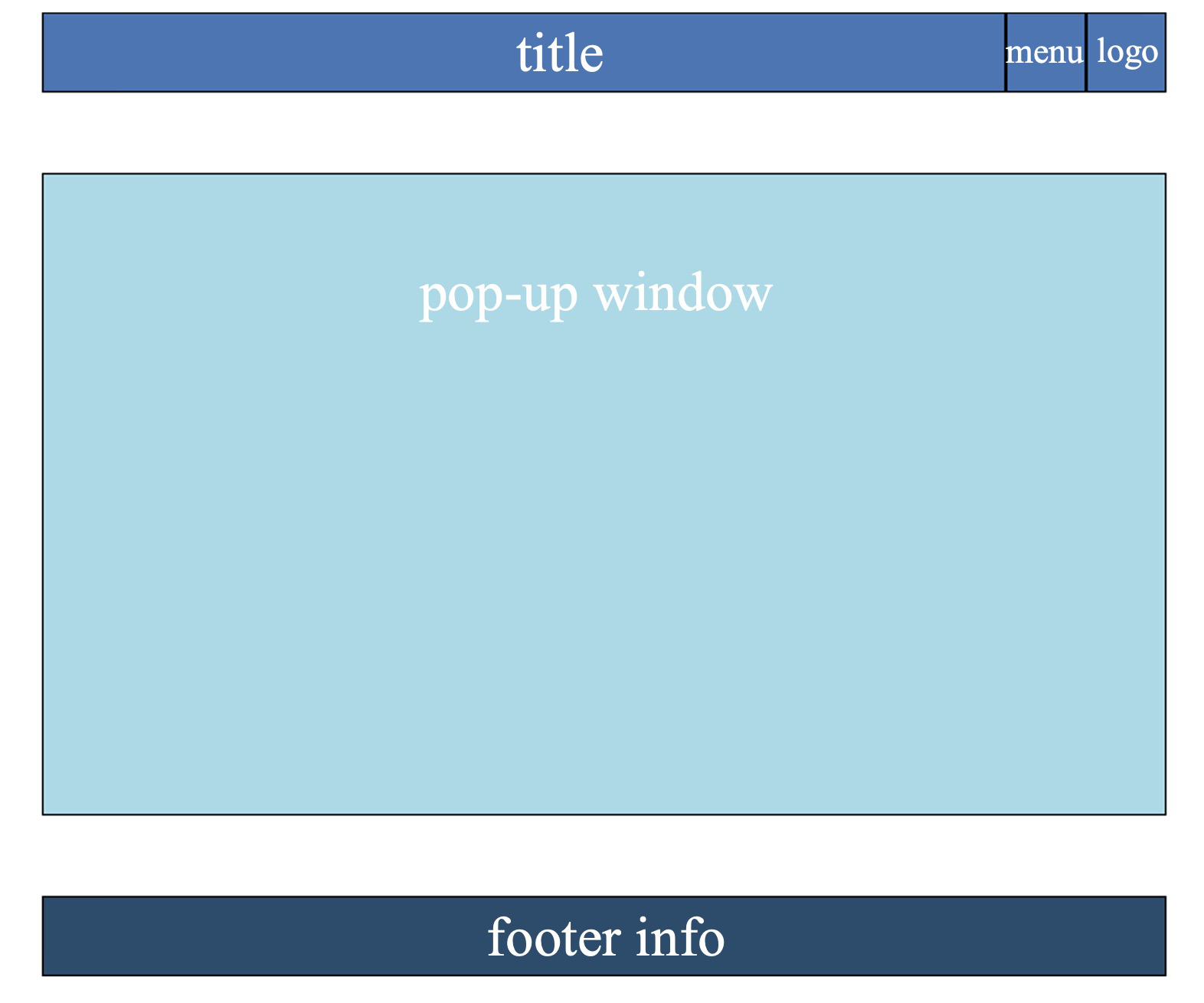}
        \vspace{3pt}
        {\small (b)}
        
    \end{minipage}
    \caption{In a two-stage optimization pass, our ASP solver (a) first generates a layout, and then (b) optimizes a popup window that partially covers existing elements.}
    \label{fig:app2}
\end{figure*}

Dynamically appearing interface elements, such as popups and modals, pose unique design challenges because they must coexist with an already designed base layout without undermining its structure or usability \citep{bahr2011and}.
Despite prior work on solver-based approaches for optimizing GUIs, transient overlays have received little attention.
As a solution to this dynamic design challenge, our ASP framework can implement a two-phase optimization process: in Phase~1, the solver generates a base layout under the full set of constraints, such as optimal positioning, alignment, and aesthetic balance of the static elements.
In Phase~2, the popup is introduced as a new element, with the base layout treated as fixed constraints, and optimization objectives tailored to transient overlays are applied.
These include positioning the popup near the centroid of the base elements to maintain visual balance, maintaining grid alignment with existing components, and scaling the popup to an appropriate relative size to provide salience without overwhelming the layout, as seen in Figure~\ref{fig:app2}.
This application illustrates how ASP naturally supports multi-pass optimization and selective relaxation of constraints, such as permitting overlap for a popup while maintaining alignment and balance, demonstrating the system’s flexibility for handling dynamic interface elements.

\subsection{Color Harmony}
Color selection in interface design presents a fundamental challenge, as designers must balance aesthetic appeal with functional requirements such as contrast and readability. Traditional approaches often rely on intuition or simple heuristics, which makes it difficult to systematically explore palettes that are both harmonious and tailored to specific design requirements. Our ASP-based system addresses this by integrating seven harmonic templates (i-type, V-type, L-type, I-type, T-type, Y-type, and X-type) derived from~\citet{cohenor2006color}. These templates provide computational criteria for palette generation, moving beyond subjective preference toward perceptually grounded color selection.

\begin{figure*}[htbp]
    \centering

    \begin{minipage}{0.3\textwidth}
        \centering
        \includegraphics[width=\linewidth]{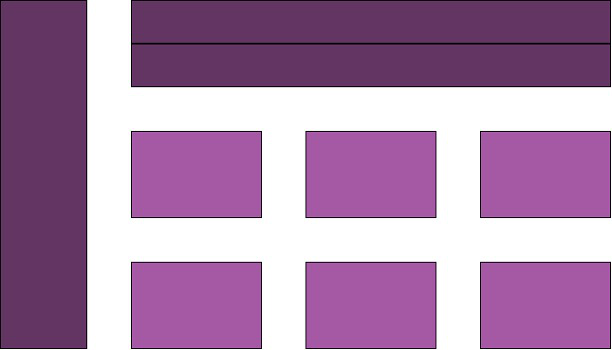}
        \vspace{3pt}
        {\small (a)}
    \end{minipage}
    \hfill
    \begin{minipage}{0.3\textwidth}
        \centering
        \includegraphics[width=\linewidth]{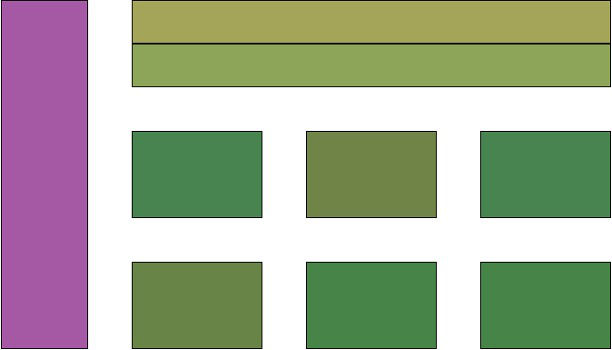}
        \vspace{3pt}
        {\small (b)}
    \end{minipage}
    \hfill
    \begin{minipage}{0.3\textwidth}
        \centering
        \includegraphics[width=\linewidth]{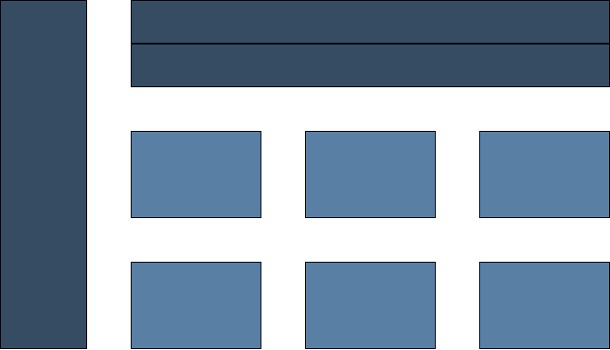}
        \vspace{3pt}
        {\small (c)}
    \end{minipage}

    \caption{Application of harmonic color templates to layout designs. (a) Layout using i-type color template with purple as the designer-selected base color, (b) Y-type template derived from the same purple base but emphasizing complementary green hues for stronger contrast, (c) Layout using i-type color template with blue as the base color.}
    \label{fig:app_color}
\end{figure*}

The optimization process begins with the designer specifying a primary base color and selecting one of the harmonic templates. The ASP solver treats these inputs as constraints, generating a complete palette in which all colors maintain harmonic consistency with the chosen base while simultaneously satisfying other layout objectives such as grouping and saliency. Figure~\ref{fig:app_color} demonstrates the application of different harmonic templates to identical layout structures. In Figure~\ref{fig:app_color}(a), the layout uses an i-type template with purple as the base color, creating subtle monochromatic variations within a narrow \(18^{\circ}\)~sector. Figure~\ref{fig:app_color}(b) shows the same layout with a Y-type template applied to the same purple base, but now most elements are drawn from the complementary green sector, combining the narrow base sector with the wider \(93.6^{\circ}\) sector to increase visual contrast. When the primary color changes to blue while maintaining the i-type template (see Figure~\ref{fig:app_color}(c)), the system automatically adapts the entire palette while preserving harmonic relationships.

\subsection{Hierarchical Grouping}

\begin{figure}[!ht]
    \centering
    \includegraphics[width=0.9\linewidth]{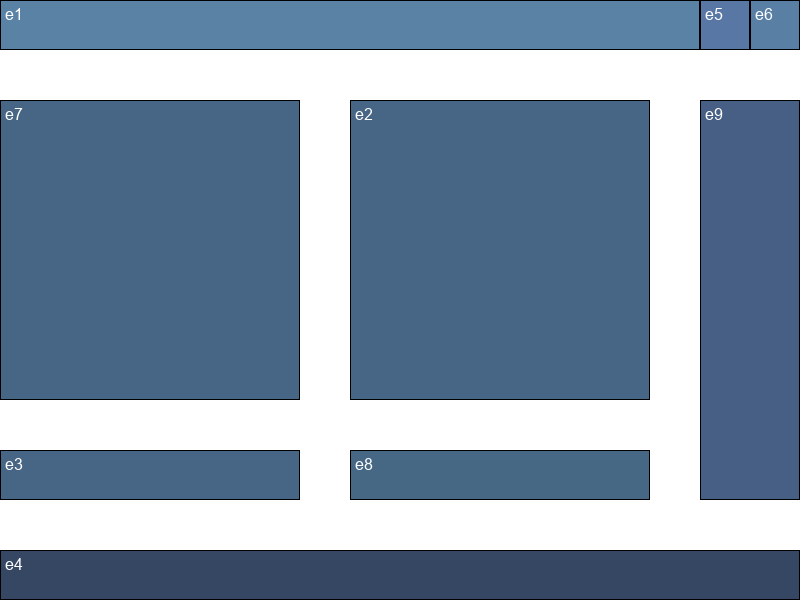}
    \caption{A hierarchical layout generated by the solver, where the central content region (parent group) organizes two nested subgroups (e3, e7 and e2, e8, e9) into compact clusters.}
    \label{fig:app_group}
\end{figure}

Hierarchical visual structures are essential for organizing complex information. To address this, we extended our grouping module to support recursive hierarchical relationships. As illustrated in Figure~\ref {fig:app_group}, the central content region is defined as a parent group containing two nested subgroups: elements e7 and e3 form one cohesive cluster, while e2, e8, and e9 comprise a second cluster. The ASP solver automatically balances competing objectives, maintaining global alignment—such as aligning the left and right edges of the title group with the middle content group—while simultaneously preserving the local compactness of each subgroup. This demonstrates our model's ability to naturally handle the multi-level visual hierarchies common in information-dense interfaces.

\subsection{Individualizing GUIs}
\label{sec:who}

In GUI optimization, objectives based on cognitive models with psychologically grounded individual parameters can be adapted to account for user differences.
To demonstrate this, we conducted a computational experiment optimizing a layout for pointing accuracy.
In addition to the models reported above, we implemented the WHo model of pointing~\citep{guiard2015mathematical}, an extension of 
Fitts’ law that incorporates endpoint variability and accuracy requirements across user populations.
By varying the parameter $k_\alpha$, we simulated users with different capacities for fast and accurate pointing.
We set $k_\alpha = 0.06$ for accurate users and $k_\alpha = 0.5$ for users with motor difficulties, such as essential tremor~\citep{sarcar2018ability}.

\begin{figure*}[htbp]
    \centering
    \begin{minipage}{0.43\textwidth}
        \centering
        \includegraphics[width=\linewidth]{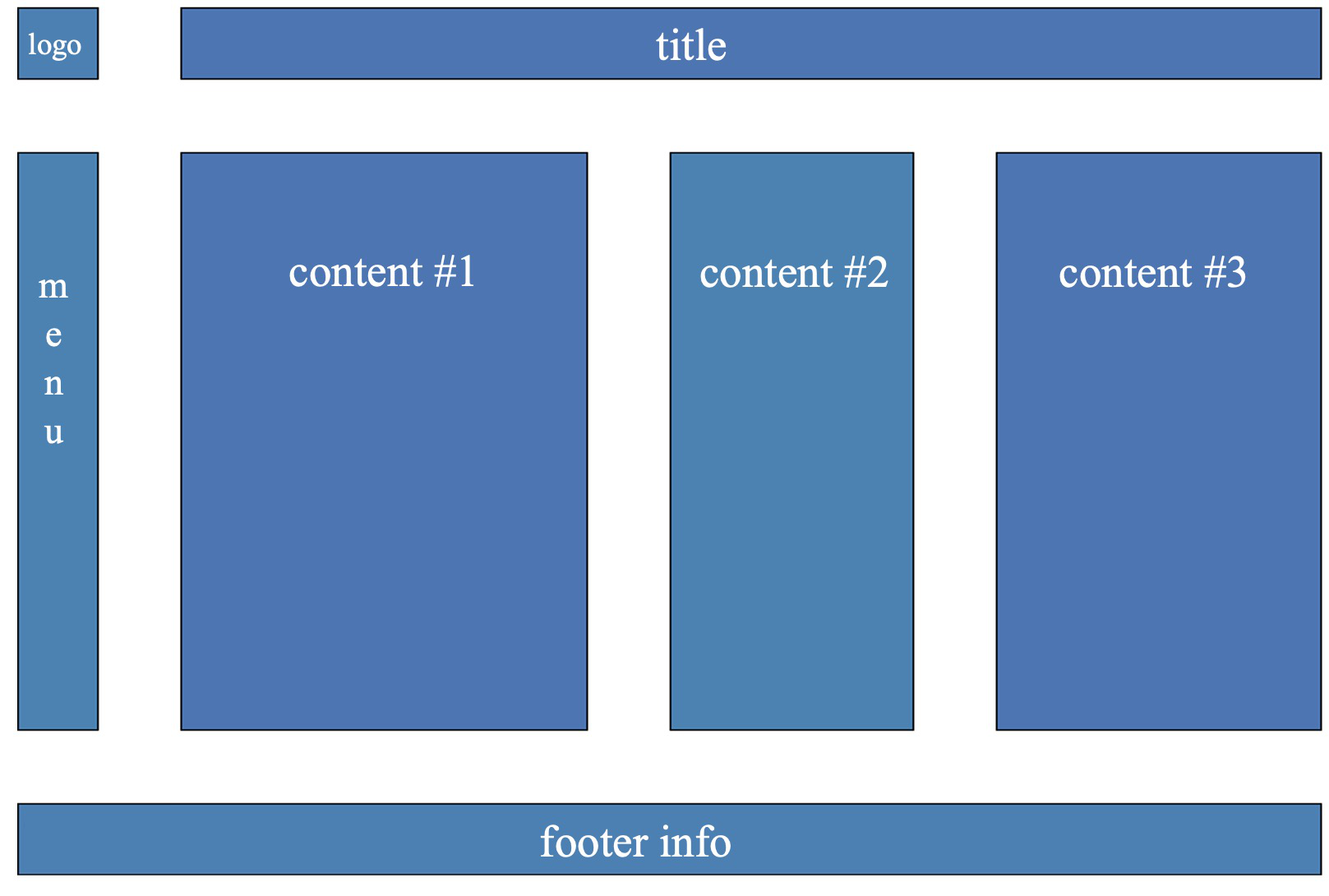}
        \vspace{3pt}
        {\small (a) Layout optimized for accurate users.}
    \end{minipage}\hspace{0.05\textwidth}
    \begin{minipage}{0.43\textwidth}
        \centering
        \includegraphics[width=\linewidth]{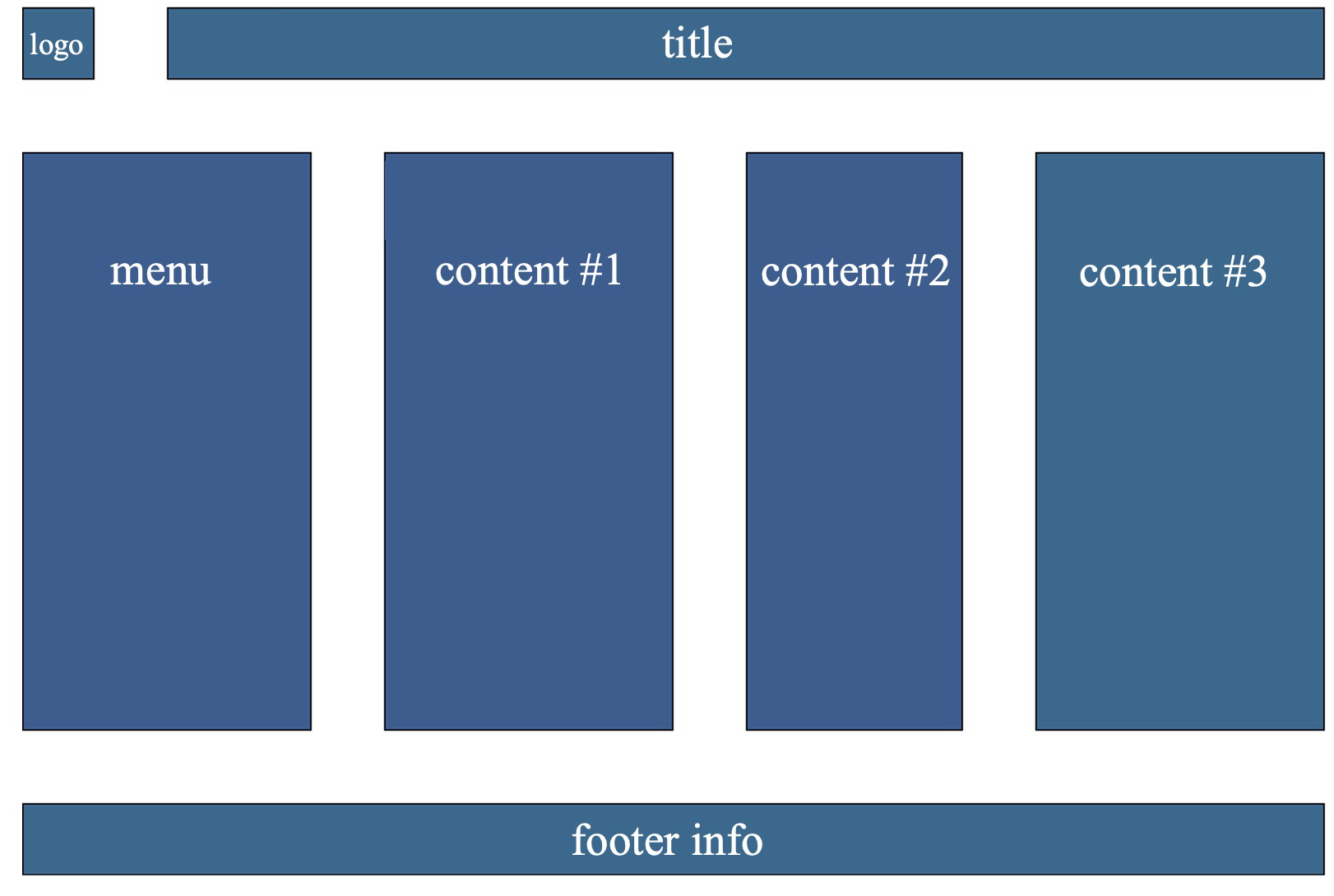}
        \vspace{3pt}
        {\small (b) Layout optimized for users with motor difficulties.}
    \end{minipage}
    \caption{Individually adapted layouts focusing on the size of the ``menu'' element. (a) For accurate users, the WHo model predicts reliable pointing performance even with a relatively small width, allowing the optimizer to maintain strict grid alignment. (b) For users with reduced pointing accuracy, the optimizer prioritizes a larger target, trading off grid alignment to improve accessibility.}
    \label{fig:app4}
\end{figure*}

In our model, we added an objective to minimize pointing error for a specified element.
This was implemented by calling an external Python function from the ASP model, passing the current design candidate as input and returning the WHo-predicted error.
The optimizer integrated this information with existing objectives such as grid alignment, visual grouping, and spatial balance.
We specified that the model should minimize pointing error to the element labeled ``menu.''
With different values of $k_\alpha$, distinct designs emerged.
As shown in Figure~\ref{fig:app4}(a), the optimal layout for accurate users follows strict grid alignment, producing a ``menu'' element of modest width.
Accurate users can reliably hit this target, but according to the WHo model, users with motor difficulties would find the small width cumbersome.
In response, the optimizer relaxes the grid alignment objective to enlarge the ``menu'' element, producing a layout better suited to users with reduced pointing accuracy. \rv{This example demonstrates the model's extensibility; whether such adaptation improves performance or experience for users with motor impairments requires empirical evaluation.}

\subsection{Design for Different Screen Sizes and Platforms}
\begin{figure*}[htbp]
    \centering
    \begin{minipage}{0.4\textwidth}
        \centering
        \includegraphics[width=\linewidth]{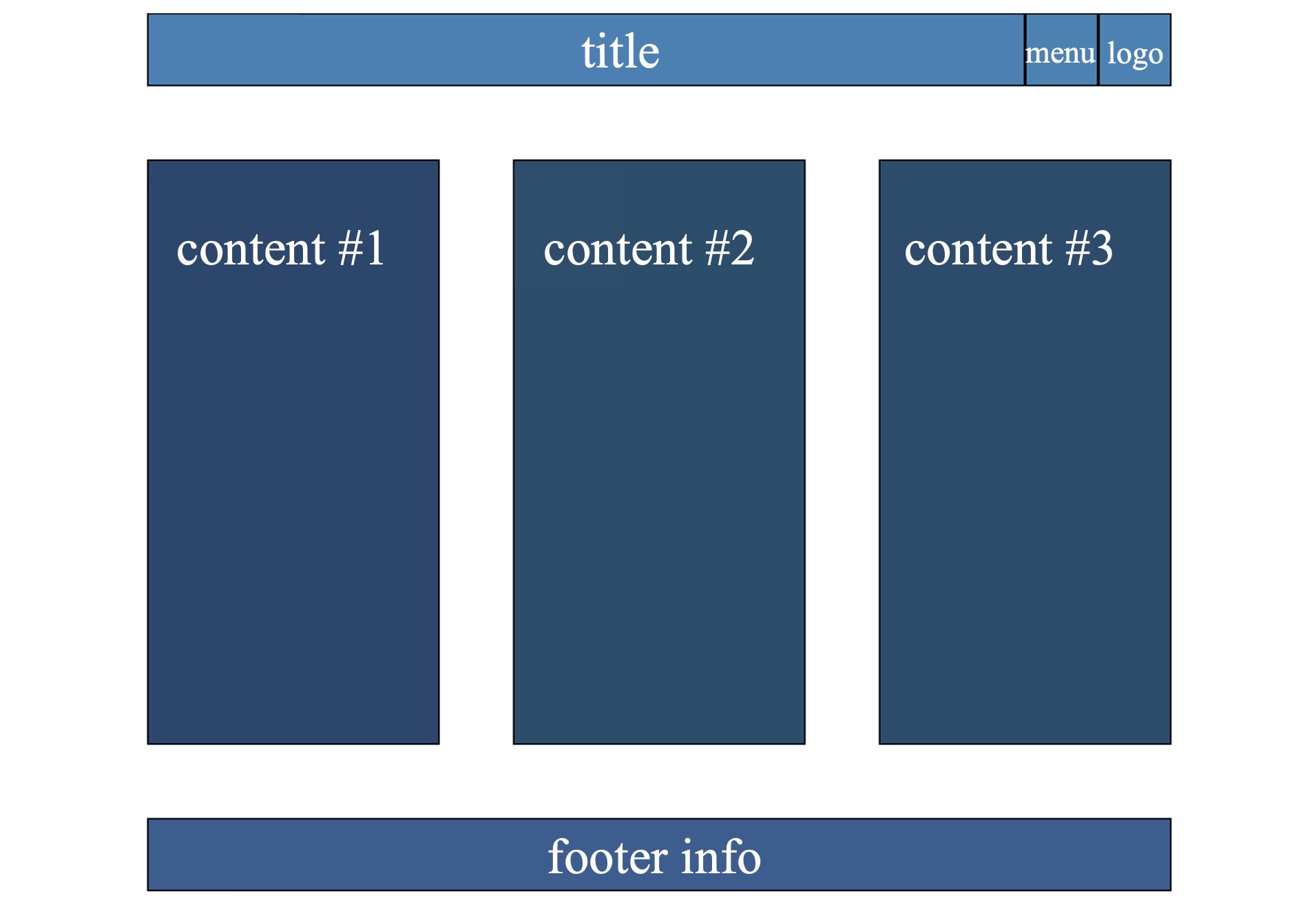}
        \vspace{3pt}
        {\small (a)}
    \end{minipage}\hspace{0.05\textwidth}
    \begin{minipage}{0.25\textwidth}
        \centering
        \includegraphics[width=\linewidth]{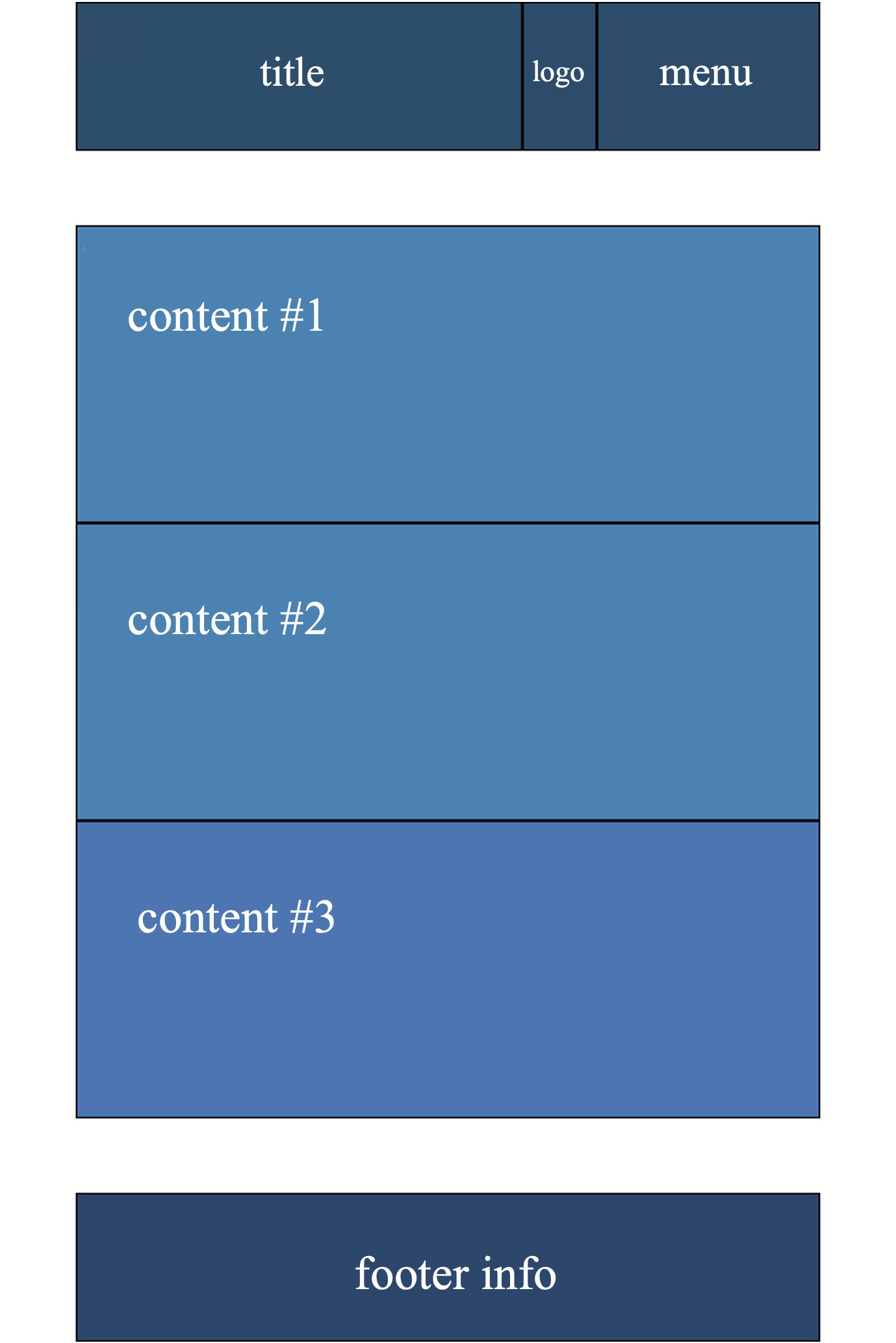}
        \vspace{3pt}
        {\small (b)}
    \end{minipage}
    \caption{Multi-platform layout adaptation. (a) Desktop layout exploits landscape orientation with horizontal content arrangement for efficient screen utilization. (b) Mobile layout reconfigures elements into a vertical stack suitable for portrait viewing, with more tightly packed content areas and an enlarged menu target.}
    \label{fig:app5}
\end{figure*}

Interface development often requires designing for different screen sizes and orientations while preserving both usability and visual coherence across platforms. Our ASP-based approach naturally supports multi-platform optimization by treating screen dimensions as input constraints, allowing the same set of optimization objectives to be seamlessly reapplied across different scenarios. To demonstrate cross-platform adaptation, we optimized a 7-element interface for both desktop (landscape) and mobile (portrait) orientations. As shown in Figure~\ref{fig:app5}, the desktop configuration arranges content elements horizontally to exploit available width, positioning the three main content areas side-by-side below the title bar.
 
In contrast, the mobile layout reorganizes these same elements into a vertical stack that accommodates the narrower viewport while preserving both the header and footer structures. The three content areas become more tightly packed to optimize the limited vertical space while maintaining readability. Notably, the “menu” element appears larger—a direct consequence of the WHo model of pointing introduced in Subsection \ref{sec:who}. This enlargement ensures adequate target size to maintain pointing accuracy and account for increased motor control demands on touchscreen devices. This application demonstrates the value of declarative constraint specification in addressing multi-platform design challenges. By modifying screen dimensions and adding platform-specific constraints, the solver can generate optimal spatial arrangements for each target configuration using the same core set of optimization objectives.

\subsection{Visual Search Simulation}
\label{sec:visualsearch}
This example illustrates the capability introduced abstractly in Subsection \ref{sec:simulators}: invoking external cognitive simulators as optimization objectives within the optimizer.
Instead of formalizing a new explicit objective in ASP, we employ a computational model of visual search to evaluate candidate layouts based on predicted interaction performance~\citep{sourulahti2026modeling}.
The purpose is to demonstrate that psychologically grounded simulators can substitute for hand-crafted design rules inside a solver-based optimization loop.
The demonstration concentrates on visual grouping, a well-established design principle that is traditionally implemented as an explicit geometric constraint derived from Gestalt principles~\citep{wagemans2012gestalt,graver2012best}.

The demonstration compares two optimization runs with identical inputs, constraints, and priorities.
The only difference is whether a simulator-based visual search objective is included.
In both conditions, the explicit solver-based grouping objective is disabled.
In the first condition, the standard optimization is augmented with a minimization objective based on visual search time predicted by an external simulator.
In the second condition, this external objective is omitted.

\begin{figure*}[!ht]
\centering
\setlength{\tabcolsep}{5pt}   
\renewcommand{\arraystretch}{1.2} 

\begin{tabular}{
  >{\raggedright\arraybackslash}m{0.22\textwidth}
  >{\centering\arraybackslash}m{0.36\textwidth}  
  >{\centering\arraybackslash}m{0.36\textwidth}  
}
 & \textbf{Simulator used} & \textbf{Simulator not used} \\[2mm]

\textbf{Optimized layout} &
\includegraphics[width=0.94\linewidth]{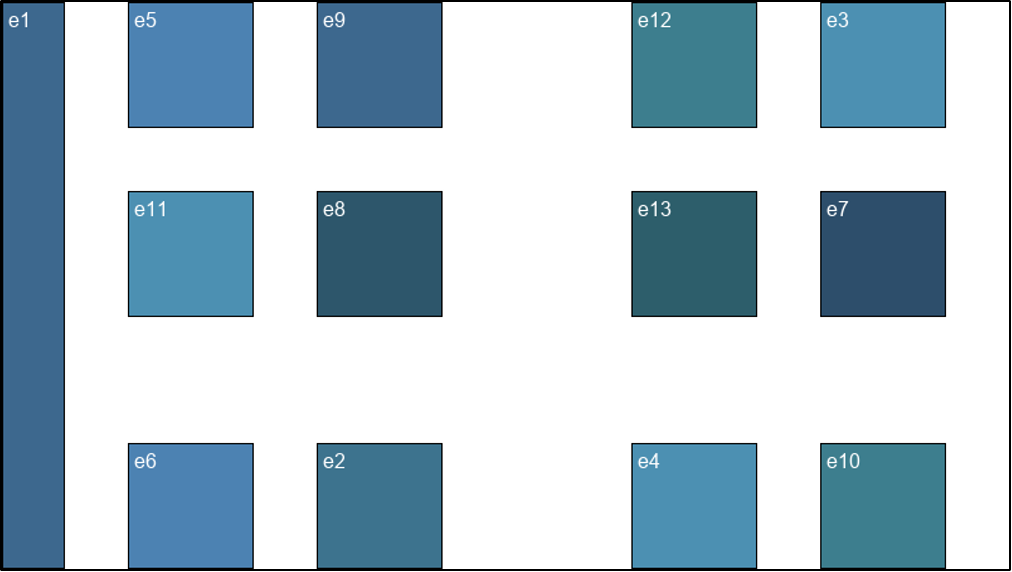} &
\includegraphics[width=0.94\linewidth]{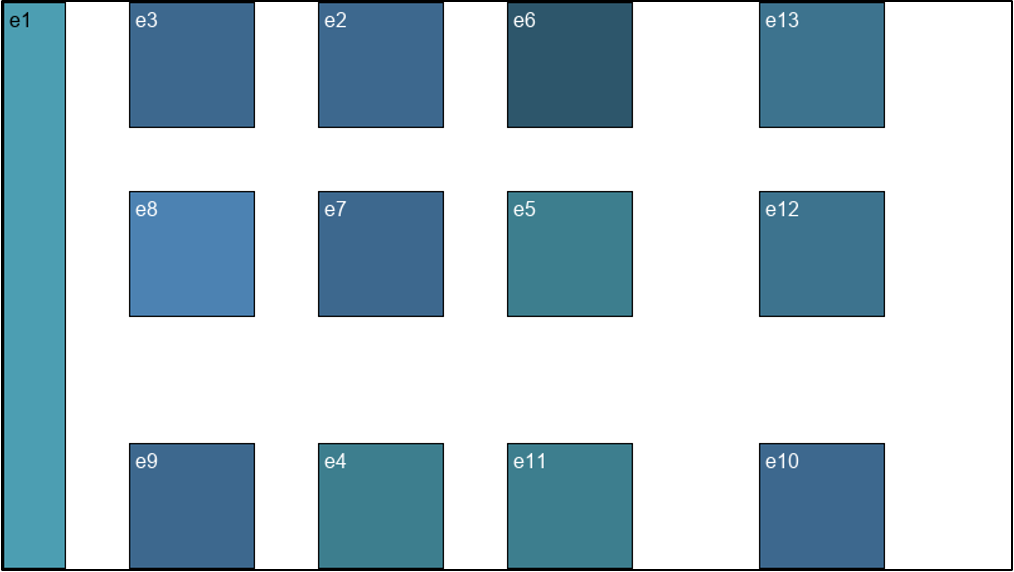} \\[5mm]

\textbf{Simulation eye path} &
\includegraphics[width=\linewidth]{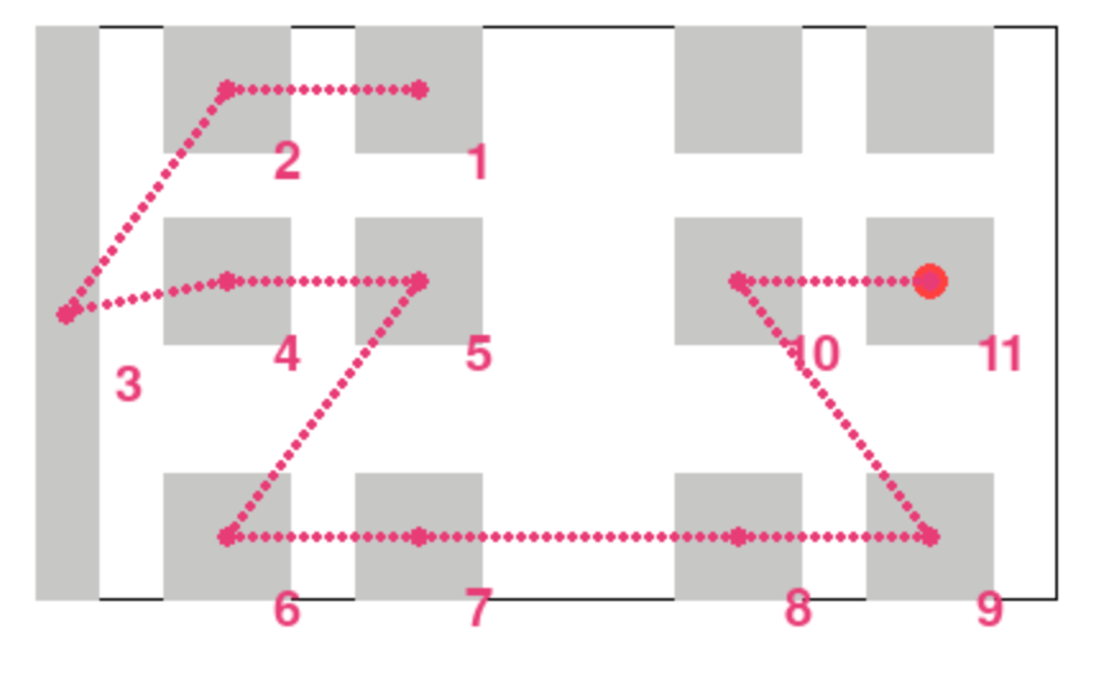} &
\includegraphics[width=\linewidth]{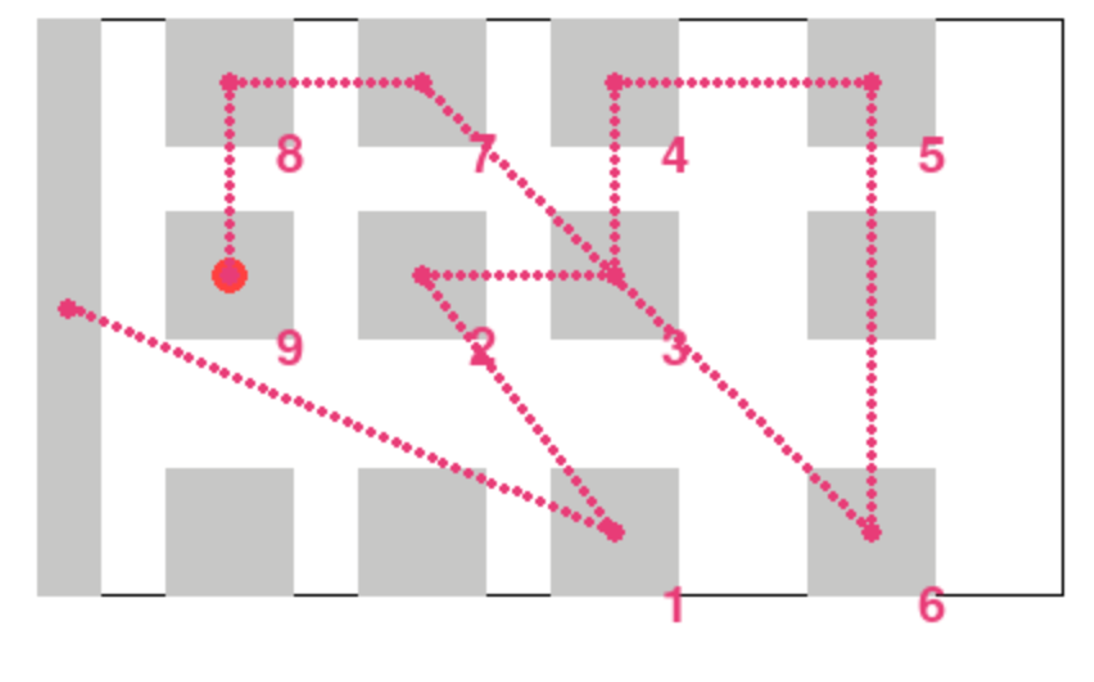} \\
\end{tabular}

\caption{The optimizer can use an external cognitive simulator, such as a visual search model, during layout optimization.
Fixation paths show how the simulator, grounded in psychological accounts of visual grouping, exploits visual structures that emerge when minimizing simulator-based visual search time.
In contrast, post hoc simulation of layouts not optimized with the simulator shows less structured and slower search.}

\label{fig:visualsearch}
\end{figure*}

Results of the optimization are shown in Figure \ref{fig:visualsearch}.
The optimizer converges to an example layout (left) that exhibits spatial grouping, despite the absence of any explicit grouping rules in the ASP model.
This occurs because the visual search model exploits emerging visual groups, which mitigate visual memory limitations.
In contrast, the scanpath for the second layout, which is not optimized for the visual search model, is less structured.
On average (over 5000 simulation runs on the optimized layouts), the predicted search time for the first layout was $2.5\,\mathrm{s}$, compared to $3.0\,\mathrm{s}$ for the second.
To verify that the advantage of the search-model-based optimization was not due to a specific design input, we conducted additional comparisons with varying element specifications and numbers of elements.
Across all comparisons, using the search model as an optimization objective produced visible grouping, and the predicted search times were consistently lower for these layouts.



This demonstration shows how intuitive design principles can be grounded in executable cognitive models rather than encoded as fixed heuristics.
Gestalt grouping is no longer treated as a prescriptive rule; instead, it emerges as a consequence of optimizing predicted behavior under a cognitive model of the human visual system.
The example indicates that solver-based and simulator-based objectives are not competing approaches, but can be systematically substituted or combined.
This integration allows design optimization to function as a bridge between interface design and computational psychology, where theories of perception and interaction can be examined through their design implications.
Our framework permits arbitrary simulators to participate in optimization, supporting a shift from rule-based design guidance toward model-based, testable, and theoretically grounded design reasoning.



%% file: 07-discussion.tex
\subsection{General Discussion}

This paper demonstrates that Answer Set Programming (ASP) provides a strong foundation for formalizing GUI design based on established principles.
Our declarative framework encodes alignment, grouping, saliency, compactness, balance, color harmony, and element dimensions and relations as explicit optimization objectives, while remaining extensible to black-box cognitive models and domain-specific constraints.
Across empirical and application-based evaluations, we show that the approach is effective: end users consistently rated layouts generated with the full model as visually superior to those produced under the \emph{Grid} and \emph{Random} ablation conditions, and designers report that the system supports early-stage ideation, provides objective design feedback, and integrates naturally into prototyping workflows.

The application demonstrations illustrate the flexibility of our approach: it supports adaptations for individual motor abilities, optimization of menus with custom associations, incremental updates for dynamically appearing elements, color design, and layouts for different screen sizes and orientations.
Because the objectives in our framework are explicit, the system can also explain why particular design choices are made, offering designers a transparent and interrogable optimization pipeline.
Together, these contributions advance computational design beyond heuristic layout engines and purely data-driven methods, positioning ASP as a next-generation paradigm for design automation and evaluation.

\subsection{Limitations and Future Work}

In this paper, our tool remains a research prototype.
Although the designer experiences reported here were promising, further work is required to support widespread and expert adoption.
Future development should prioritize richer grouping controls, better communication of some of the objectives, shape diversity, and direct manipulation features to complement constraint specification.
We also plan to extend optimization objectives beyond static layouts into dynamic interaction, content changes, and personalization.
This will involve incorporating more advanced cognitive models capable of simulating step-by-step interaction with generated layouts.
A logical next step is integration with front-end technologies, such as generating responsive HTML and CSS directly from solver outputs.
Longer-term studies with professional designers and educators will be needed to validate the approach at scale and refine workflows that integrate declarative modeling with creative design practices.

\rv{
Individual differences relevant to interface optimization extend beyond perceptual and motor abilities to personality-related preferences and interaction patterns.
Personality has been associated with preferences for interface properties such as color, information density, navigation, typography, and layout, but available evidence remains fragmented and does not yet support comprehensive design guidelines~\citep{alves2020incorporating}.
Some recent work has sought to derive such guidance empirically: one study clustered users according to Five-Factor Model profiles and applied association-rule mining to identify profile-specific preferences for GUI features \citep{alves2022examining}.
Another used clustering and the Apriori algorithm to construct interfaces for different personality profiles and found, using eye-tracking measures, that matching the interface to these profiles improved users' visual experience \citep{sarsam2018towards}.
Our declarative framework provides a natural way to operationalize this line of research: empirically derived associations between user profiles and design features could be represented as conditional preferences or weighted optimization objectives, allowing the solver to generate different layouts from the same functional specification.
Demonstrations in this paper instantiate individual adaptation through a computational model of motor ability rather than personality, and our empirical evaluations assess population-level judgments; consequently, whether personality-conditioned ASP optimization improves preference, performance, or satisfaction remains an important question for future evaluation.
}

\rv{
A further direction concerns how the scope and selection of optimization objectives should be determined.
Rather than merely expanding the model through an ad hoc accumulation of individual metrics, comprehensive design systems such as Material Design~\citep{materialdesign} could provide a structured foundation for reusable ASP knowledge bases.
Such systems already distinguish reusable design tokens, component specifications, interaction states, accessibility requirements, and more contextual recommendations.
In an ASP formalization, tokens and component properties could be represented as facts, requirements such as minimum target sizes as hard constraints, and defeasible recommendations concerning spacing, hierarchy, or emphasis as defaults or weighted objectives.
Different subsets and priorities could then be activated for particular platforms, brands, or user profiles, avoiding the assumption that every available objective should be optimized simultaneously.
However, porting a design system would be a considerable knowledge-engineering task because many guidelines combine quantitative specifications with examples, contextual qualifications, and natural-language rationale. Large language models (LLMs) could assist by extracting candidate concepts and translating such guidance into preliminary ASP rules.
This approach is supported by prior work showing that LLMs can generate nontrivial answer-set programs from natural-language descriptions~\citep{ishay2023leveraging}.
Crucially, the LLM should serve as a rule-authoring assistant rather than an authority: generated rules could retain links to their source passages and be subjected to expert review, consistency checking, and solver-based tests using positive and counterexample layouts. This combination would exploit the linguistic coverage of LLMs while preserving the explicitness, inspectability, and formal guarantees that motivate the ASP approach.
}

\subsection{Conclusion}

We present the first application of ASP to GUI layout optimization, demonstrating how symbolic modeling and solver guarantees can be combined to generate layouts informed by psychological principles.  
The full system, including solver code, example constraints, and experimental materials, will be released as open source upon publication to support reproducibility and community-driven development.  
We anticipate that our work stimulates further exploration of declarative, explainable, and cognitively grounded optimization tools for interface design.